\documentclass[bibyear]{aa}
\usepackage{psfig}
\usepackage{graphicx,lscape,longtable}
\usepackage{subfigure}
\usepackage{rotating}
\usepackage{natbib}

\usepackage{appendix}

\begin{document}

\title{The nature of fifty Palermo \textit{Swift}-BAT hard X-ray objects through optical spectroscopy
\thanks{Based on observations obtained from the following observatories: 
Cerro Tololo Interamerican Observatory (Chile);  
Astronomical Observatory of Bologna in Loiano (Italy);  
Observatorio Astron\'omico Nacional (San Pedro M\'artir, Mexico); 
Radcliffe telescope of the South African Astronomical Observatory (Sutherland, South Africa); 
Sloan Digital Sky Survey;  
Observatorio del Roque de los Muchachos of the Instituto de Astrof\'isica de Canarias (Canary Islands, Spain) and 
New Technology Telescope (NTT) of La Silla Observatory, Chile.}}


\author{A.F. Rojas\inst{1}, N. Masetti\inst{2,}\inst{1}, D. Minniti\inst{1,}\inst{3,}\inst{4}, E. Jim\'enez-Bail\'on\inst{5}, V. Chavushyan\inst{6}, G. Hau\inst{7}, V.A.  McBride\inst{8,}\inst{9}, L. Bassani\inst{2}, A. Bazzano\inst{10}, A.J. Bird\inst{11}, G. Galaz\inst{12}, I. Gavignaud\inst{1}, R. Landi\inst{2}, A. Malizia\inst{2}, L. Morelli\inst{13,}\inst{14}, E. Palazzi\inst{2}, V. Patiño-\'Alvarez\inst{15,}\inst{6}, J.B. Stephen\inst{2} and P. Ubertini\inst{10}
}
\institute{ 
Departamento de Ciencias F\'isicas, Universidad Andr\'es Bello, Campus La Casona, Fern\'andez Concha 700, Santiago, Chile
\and 
INAF -- Istituto di Astrofisica Spaziale e Fisica Cosmica di Bologna, Via Gobetti 101, I-40129 Bologna, Italy 
\and 
Millennium Institute of Astrophysics, Av. Vicu\~na Mackenna 4860, 782-0436 Macul, Santiago, Chile
\and 
Vatican Observatory, I-00120, Vatican City State
\and  
Universidad Nacional Aut\'onoma de M\'exico, Apartado Postal 70-264, 04510 M\'exico D.F., Mexico
\and  
Instituto Nacional de Astrof\'isica, \'Optica y Electr\'onica, Apartado Postal 51-216, 72000 Puebla, Mexico
\and  
European Southern Observatory, Ave. Alonso de Cordoba 3107, Casilla 19001, Santiago, Chile
\and 
Department of Astronomy, Astrophysics, Cosmology and Gravity Centre, University of Cape Town, Private Bag X3, Rondebosch 7701, South Africa
\and  
South African Astronomical Observatory, PO Box 9, Observatory 7935, South Africa
\and  
INAF-- Instituto di Astrofisica e Planetologia Spaziali, via Fosso del Cavaliere 100, I-00133 Roma, Italy
\and  
School of Physics \& Astronomy, University of Southampton, Southampton, SO17 1BJ, UK
\and  
Instituto de Astrof\'isica, Facultad de F\'isica, Pontificia Universidad Cat\'olica de Chile, Casilla 306, Santiago 22, Chile
\and  
Dipartimento di Fisica ed Astronomia "G. Galilei", Universit\`a di Padova, vicolo dell'Osservatorio 3, 35122 Padova, Italy
\and 
INAF - Osservatorio Astronomico di Padova, Vicolo dell'Osservatorio 5, 35122 Padova, Italy
\and
Max-Planck-Institut für Radioastronomie, Auf dem Hügel 69, 53121 Bonn, Germany
}

\offprints{A.F. Rojas (\texttt{ale.rojaslilayu@gmail.com)}}


\abstract{We present the nature of 50 hard X-ray emitting objects unveiled through an optical spectroscopy campaign performed at seven telescopes in the northern and southern hemispheres. These objects were detected with \textit{Swift}-BAT and listed as of unidentified nature in the 54-month Palermo BAT catalogue. In detail, 45 sources in our sample are identified as active galactic nuclei of which, 27 are classified as type 1 (with broad and narrow emission lines) and 18 are classified as type 2 (with only narrow emission lines). Among the broad-line emission objects, one is a type 1 high-redshift quasi-stellar object, and among the narrow-line emission objects, one is a starburst galaxy, one is a X-ray bright optically normal galaxy, and one is a low ionization nuclear emission line region. We report 30 new redshift measurements, 13 confirmations and 2 more accurate redshift values. The remaining five objects are galactic sources: three are Cataclismic Variables, one is a X-ray Binary probably with a low mass secondary star, and one is an active star.}


\titlerunning{The nature of 50 sources in the Palermo \textit{Swift}-BAT catalogue}
\authorrunning{A.F. Rojas et al.}

\maketitle

\fancyhead[L]{\thepage}

\section{Introduction}

X-ray and $\gamma$-ray satellites have detected a large number of high energy sources, allowing us to explore and study in detail the hard X-ray band, above 15 keV. 
Specifically, the \textit{INTEGRAL} (Winkler et al. 2003) and \textit{Swift} (Gehrels et al. 2004) satellites are able to detect hard X-ray sources with positional accuracy better than 5 arcmin with their instruments IBIS (Ubertini et al. 2003) and BAT (Barthelmy 2004), respectively.
There are more than 1200 hard X-Ray sources detected by these satellites, mostly Active Galactic Nuclei (AGN, $\sim$ 60\%) followed by X-Ray Binaries (XRBs, $\sim$ 15\%), Cataclismic Variables (CVs, $\sim$ 5\%) and finally active stars (e.g. Baumgartner et al. 2013), but more than 20\% of the whole set of detections has no precise classification or has no obvious counterpart at other wavelengths and therefore they cannot be associated with any known class of high-energy emitting objects.  

Although the cross-correlation with catalogues of surveys at other wavelengths (soft X-ray, radio or optical bands) is important for improving the positional accuracy of detected objects and poinpointing longer-wavelength candidate counterparts of these sources, only with the detection of emission spectral features especially in the optical we can confirm these associations and reveal the actual nature of these unclassified objects. 

The identification and classification of these newly detected objects is important as it is useful for instance for binary system studies (Coleiro \& Chaty 2013; Scaringi et al. 2010; Landi et al. 2009) and for population analysis of extragalactic objects (e.g. Ricci et al. 2016; Ajello et al. 2008). Moreover, one can find interesting classes of X-ray emitting objects, for example types of unusual AGN obscured by local absorption in their galaxies and Compton-thick active galactic nuclei with a line-of-sight hydrogen column $N_{H}$ $>$ $10^{24}$ $cm^{-2}$ due to the torus of dust surronding the supermassive black hole (Comastri et al. 2002; Malizia et al. 2009; Lutovinov et al. 2012; Balokovi\'c et al. 2014; Koss et al. 2016a; Ricci et al. 2016). Also, new types of hard X-ray binaries have been reported namely the \textit{INTEGRAL} discovered Supergiant Fast X-ray Transients (SFXTs, Sguera et al. 2006) and the Gamma-ray binary in de Martino et al. 2013. Due to local or Galactic absorption, many of these objects are optically faint, so one needs relatively large telescopes to identify their nature and get information about them, as the study of the stellar populations of their central regions (Morelli et al. 2013).

Many efforts have been made to characterize and analyse the properties of the \textit{Swift} BAT sample using optical spectroscopy (e.g. Landi et al. 2007; Winter et al. 2010; Berney et al. 2015; Halpern \& Thorstensen 2015; Ueda et al. 2015), but the identification and precise classification of unassociated or poorly studied objects remains incomplete.
The work presented here extends to fainter objects the work started in Parisi et al. (2009, 2012, 2014; hereafter PI, PII and PIII), and focuses on the identification and classification of unassociated hard X-ray sources detected by BAT instrument onboard \textit{Swift}, that are listed in the Second Palermo \textit{Swift}/BAT Catalogue (2PBC, Cusumano et al. 2010).

The paper is structured as follows: in Section 2, we provide information about the sample selection; in Section 3, we describe the optical observations, telescopes employed and reduction method used; in Section 4, the results of optical classification of different classes (AGNs, XRBs, CVs and active stars) are presented and a statistical analysis of the sample is given; finally, in Section 5 we summarize the main conclusions of our work.

\begin{table}[htbp]
\caption[]{Observational information of finding charts of the soft X-ray counterpart identifications to the present sample of {\it 2PBC} sources.}
\scriptsize
\setlength{\tabcolsep}{4pt} 
\begin{center}
\begin{tabular}{lccc}
\noalign{\smallskip}
\hline
\hline
\noalign{\smallskip}
\multicolumn{1}{c}{Object} & Obs Date & Start Time & Exposure \\
\noalign{\smallskip}
  & & (UT) & Time (s)  \\
\noalign{\smallskip}
\hline
\noalign{\smallskip}					
2PBC J0057.2$+$6401   &  Aug 29, 2010  &  00:19:00  &  9807  \\
 					
2PBC J0116.5$-$1235   &  May 19, 2011  &  00:45:00  &  7372  \\
 					
2PBC J0128.5$+$1628   &  Jun 04, 2009  &  01:33:00  &  10643  \\
 					
2PBC J0154.1$-$5034   &  Nov 14, 2012  &  00:29:00  &  2379  \\
 					
2PBC J0217.0$-$7250   &  Nov 29, 2014  &  23:59:00  &  1743  \\
 					 					
2PBC J0252.3$+$4309  &  Mar 28, 2011  &  10:54:00  &  4189  \\
 					 					
2PBC J0356.6$-$6252   &  Aug 05, 2008  &  01:32:00  &  5659  \\
 					 					
2PBC J0440.6$-$6507   &  Nov 18, 2012  &  16:38:00  &  4375  \\
 										
2PBC J0505.4$-$6734   &  Oct 21, 2013  &  13:00:00  &  2071  \\
 					
2PBC J0550.7$-$2304  &  Nov 27, 2012  &  07:23:00  &  3520  \\
 					 					 					 					 					 					 									
2PBC J0608.0$+$5749   &  Dec 04, 2014  &  03:28:00  &  1093  \\
 					
2PBC J0620.8$-$2932  &  Jul 03, 2013  &  00:14:00  &  4689  \\
 					
2PBC J0653.1$-$1227  &  Nov 21, 2012  &  12:14:00  &  3139  \\
 					
2PBC J0709.5$-$3538   &  Jul 23, 2013  &  16:58:02  &  2590  \\
 					
2PBC J0714.6$-$2521  &  Nov 22, 2012  &  18:46:00  &  2724  \\
 					
2PBC J0751.6$+$6450 &  Feb 28, 2013  &  05:04:00  &  1114  \\
 										
2PBC J0757.9$+$0113   &  Jun 05, 2014  &  20:01:29  &  804  \\
 					
2PBC J0819.2$-$2508 &  Jul 11, 2013  &  02:02:00  &  3814  \\

2PBC J0838.7$+$2612  &  Dec 22, 2013  &  11:25:41  &  2377  \\
 					
2PBC J1020.5$-$0235   &  Jul 07, 2010  &  18:18:00  &  3144  \\
 										
2PBC J1042.2$+$0043   &  Jun 25, 2010  &  09:12:50  &  7537  \\
 					 					
2PBC J1228.1$-$0925   &  Nov 19, 2013  &  09:29:04  &  4173  \\
 					
2PBC J1332.1$-$7751  &  Feb 26, 2009  &  08:49:00  &  9264  \\
 					 					
2PBC J1419.2$+$6804$^{*}$  &  Jun 11, 2005  &  17:27:54  &  9851  \\
 					
2PBC J1520.2$-$0433  &  Jan 21, 2011  &  02:55:00  &  3344  \\
 					 					
2PBC J1548.5$-$3208   &  Feb 01, 2009  &  00:23:06  &  2849  \\
 					
2PBC J1555.0$-$6225 &  Jul 04, 2011  &  05:44:00  &  2430  \\
 										 					 					 					 									
2PBC J1649.3$-$1739  &  Sep 07, 2010  &  00:19:21  &  2720  \\
 					
2PBC J1809.7$-$6555  &  Jul 09, 2011  &  01.39:52  &  2979  \\
 					
2PBC J1832.8$+$3124  &  Jan 27, 2014  &  06:06:00  &  3494  \\
 					
2PBC J1911.4$+$1412  &  Nov 28, 2012  &  06:48:23  &  1239  \\
 					
2PBC J2029.4$-$6146  &  Jun 20, 2010  &  00:03:21  &  6246  \\
 					
2PBC J2030.7$-$7530  &  Jul 02, 2010  &  17:23:32  &  6547  \\
 								
2PBC J2045.9$+$8321  &  Jun 18, 2010  &  01:11:42  &  5584  \\
 					 					
2PBC J2136.3$+$2003  &  Jun 20, 2013  &  13:55:34  &  1573  \\
 					 					 					 					 					 									
2PBC J2155.1$+$6205  &  Jun 23, 2013  &  02:25:26  &  4240  \\
 					
2PBC J2238.9$+$4050  & Dec 23, 2012   &  11:24:04  &  1556  \\
 					
2PBC J2322.6$+$2903  &  Nov 11, 2012  &  01:45:09  &  4373  \\
 					
2PBC J2348.9$+$4153  &  Nov 12, 2013  &  00:13:14  &  2275  \\
  	
\noalign{\smallskip} 
\hline
\noalign{\smallskip} 
\multicolumn{4}{l}{Note: Soft X-ray observations were obtained from ASI Science } \\
\multicolumn{4}{l}{XRT database (\textit{http://www.asdc.asi.it/}). } \\ 
\multicolumn{4}{l}{$^{*}$: soft X-ray observation from HEASARC database } \\
\multicolumn{4}{l}{(\textit{http://heasarc.gsfc.nasa.gov}).} \\
\noalign{\smallskip} 
\hline
\hline
\end{tabular}
\end{center}
\end{table}

\section{Sample selection}

In this work we focus on the optical follow-up of objects with unknown classification and/or redshift that are reported in the Second Palermo \textit{Swift}-BAT hard X-ray catalogue obtained by analysing data acquired in the first 54 months of the \textit{Swift} mission (Cusumano et al. 2010). 

This survey covers 90\% of the sky down to a flux limit of 1.1 x $10^{-11}$ \textit{erg $cm^{-2}$ $s^{-1}$} and 50\% of the sky down to a flux limit of 0.9 x $10^{-11}$ \textit{erg $cm^{-2}$ $s^{-1}$} in the 15-150 keV band. It lists 1256 sources, of which 57\% are extragalactic, 19\% are galactic, and 24\% are of unknown type. 

From this BAT survey, we selected a sample of unknown type objects and used the available soft X-ray data ($<$ 10 keV) to reduce the source positional uncertainty from arcmin to arcsec sized radii, using ROSAT (Voges et al. 1999), \textit{Swift}/XRT (0.3-10 keV, Evans et al. 2014), XMM-Newton (0.2-12 keV, Watson et al. 2009) catalogues or Chandra\footnote{\tt http://cxc.harvard.edu} (0.3-10 keV, Aldcroft et al. 2000) and XRT\footnote{\tt http://asdc.asi.it} databases. This approach was proven by Stephen et al. (2006) to be very effective in associating hard X-ray sources with a strong, softer X-ray counterpart within the high-energy error circle with a high degree of probability, which in turn drastically reduces their positional error circles to a few arcsec in radius, thus shrinking the search area by a factor of $\sim$ $10^{4}$.

More specifically, for the present sample, we analysed the soft X-ray ($<$10 keV) images with longest available exposure of the field of each BAT object in Table 1 to search for sources detected (at confidence level $>$ 3$\sigma$) within the 90\% \textit{Swift}/BAT error circles. This choice also allowed us to investigate the source hardness in case of multiple objects within the hard X-ray error box, which permitted us to pinpoint the hardest sources, hence the most likely counterparts to the BAT objects.

Nearly all soft X-ray images were extracted from the ASI Science Data XRT database (see footnote 2); that of 2PBC J1419.2$+$6804 was instead obtained from an XMM-Newton observation downloaded from the HEASARC database\footnote{{\tt http://heasarc.gsfc.nasa.gov}}. The information about these images can be found in Table 1 and the coordinates of the most likely counterparts are reported in Table 2 (second and third columns) together with their relative uncertainties (fourth column).

In most of the cases, that is for 33 objects, we found a single bright soft X-ray object within the BAT 90\% confidence level error box with emission above 3 keV, and the soft X-ray counterpart positions of these BAT objects are provided in Figures A.1, A.2 and A.3 in the Appendix A, where for each image the BAT error circle is shown in blue and the black arrow indicates the soft X-ray counterpart for which the optical spectra were acquired. On the other hand, we found that 11 of the proposed soft X-ray counterparts to BAT objects in this paper are coincident with the proposed ROSAT or XRT counterpart reported in the Cusumano et al. (2010) and/or Baumgartner et al. (2013) catalogues. We list them in Table 2.

For the remaining 6 objects we needed to consider additional criteria as we detailed below.

 When no soft X-ray counterpart was found inside the BAT error circle at 90\% position confidence level, we expanded the search area by considering the 99\% BAT error circle assuming a gaussian uncertainty distribution for the error radius. In our sample, the putative counterparts of three objects were selected considering the 99\% BAT error circle: those of 2PBC J0550.7$-$2304, 2PBC J0620.8$-$2932 and 2PBC J1832.8$+$3124. Since no other object is detected inside the 99\% positional uncertainty we assumed these associations to be correct (see Figure A.4 in Appendix A). We remark here that the soft X-ray counterpart indicated with the black arrow for the object 2PBC J0550.7$-$2304 corresponds to the ROSAT object proposed as a soft X-ray counterpart in the Cusumano et al. (2010) catalogue and lies at the edge of the 99\% BAT error circle.

In one single case, we included in our sample an object reported by PIII and for which no soft X-ray counterpart was found within its 99\% BAT error circle. Actually, PIII found two nearby soft X-ray sources as possible associations of the object PBC J1020.5-0235, which are the same sources labeled as SWIFT J1020.5$-$0237A and SWIFT J1020.5$-$0237B in the 70-month BAT Catalogue of Baumgartner et al. (2013). In PIII the object with higher 3-10 keV flux was considered for optical spectroscopy (object B) and identified as Sy 1 galaxy at z = 0.060. We thus decided to complete the information about this field by analysing the optical spectrum of the object SWIFT J1020.5$-$0237A (See Figure A.5 in Appendix A). 

Finally for two cases, 2PBC J0819.2$-$2508 and 2PBC J1520.2$-$0433, the soft X-ray counterparts proposed in this work do not coincide with those reported in the Cusumano et al. (2010) and/or Baumgartner et al. (2013) catalogues (see Figure A.6 in Appendix A).

\begin{table*}[th!]
\caption[]{Log of the spectroscopic observations presented in this paper
(see text for details).}
\scriptsize
\begin{center}
\resizebox{18cm}{!}{
\begin{tabular}{lllcllcccclr}
\noalign{\smallskip}
\hline
\hline
\noalign{\smallskip}
\multicolumn{1}{c}{{\it (1)}} & \multicolumn{1}{c}{{\it (2)}} & \multicolumn{1}{c}{{\it (3)}} & \multicolumn{1}{c}{{\it (4)}} & 
\multicolumn{1}{c}{{\it (5)}} & \multicolumn{1}{c}{{\it (6)}} & \multicolumn{1}{c}{{\it (7)}} & \multicolumn{1}{c}{{\it (8)}} & 
\multicolumn{1}{c}{{\it (9)}} & \multicolumn{1}{c}{{\it (10)}} & \multicolumn{1}{c}{{\it (11)}} & \multicolumn{1}{c}{{\it (12)}}  \\
\multicolumn{1}{c}{Object} & \multicolumn{1}{c}{RA X-ray} & \multicolumn{1}{c}{Dec X-ray} & \multicolumn{1}{c}{Error radius} & \multicolumn{1}{c}{RA Opt.} & \multicolumn{1}{c}{Dec Opt.} & \multicolumn{1}{c}{Telescope+instrument} & $\lambda$ range & Disp. & Resolution & \multicolumn{1}{c}{UT Date $\&$ Time}  & \multicolumn{1}{c}{Exposure} \\
  & \multicolumn{1}{c}{(J2000)} & \multicolumn{1}{c}{(J2000)} & \multicolumn{1}{c}{(arcsec)} & \multicolumn{1}{c}{(J2000)} & \multicolumn{1}{c}{(J2000)} & & (\AA) & (\AA/pix) & FWHM (\AA) & \multicolumn{1}{c}{at mid-exposure} & \multicolumn{1}{c}{time (s)}  \\
  
\noalign{\smallskip}
\hline
\noalign{\smallskip}

 2PBC J0057.2$+$6401$^{\vartriangle}$            & 00:57:12.77               & $+$63:59:41.3     & 3.6         & 00:57:12.85 & $+$63:59:42.9 & TNG$+$DOLoReS           & 3700-8000  & 2.5 &   10    & 02 Aug. 2011, 04:48 & 2$\times$1200 \\ 
 
 2PBC J0116.5$-$1235$^{\vartriangle}$            & 01:16:31.05               & $-$12:36:19.0         & 3.5      & 01:16:31.14 & $-$12:36:17.0 & TNG$+$DOLoReS           & 3700-8000  & 2.5  &   15    & 19 Aug. 2011, 04:29 & 2$\times$900  \\ 

 2PBC J0128.5$+$1628$^{\vartriangle}$            & 01:28:24.46               & $+$16:27:30.4       & 3.9     & 01:28:24.45 & $+$16:27:33.5 & SPM 2.1m$+$B$\&$C Spec. & 3300-7900  & 4.0  &    18   & 28 Sep. 2011, 07:36 & 2$\times$1800 \\ 

 2PBC J0154.1$-$5034            &         01:53:50.90$^{\centerdot}$       & $-$50:31:36.8$^{\centerdot}$     & 6.0           &      01:53:51.51       & $-$50:31:37.7           & NTT$+$EFOSC2            &    3685-9315        &  2.8  &   21      & 29 Aug. 2015, 06:30 & 2$\times$600  \\ 

 2PBC J0217.0$-$7250            &      02:17:36.10$^{\centerdot}$       & $-$72:51:27.7$^{\centerdot}$        & 6.0           &       02:17:35.72      & $-$72:51:28.1          & NTT$+$EFOSC2            &     3685-9315       &  2.8   &    21    & 28 Aug. 2015, 10:09 & 2$\times$300  \\ 

 2PBC J0252.3$+$4309$^{\vartriangle}$            & 02:52:34.02               & $+$43:10:01.8     & 4.1        & 02:52:34.02 & $+$43:10:02.8 & SPM 2.1m$+$B$\&$C Spec. & 3300-7900  & 4.0  &    18   & 29 Sep. 2011, 09:04 & 2$\times$1800 \\ 

 2PBC J0356.6$-$6252$^{\vartriangle}$            & 03:56:19.66               & $-$62:51:36.6    & 4.1        & 03:56:19.97 & $-$62:51:39.1 & CTIO 1.5m$+$RC Spec.    & 3300-10500 & 5.7  &   14    & 24 Sep. 2011, 05:40 & 2$\times$1800 \\ 

 2PBC J0440.6$-$6507            &    04:40:35.30$^{\centerdot}$       & $-$65:07:50.8$^{\centerdot}$    & 6.0      &       04:40:35.63      & $-$65:07:48.7           & NTT$+$EFOSC2            &     3685-9315       &  2.8  &   21     & 29 Aug. 2015, 06:57 & 2$\times$600  \\ 

 2PBC J0505.4$-$6734$^{\vartriangle}$            & 05:05:24.33$^*$           & $-$67:34:35.5$^*$    & 2.0       &    05:05:24.50$^{U}$          & $-$67:34:36.2$^{U}$         & CTIO 1.5m$+$RC Spec.    & 3300-10500 & 5.7  &    14   & 24 Sep. 2011, 07:34 & 2$\times$1500 \\ 

 2PBC J0550.7$-$2304$^\ddagger$ & 05:50:39.10$^{\centerdot}$ & $-$23:11:16.4$^{\centerdot}$ & 6.0     &     05:50:39.51        & $-$23:11:17.7           & CTIO 1.5m$+$RC Spec.    & 3300-10500 & 5.7  &   14    & 02 Oct. 2011, 08:40 & 2$\times$400  \\ 

 2PBC J0608.0$+$5749            & 06:08:09.40$^{\centerdot}$ & $+$57:51:07.9$^{\centerdot}$ & 6.0     &    06:08:09.67         & $+$57:51:04.5           & SPM 2.1m$+$B$\&$C Spec. & 3300-7900  & 4.0  &   18    & 27 Feb. 2015, 04:09 & 3$\times$1200 \\ 

 2PBC J0620.8$-$2932$^{\ddagger ,\circ}$   &    06:20:59.90$^{\centerdot}$   &  $-$29:27:31.6$^{\centerdot}$  & 6.0      &      06:21:00.05       & $-$29:27:34.6    & SPM 2.1m$+$B$\&$C Spec. & 3300-7900  & 4.0 &   18    & 04 Dec. 2013, 08:37 & 1200+900      \\ 

 2PBC J0640.1$-$4740$^{\vartriangle}$            & 06:40:13.46               & $-$47:41:33.4      & 3.9         & 06:40:13.53 & $-$47:41:34.3 & CTIO 1.5m$+$RC Spec.    & 3300-10500 & 5.7  &    14   & 17 Nov. 2011, 04:43 & 2$\times$1800 \\ 
(= SWIFT J064013.50$-$474132.9) & & & & & & & & & & \\ 

 2PBC J0653.1$-$1227            & 06:53:16.40$^{\centerdot}$ & $-$12:29:37.3$^{\centerdot}$ & 6.0     & 06:53:16.60$^{\wr}$   & $-$12:29:37.0$^{\wr}$   & TNG$+$DOLoReS           & 3700-8000  & 2.5  &   15    & 03 Feb. 2013, 22:55 & 2$\times$1800 \\ 

 2PBC J0658.0$-$1746            & 06:58:06.08               & $-$17:44:21.0       & 3.6        &     06:58:05.87$^{U}$     &  $-$17:44:16.3$^{U}$           & Radcliffe$+$Gr. Spec.   & 3750-7750  & 2.3  &   6    & 20 Nov. 2011, 07:20 & 1310          \\ 
 (= 1RXS J065806.3$-$174427) & & & & & & & & & & \\ 

 2PBC J0709.5$-$3538            & 07:09:31.80$^{\centerdot}$ & $-$35:37:46.2$^{\centerdot}$ & 6.0     &     07:09:32.05    & $-$35:37:46.4           & SPM 2.1m$+$B$\&$C Spec. & 3300-7900  & 4.0  &   18    & 05 Dec. 2013, 11:00 & 2$\times$1800 \\ 

 2PBC J0714.6$-$2521$^{\vartriangle}$            & 07:14:37.10$^{\centerdot}$ & $-$25:17:22.9$^{\centerdot}$ & 6.0     & 07:14:36.98 & $-$25:17:25.2 & TNG$+$DOLoReS           & 3700-8000  & 2.5   &    15  & 04 Feb. 2013, 00:38 & 2$\times$1200 \\ 

 2PBC J0717.8$-$2156$^{\vartriangle}$            & 07:17:48.28               & $-$21:53:03.0      & 3.6         &      07:17:48.28$^{U}$   & $-$21:53:05.1$^{U}$      & TNG$+$DOLoReS           & 3700-8000  & 2.5  &  20     & 22 Oct. 2012, 04:11 & 1200          \\ 
 (= 1RXS J071748.9$-$215306) & & & & & & & & & & \\ 

 2PBC J0751.6$+$6450            & 07:51:44.70$^{\centerdot}$ & $+$64:48:59.7$^{\centerdot}$ & 6.0     &      07:51:45.41       & $+$64:49:03.7           & Cassini$+$BFOSC         & 3500-8700  & 4.0  &  12     & 21 Mar. 2013, 21:37 & 1800          \\ 

 2PBC J0757.9$+$0113            & 07:57:44.10$^{\centerdot}$ & $+$01:13:42.9$^{\centerdot}$ & 6.0     &       07:57:44.00      & $+$01:13:42.7           & TNG$+$DOLoReS           & 3700-8000  & 2.5  &   15   & 04 Jan. 2014, 00:24 & 2$\times$1200 \\ 

 2PBC J0800.5$-$4306$^{\blacktriangle ,\vartriangle}$            & 08:00:39.99               & $-$43:11:06.4       & 3.5        & 08:00:39.98 & $-$43:11:07.6 & CTIO 1.5m$+$RC Spec.    & 3300-10500 & 5.7 &   14    & 07 Dec. 2011, 03:48 & 2$\times$1800 \\ 
 (= SWIFT J080040.2$-$431107) & & & & & & & & & & \\ 

 2PBC J0812.3$-$4003$^{\vartriangle}$            & 08:12:14.16               & $-$40:03:23.7     & 3.5          & 08:12:14.01 & $-$40:03:23.9 & CTIO 1.5m$+$RC Spec.    & 3300-10500 & 5.7 &   14    & 24 Nov. 2011, 06:58 & 3$\times$1800 \\ 
(= 1RXS J081215.2$-$400336) & & & & & & & & & & \\ 

2PBC J0819.2$-$2508           & 08:19:14.80$^{\centerdot}$ & $-$25:11:17.1$^{\centerdot}$ & 6.0     &      08:19:14.74       & $-$25:11:16.6           & CTIO 1.5m$+$RC Spec.    & 3300-10500 & 5.7 &    14   & 17 Nov. 2011, 05:45 & 2$\times$1000 \\ 

 2PBC J0838.7$+$2612            & 08:38:57.00$^{\centerdot}$ & $+$26:10:37.5$^{\centerdot}$ & 6.0     &   08:38:56.88      & $+$26:10:37.5           & SDSS$+$CCD Spc.         & 3800-9200  & 1.0&    3   & 08 Dec. 2004, 00:00 & 4300          \\ 

 2PBC J0854.3$-$0826$^{\vartriangle}$            & 08:54:29.20               & $-$08:24:28.6      & 2.7        & 08:54:29.27 & $-$08:24:27.6 & TNG$+$DOLoReS           & 3700-8000  & 2.5&    15   & 01 Dec. 2011, 02:41 & 2$\times$900  \\ 
 (= SWIFT J085429.35$-$082428.6) & & & & & & & & & & \\ 

 2PBC J1020.5$-$0235$^{\wr, \vartriangle}$            & 10:21:02.90$^{\centerdot}$               & $-$02:36:46.0$^{\centerdot}$  & 6.0         &      10:21:03.09$^{U}$   & $-$02:36:42.6$^{U}$       & TNG$+$DOLoReS           & 3700-8000  & 2.5&   15    & 04 Jan. 2014, 05:34 & 2$\times$1200 \\ 

 2PBC J1042.2$+$0043$^{\vartriangle}$            & 10:42:08.33               & $+$00:42:06.8      & 3.6         & 10:42:08.36 & $+$00:42:06.0 & TNG$+$DOLoReS           & 3700-8000  & 2.5 &   20   & 23 Feb. 2012, 05:40 & 2$\times$1500 \\ 

 2PBC J1228.1$-$0925            & 12:28:10.00$^{\centerdot}$ & $-$09:27:03.9$^{\centerdot}$ & 6.0     &    12:28:10.14     & $-$09:27:03.5           & SPM 2.1m$+$B$\&$C Spec. & 3300-7900  & 4.0&   18    & 26 Feb. 2015, 11:18 & 2$\times$1800 \\ 

 2PBC J1251.8$-$5127            & 12:51:44.04               & $-$51:28:04.3      & 3.6         & 12:51:44.07 & $-$51:28:05.2 & CTIO 1.5m$+$RC Spec.    & 3300-10500 & 5.7&   14    & 22 Dec. 2011, 08:28 & 1620          \\ 
(= 1RXS J125144.2$-$512809) & & & & & & & & & & \\ 

 2PBC J1332.1$-$7751$^{\vartriangle}$            & 13:32:40.64               & $-$77:50:39.0     & 4.1          & 13:32:40.59 & $-$77:50:40.5 & CTIO 1.5m$+$RC Spec.    & 3300-10500 & 5.7 &   14   & 03 Feb. 2012, 07:52 & 2$\times$1000 \\ 

 2PBC J1419.2$+$6804            & 14:18:49.94$^*$           & $+$68:04:09.9$^*$      & 2.0      &   14:18:49.91    & $+$68:04:09.7    & Cassini$+$BFOSC         & 3500-8700  & 4.0 &  12    & 26 Feb. 2012, 02:00 & 2$\times$1800 \\ 

 2PBC J1520.2$-$0433            & 15:20:14.98               & $-$04:35:52.6      & 3.6       & 15:20:15.07 & $-$04:35:53.0 & Cassini$+$BFOSC         & 3500-8700  & 4.0&    12   & 12 Apr. 2015, 01:25 & 2$\times$1800 \\ 

 2PBC J1548.5$-$3208            & 15:48:43.35               & $-$32:07:11.0     & 2.7       & 15:48:43.39 & $-$32:07:13.3 & CTIO 1.5m$+$RC Spec.    & 3300-10500 & 5.7 &   14   & 05 Apr. 2012, 08:22 & 2$\times$1500 \\ 

 2PBC J1555.0$-$6225            & 15:54:59.25               & $-$62:24:30.0      & 4.2        &   15:54:58.95    & $-$62:24:28.3           & CTIO 1.5m$+$RC Spec.    & 3300-10500 & 5.7 &   14   & 19 May. 2012, 07:14 & 2$\times$1500 \\ 

 2PBC J1649.3$-$1739            & 16:49:20.92               & $-$17:38:40.7     & 3.6        &   16:49:21.02   & $-$17:38:40.6     & SPM 2.1m$+$B$\&$C Spec. & 3300-7900  & 4.0&   18    & 22 Jun. 2012, 06:34 & 2$\times$1800 \\ 

 2PBC J1709.7$-$2349            & 17:09:44.69$^*$           & $-$23:46:53.1$^*$     & 2.0         &  17:09:44.70   & $-$23:46:53.2           & SPM 2.1m$+$B$\&$C Spec. & 3300-7900  & 4.0 &  18    & 25 Jun. 2012, 05:59 & 2$\times$1200 \\ 
 (= 1RXS J170944.9$-$234658) & & & & & & & & & & \\ 

 2PBC J1742.0$-$6053$^{\vartriangle}$            & 17:42:01.74               & $-$60:55:13.6      & 3.5         & 17:42:01.50 & $-$60:55:12.2 & CTIO 1.5m$+$RC Spec.    & 3300-10500 & 5.7 &   14    & 26 Oct. 2011, 00:24 & 2$\times$1200 \\ 
(= 1RXS J174201.5$-$605514) & & & & & & & & & & \\ 

 2PBC J1809.7$-$6555            & 18:09:52.19               & $-$65:56:11.9      & 3.5       & 18:09:52.28 & $-$65:56:13.8 & CTIO 1.5m$+$RC Spec.    & 3300-10500 & 5.7  &   14   & 08 Oct. 2011, 02:32 & 2$\times$1800 \\ 

 2PBC J1832.8$+$3124$^\ddagger$ & 18:32:26.20$^{\centerdot}$ & $+$31:29:04.2$^{\centerdot}$ & 6.0     &   18:32:26.19     & $+$31:29:04.8    & TNG$+$DOLoReS           & 3700-8000  & 2.5 &   15    & 04 May. 2014, 02:00 & 3$\times$1200 \\ 

 2PBC J1911.4$+$1412            & 19:11:24.70$^{\centerdot}$ & $+$14:11:44.8$^{\centerdot}$ & 6.0     &     19:11:24.87$^{U}$   & $+$14:11:44.9$^{U}$     & TNG$+$DOLoReS           & 3700-8000  & 2.5  &   15    & 11 Apr. 2013, 03:20 & 2$\times$1800 \\ 

 2PBC J2010.3$-$2522$^{\vartriangle}$            & 20:10:19.74               & $-$25:23:59.8        & 3.6        &   20:10:19.76$^{U}$  & $-$25:23:59.4$^{U}$     & TNG$+$DOLoReS           & 3700-8000  & 2.5  &   15    & 20 Jun. 2012, 04:56 & 2$\times$1800 \\ 
(= 1RXS J201020.0$-$252356) & & & & & & & & & & \\ 

 2PBC J2029.4$-$6146$^{\vartriangle}$            & 20:29:31.08               & $-$61:49:07.7        & 3.6       & 20:29:31.23 & $-$61:49:08.7 & CTIO 1.5m$+$RC Spec     & 3300-10500 & 5.7  &   14    & 22 Aug. 2011, 06:09 & 2$\times$600  \\ 

 2PBC J2030.7$-$7530$^{\vartriangle}$            & 20:30:41.80               & $-$75:32:41.2      & 3.6        & 20:30:41.71 & $-$75:32:43.0 & CTIO 1.5m$+$RC Spec     & 3300-10500 & 5.7  &   14    & 22 Aug. 2011, 05:14 & 2$\times$1800 \\ 

 2PBC J2045.9$+$8321            & 20:44:13.29               & $+$83:19:40.1      & 3.6       & 20:44:14.22$^{U}$    & $+$83:19:42.4$^{U}$           & TNG$+$DOLoReS           & 3700-8000  & 2.5  &   15    & 04 Aug. 2011, 02:14 & 2$\times$1200 \\ 

 2PBC J2048.3$+$3812$^{\vartriangle}$            & 20:48:27.08               & $+$38:11:23.5     & 3.6      & 20:48:27.06 & $+$38:11:24.5 & TNG$+$DOLoReS           & 3700-8000  & 2.5  &   15    & 04 Aug. 2011, 01:09 & 2$\times$1800 \\ 
 (= 1RXS J204826.8$+$381120) & & & & & & & & & & \\ 

 2PBC J2136.3$+$2003            & 21:36:15.10$^{\centerdot}$ & $+$20:02:03.4$^{\centerdot}$ & 6.0     &  21:36:15.31    & $+$20:02:07.2           & SPM 2.1m$+$B$\&$C Spec. & 3300-7900  & 4.0  &   18    & 12 Jun. 2013, 08:58 & 2$\times$1800 \\ 

 2PBC J2155.1$+$6205            & 21:55:15.30$^{\centerdot}$ & $+$62:06:50.0$^{\centerdot}$ & 6.0     &  21:55:15.37   & $+$62:06:46.4           & TNG$+$DOLoReS           & 3700-8000  & 2.5  &   15    & 19 Sep. 2013, 04:56 & 2$\times$1200 \\ 

 2PBC J2238.9$+$4050            & 22:38:56.50$^{\centerdot}$ & $+$40:51:38.5$^{\centerdot}$ & 6.0     &   22:38:57.02    & $+$40:51:42.0           & Cassini$+$BFOSC         & 3500-8700  & 4.0  &    12   & 05 Nov. 2013, 21:40 & 2$\times$1800 \\ 

 2PBC J2322.6$+$2903            & 23:22:45.70$^{\centerdot}$ & $+$29:08:12.8$^{\centerdot}$ & 6.0     &   23:22:45.86     & $+$29:08:16.2           & Cassini$+$BFOSC         & 3500-8700  & 4.0  &   12    & 03 Dec. 2012, 17:52 & 2$\times$1800 \\ 

 2PBC J2348.9$+$4153$^{\circ}$            &     23:48:53.80$^{\centerdot}$         & $+$41:54:28.0$^{\centerdot}$    & 6.0        &     23:48:53.46        & $+$41:54:28.6           & SPM 2.1m$+$B$\&$C Spec. & 3300-7900  & 4.0  &   18    & 06 Dec. 2013, 03:19 & 2$\times$1800 \\ 

\noalign{\smallskip}
\hline
\noalign{\smallskip}
\multicolumn{12}{l}{Notes: If not indicated otherwise, soft X-ray source coordinates were extracted from XRT ($http://www.swift.ac.uk/1SXPS/$ or interactive mode ($^{\centerdot}$) in $http://www.asdc.asi.it/mmia/index.php?mission=swiftmastr$,} \\
\multicolumn{12}{l}{ with an accuracy $\sim$ 6$''$) or from XMM-Newton observations  ($^*$) with an accuracy better than 2$''$. Coordinates from the 2MASS catalogue have an accuracy better than 0$\farcs$1 whereas from USNO-A2.0 catalogue ($^{U}$) have an } \\ 
\multicolumn{12}{l}{accuracy of about 0$\farcs$2. } \\
\multicolumn{12}{l}{$^\ddagger$: sources for which the 99\% BAT error circle was considered to search for a soft X--ray counterpart. } \\
\multicolumn{12}{l}{$^{\wr}$: source outside the corresponding 99\% XRT error circle.} \\
\multicolumn{12}{l}{$^{\circ}$: proposed soft X--ray counterpart with no detected emission above 3 keV. } \\
\multicolumn{12}{l}{$^{\vartriangle}$: object listed also in the 70-month \textit{Swift}/BAT catalogue (Baumgartner et al. 2013).}  \\
\multicolumn{12}{l}{$^{\blacktriangle}$: object listed also in the IBIS soft gamma-ray sky after 1000 \textit{INTEGRAL} orbits (Bird et al. 2016).}  \\

\noalign{\smallskip}
\hline
\hline
\noalign{\smallskip}
\end{tabular} } 
\end{center}
\end{table*}

Then, within the soft X-ray error boxes of each source of our sample, we identified the putative optical counterpart(s) to the BAT object and performed optical spectroscopic follow-up work with long-slit spectroscopy. Following the method applied by Masetti et al. (2004, 2006a-d, 2008, 2009, 2010, 2012, 2013; hereafter MI to MX respectively) and by PI, PII and PIII we determined the nature of 50 selected objects, estimating also redshifts, distances, Galactic and local extinction, and central black hole masses for type 1 AGNs. The observational information of the 50 objects are listed in Table 2.

\section{Optical observations} 

The data presented in this work involved the use of the following telescopes:

\begin{itemize}
\item the 1.5m at the Cerro Tololo Interamerican Observatory (CTIO), Chile;
\item the 1.52m Cassini telescope of the Astronomical Observatory of Bologna, in Loiano, Italy; 
\item the 2.1m telescope of the Observatorio Astron\'omico Nacional in San Pedro M\'artir (SPM), Mexico;
\item the 3.58m Telescopio Nazionale Galileo (TNG) at the Roque de Los Muchachos Observatory in La Palma, Spain;
\item the 1.9m Radcliffe telescope of the South African Astronomical Observatory (Sutherland, South Africa);
\item the 2.5m SDSS at the Apache Point Observatory (APO), Sunspot, New Mexico;
\item the 3.5m New Technology Telescope (NTT) of La Silla Observatory, Chile.
\end{itemize}

In Table 2, we show a detailed log of all the spectroscopic observations presented in this work. Column 1 indicates the names of the observed 2PBC 
sources. In Cols. 2 and 3, we list the soft X-ray coordinates of the proposed counterpart (with their respective positional uncertainties in col. 4)
extracted from the XMM-Newton (with an uncertainty better than 2$''$, Watson et al. 2009) and the XRT catalogues (with an uncertainty of $\sim$6$''$ for the PC interactive mode according to Evans et al. 2014). In Cols. 5 and 6, we list the equatorial coordinates of the proposed counterpart, mostly extracted from the 2MASS (with an 
uncertainty of 0$\farcs$1: Skrutskie et al. 2006) or USNO-A2.0 catalogues (with an uncertainty of 0$\farcs$2, Deutsch 1999; Assafin et al. 2001).

The telescope and instrument used for the observations are reported in Col. 7, while the main characteristics of each spectroscopic setup 
are presented in Cols. 8, 9 and 10. Column 11 reports the observation date and the UT time at mid-exposure, and Col. 12 provides the 
exposure times and the number of observations for each source.

The instrumental resolutions (in \AA) for the different setups of the optical spectra presented here are listed in Tables 2 and 3. The spectroscopic data acquired have been extracted (Horne 1986) and reduced following standard procedures using IRAF\footnote{IRAF is the Image Reduction and Analysis Facility made available to the astronomical community by the National Optical Astronomy Observatories, which are 
operated by AURA, Inc., under contract with the U.S. National Science Foundation. It is available at {\tt http://iraf.noao.edu/}}. 
Calibration frames (flat fields and bias) were taken on the day preceeding or following the observing night. 
The wavelength calibration was performed using lamp data acquired soon after each on-target spectroscopic acquisition; the uncertainty in this 
calibration was $\sim$0.5~\AA~ in all cases according to our checks that used the positions of background night sky lines. 
Flux calibration was obtained using catalogued spectrophotometric standards. Finally, the data of a given object were stacked together to increase the S/N ratio when multiple spectra were acquired. The SDSS spectrum of 2PBC J0838.7$+$2612 was acquired from the corresponding online archive\footnote{{\tt http://www.sdss.org}}, already reduced and calibrated.

\begin{table}[htbp]
\begin{center}
\caption[]{Instrumental setups of the telescopes used to gather the observations presented in this paper.}
\scriptsize
\begin{tabular}{lcccc}
\noalign{\smallskip}
\hline
\hline
\noalign{\smallskip}
\multicolumn{1}{c}{Telescope} & Instrument & Grating & Slit Width & Resolution  \\
\noalign{\smallskip}
 & &  & (") & FWHM (\AA)  \\
\noalign{\smallskip}
\hline
\noalign{\smallskip}
TNG			&		LRS		&	LR-B 	 &	1.0		 &		10  \\
				&		 			&		 	 	&	1.5		 &		15  \\
		 		&		 			&	 		 	&	2.0		 &		20   \\ \hline
SPM	 		&		B$\&$C	 &	300		 	&	2.5		 &		18   \\ \hline
NTT	 		&		EFOSC2   &	Gr13     &	 1.0	 	 &		21   \\ \hline
CTIO 1.5m &		R-C  	 	&	13/I	 		&	1.5		 &		14   \\ \hline
SAAO 1.9m  &	  300	 	&	Gr7			&	2.0		 &		6   \\ \hline
SDSS  		&	SDSS Spec.		&	Blue+Red	 	&	2.3		 &		3  \\ \hline
Cassini	 	&	BFOSC   	&	 Gr4	 		&	2.0		 &		12   \\
\noalign{\smallskip}  
\hline
\hline
\end{tabular}
\end{center}
\end{table}

\section{Results}

The spectra of fifty objects were analysed according to the above steps. Forty five of them were classified as extragalactic objects and the remaining five objects presented features in their spectra consistent with those of galactic sources.

The main criteria used for the classification of the extragalactic sources spectra are reported hereafter. 

In order to distinguish between type 1 and type 2 emission-line AGN, we used the criteria of Veilleux \& Osterbrock (1987) 
and we used the line ratio diagnostics of Ho et al. (1993, 1997) and Kauffmann et al. (2003) to distinguish among the Seyfert 2, low-ionization nuclear emission-line regions (LINERs; Heckman 1980), HII regions and transition objects (LINERs whose integrated spectra are diluted or contamined by neighbouring HII regions, Ho et al. 1997). 

For the subclass assignation to Seyfert 1 galaxies, we used the $H_{\beta}$/[OIII]$\lambda$5007 line flux ratio criterion of Winkler et al. (1992).

To estimate the E(B-V) local optical absorption in our type 2 AGN sample, when possible, we first dereddened the spectrum by applying a correction for the Galactic absorption along the line of sight to the source. This was done using the galactic colour excess E(B-V)$_{Gal}$ given by Schlegel et al. (1998) and the Galactic extinction law obtained by Cardelli et al. (1989). We then estimated the color excess E(B-V)$_{AGN}$ local to the AGN host galaxy by using the resulting $\frac{F_{H_{\alpha}}}{F_{H_{\beta}}}$ line ratio and corrected that assuming a Milky Way reddening using the Balmer decrement flux relation of Osterbrock (1989) that for normal density and temperature conditions implies $ \left(\frac{F_{H_{\alpha}}}{F_{H_{\beta}}}\right) _{*}$ = 2.86.

For extragalactic sources, we assume a cosmology with $H_{0}$ = 70 km s$^{-1}$ Mpc$^{-1}$, $\Omega_{\Lambda}$ = 0.7, and $\Omega_{m}$ = 0.3; 
the luminosity distances of the extragalactic objects presented in this paper were determined for these parametres using the Cosmology Calculator of Wright (2006).

For most of our type 1 AGN sample we also estimated the mass of the central black hole. We generally used the H$_{\beta}$ emission line flux, corrected for the Galactic color excess (Schlegel et al. 1998), and its Full Width Half Maximmum (FWHM) to measure the broad-line region (BLR) gas velocity (v$_{FWHM}$). Using Eq. (2) of Wu et al. (2004), we estimated the BLR size, which is used with v$_{FWHM}$ in Eq. (5) of Kaspi et al. (2000) to calculate the AGNs black hole mass. For the single QSO of our sample we apply the formulae from McLure \& Jarvis (2002), which use the information afforded by the Mg {\sc ii} broad emission. 
As already remarked in our previous Papers, the main sources of error in these mass estimates generally come from the determination of the flux of the employed emission lines, which is $\sim$15$\%$ in the sample and from the scatter in the scaling relation between the size of the BLR and the diagnostic line luminosity (Vestergaard 2004). Therefore, we expect the typical error to be around 50$\%$ of the black hole mass value.  

The results of the BH masses determination are reported in subsection (4.4) where we also list the Eddington luminosities which were computed using the observed X-ray fluxes or upper limits in the 15-150 keV band and the median bolometric correction according to Ho (2009). For that, we first convert the observed X-ray luminosity measured in the \textit{Swift} BAT survey in the energy range 15-150 keV to the intrinsic 2-10 keV luminosity ($L_{2-10 keV}$) by dividing by 2.67 following Rigby et al.(2009) and Marconi et al. (2004), and then we applied the bolometric correction factor of $C_{X}$=15.8 between the bolometric luminosity ($L_{BOL}$) and the X-ray luminosity in the 2-10 keV band according to Ho (2009). We would like to stress that our estimation of the bolometric luminosity implies an approximation because Rigby et al. (2009) and Marconi et al. (2004) used a conversion from 14-195 keV to 2-10 keV energy band whereas our reference hard X-ray band is 15-150 keV.


To evaluate the distances of the compact Galactic X--ray sources of our sample, we assumed an absolute magnitude M$_V$ $\sim$ +9 and an intrinsic 
color index $(V-R)_0$ $\sim$ 0 mag (Warner 1995) for cataclysmic variables (CVs), whereas we assumed $M_{V}$ $\sim$ 0 and $(V - R)_{0}$ $\sim$ 0 for the X--ray binary (van Paradijs \& McClintock 1995).

Although these approaches basically provide an approximate value for the distance of Galactic sources, our past experience (MI-X and PI-III) tells us that these estimates can be considered correct to within 50\% of the refined value subsequently determined with more precise approaches. 

The values for the absorption in the optical V band along the line of sight of the galactic objects for the determination of distances were again estimated from the $H_{\alpha}$/$H_{\beta}$ flux ratio following Osterbrock (1989). 

The X--ray luminosities reported in Tables 4, 5, 6, 7, 8 and 9 are associated with a letter indicating the satellite and/or the instrument, 
namely {\it Swift}/BAT ({\it C}) from Cusumano et al. (2010), {\it Swift}/BAT ({\it B}) from Baumgartner et al. (2013) and  \textbf{{\it INTEGRAL}}/IBIS ({\it I}) from Bird et al. (2016) catalogues, and from  {\it Swift}/XRT ({\it X}) and {\it XMM-Newton} ({\it N}) databases, with which the measurement of the corresponding X--ray flux was obtained. 

For the cases in which the soft X-ray information was obtained from the ASDC\footnote{{\tt http://www.asdc.asi.it/}} archive, we derived the 0.3-10 keV flux by assuming a power law model with photon index $\Gamma$ = 1.8 and $N_{H}$ value formulated by Predehl \& Schmitt (1995) according to Schlegel et al. (1998) using the webpimms tool\footnote{{\tt https://heasarc.gsfc.nasa.gov/cgi-bin/Tools/ \\ w3pimms/w3pimms.pl}} in order to obtain their soft X-ray luminosities. 

For galactic objects with soft X-ray counterpart from the ASDC archive, we assumed different spectral models according to their nature in order to obtain their soft X-ray fluxes. Specifically, for the probable X-ray binary 2PBC J1911.4$+$1412 we considered a power law model with $\Gamma$ = 2.1, i.e. a Crab-like spectrum, and for the active star 2PBC J0620.8$-$2932 we assumed a black body model with a typical value for these objects of kT = 1 keV (Pandey et al. 2005). 

In the following, we present the object identifications by dividing them into four broad classes (X--ray binaries, CVs, active stars and AGNs).


\begin{figure}[htbp]
\begin{center}
\includegraphics[height=6.0cm]{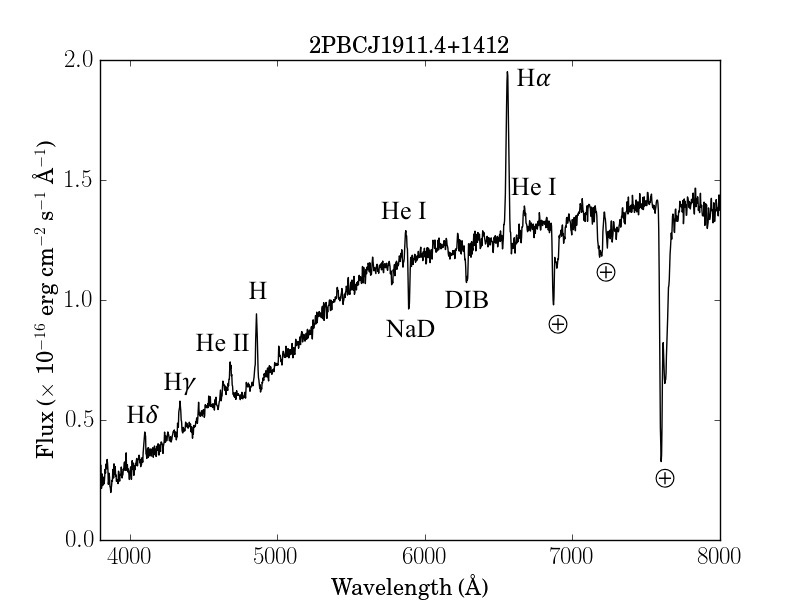}
\begin{spacing}{1}
\caption{Spectrum (not corrected for the intervening Galactic absorption) of the optical counterpart of the probably X-ray binary belonging to the sample of BAT sources presented in this paper, 2PBC J1911.4+1412. For this spectrum, the main spectral features are labeled and $\oplus$ indicates telluric absorptions.}
\end{spacing}
\end{center}
\end{figure}


\begin{table*}[htbp]
\caption[]{Main results for the probably X--ray binary identified 
in the present sample of {\it 2PBC} sources.} 
\hspace{-1.2cm}
\scriptsize
\vspace{-.5cm}
\begin{center}
\begin{tabular}{lccccccccr}
\noalign{\smallskip}
\hline
\hline
\noalign{\smallskip}
\multicolumn{1}{c}{Object} & \multicolumn{2}{c}{H$_\alpha$} & 
\multicolumn{2}{c}{H$_\beta$} &
 $R$ & $A_V$ & $d$ & Spectral & \multicolumn{1}{c}{$L_{\rm X}$} \\
\cline{2-5}
\noalign{\smallskip} 
 & EW & Flux & EW & Flux & mag & (mag) & (kpc) & type & \\
\noalign{\smallskip}
\hline
\noalign{\smallskip}
2PBC J1911.4$+$1412 & 12.3$\pm$0.6 & 1.5$\pm$0.1 & 6.3$\pm$1.5 & 0.4$\pm$0.1  & 
17.5 & $\sim$ 0.85 & $\sim$ 21 & late spectral type & 7.4 (15-150)$^{C}$ \\
 & & & & & & & & & 1.4 (0.3-10)$^{X}$ \\ 
\noalign{\smallskip} 
\hline
\noalign{\smallskip}
\multicolumn{10}{l}{Note: EWs are expressed in \AA, line fluxes are
in units of 10$^{-15}$ \textit{erg cm$^{-2}$ s$^{-1}$} } \\
\multicolumn{10}{l}{and distance was calculated assuming M$_{V}$ $\sim$ 0 mag.} \\
\multicolumn{10}{l}{X--ray luminosities
are in units of 10$^{35}$ erg s$^{-1}$, and the reference band 
(between round brackets) is expressed in keV.} \\
\multicolumn{10}{l}{In the last column, the upper case letter indicates the satellite 
and/or the instrument with which the} \\
\multicolumn{10}{l}{corresponding X--ray flux measurement was obtained (see text).} \\
\multicolumn{10}{l}{$^{C}$: from Cusumano et al. (2010).} \\
\multicolumn{10}{l}{$^{X}$: from \it{Swift}/XRT.} \\
\noalign{\smallskip}
\hline
\hline
\end{tabular} 
\end{center} 
\end{table*}

\subsection{2PBC J1911.4+1412: a X--ray binary?}

Among the objects in our sample,  2PBC J1911.4+1412 displays optical features of a X-ray binary (see Figure 1). 

In the optical spectrum of this source, we detected emissions of Balmer lines up to $H_{\delta}$ and emission of He I at a redshift consistent with 0 superimposed on a reddened continuum. Moreover, the spectrum displays an emission of He II$_{\lambda 4686}$, typical of accretion disks.
Table 4 lists the relevant optical spectral information along with the main parameters determined from available X-ray and optical data. 
The value for the absorption in the optical along the line of sight reported in that Table for the object is lower than the Galactic one along their direction according to Schlegel et al. (1998) ($A_{V}$ $\sim$ 5.4 mag), indicating that the object is indeed within the Galaxy.

A detailed study of the photometric optical/NIR information is needed to constrain the nature of the secondary star in this binary system but, due to the faintness of the optical counterpart along with the optical spectral shape and the NIR non-detection in the 2MASS catalogue (Skrutskie et al. 2006), the presence of an early type companion can be ruled out and, consequently, a classification as a HMXB is excluded.


\begin{figure*}[htbp]
\centering
\includegraphics[height=6.0cm]{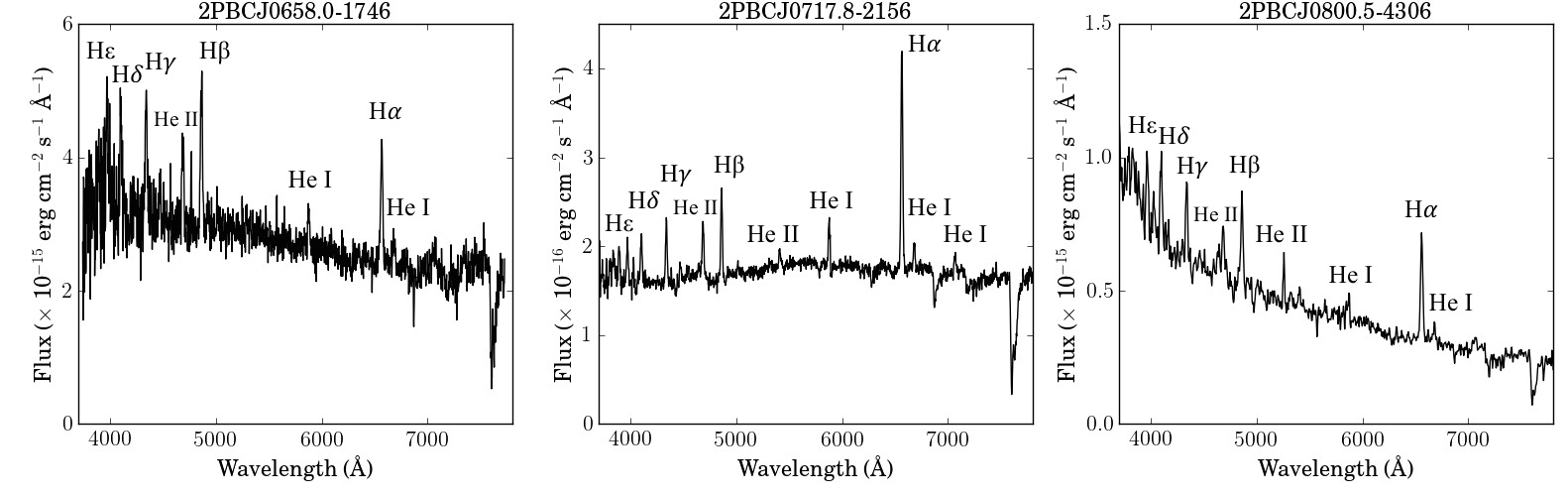}
\caption{Spectra (not corrected for the intervening Galactic absorption) of the optical counterparts of the 3 CVs belonging to the sample of BAT sources presented in this paper. For each spectrum, the main spectral features are labeled and $\oplus$ indicates telluric absorptions.}
\end{figure*}

A CV nature can also be excluded on the following basis: considering an $M_{V}$ = +9, we obtain a distance of d $\sim$ 340 pc, which is too close to justify the optical absorption displayed on the source spectrum. We thus suggest that this object is probably a Low Mass X-ray Binary with a secondary star of late spectral type at a distance of $\sim$ 21 kpc (see Table 4). This distance is quite large but is not unprecedented (see GS 1354-64, Casares et al. 2004, 2009; Cyg X-2 and XTE J1710-281, Jonker \& Nelemans 2004).
However we note that the $A_{V}$ value determined from the spectrum, and compared with the galactic $A_{V}$, is too low for that distance of 21 kpc because this distance would imply that the object is located in the far side of the Galaxy and thus the $A_{V}$ should be close to the value in Schlegel et al. (1998) and not just a minor fraction. One possibility is that the USNO magnitude is understimated: if the object is actually brighter than its tabulated optical magnitude, it would lie closer. Indeed, the USNO photometry can sometimes be offset by up to 0.3-0.4 magnitudes (e.g. Masetti et al. 2003). Alternatively, the absolute magnitudes that we assumed for the LMXB and CV cases are not correct: the object may either be a low optical luminosity LMXB or a bright CV. This, again, would alter the real distance of the object. In conclusion, we cannot give a precise identification for this object, although we deem more probable that it may be a distant LMXB.

On the other hand, in order to get clues on the measured distance to 2PBC J1911.4+1412, we noted that the object is tabulated in the proper motion catalogue (PPMXL) of Roeser et al. (2010). Unfortunately, the error in the measured proper motion is larger than the proper motion value (pmRA = 2.0 $\pm$ 6.3 mas/yr, pmDEC = -5.0 $\pm$ 6.3 mas/yr), therefore only an upper limit to the velocity ($<$ 130 km/s in case of a distance of 21 kpc) can be given. We note that this value is typical of LMXBs.

\subsection{CVs}

Three sources in our 2PBC sample display emission lines of the Balmer complex, as well as He I and He II, consistent with z = 0, indicating that these objects lie within our Galaxy (see Fig. 2). The analysis of their optical features indicates that all of them are CVs (Table 5). The $A_{V}$ of these objects are consistent with $\sim$ 0 according to their $\frac{H_{\alpha}}{H_{\beta}}$ flux ratios: when compared with the Galactic one along their direction according to Schlegel et al. (1998) ($A_{V}$ $\sim$ 1.9, $\sim$ 4.1 and $\sim$ 1.3 mag for 2PBC J0658.0-1746, 2PBC J0717.8-2156 and 2PBC J0800.5-4306, respectively) this confirms their galactic nature and their proximity to Earth. 

Through the equivalent widths (EW) of the H$_{\beta}$ and He II$_{\lambda 4686}$ lines we investigated their magnetic nature, i.e. the possibility that they are harboring a magnetized white dwarf (WD) and thus that they may be polar or intermediate polar (IP) CVs. The three objects classified as CVs have EWs of both He II$_{\lambda 4686}$ and H$_{\beta}$ larger than 10 \AA, and for 2PBC J0658.0$-$1746 and 2PBC J0717.8$-$2156 the He II$_{\lambda 4686}$/H$_{\beta}$ EW ratio is higher than 0.5, implying that both objects are magnetic CVs likely belonging to the IP subclass, whereas 2PBC J0800.5$-$4306 is likely a polar CV because its He II$_{\lambda 4686}$/H$_{\beta}$ EW ratio is higher than 1 (see Warner 1995). Optical - NIR time resolved spectrophotometric followup is thus mandatory to confirm the magnetic nature of all of these CVs. 

We note that, the object 2PBC J0717.8$-$2156 was analysed by Halpern \& Thorstensen (2015) who indicated a CV nature for it; however, the optical spectrum of this source published in that paper was not flux calibrated. Here we complete the information on this object by presenting a flux-calibrated spectrum and by confirming the (possibly magnetic) CV nature for it.

The main spectroscopic results and the main astrophysical parametres, which 
can be inferred from the available observational data, are reported in Table 5.

\begin{table*}
\caption[]{Synoptic table containing the main results concerning the 
3 CVs identified in the present sample of {\it 2PBC} sources.}
\scriptsize
\vspace{-.3cm}
\begin{center}
\begin{tabular}{lccccccccr}
\noalign{\smallskip}
\hline
\hline
\noalign{\smallskip}
\multicolumn{1}{c}{Object} & \multicolumn{2}{c}{H$_\alpha$} & 
\multicolumn{2}{c}{H$_\beta$} & \multicolumn{2}{c}{He {\sc ii} $\lambda$4686} & 
$R$  & $d$ & \multicolumn{1}{c}{$L_{\rm X}$} \\
\cline{2-7}
\noalign{\smallskip} 
 & EW & Flux & EW & Flux & EW & Flux & mag  & (pc) & \\

\noalign{\smallskip}
\hline
\noalign{\smallskip}

2PBC J0658.0$-$1746 & 20$\pm$2 & 48$\pm$3 & 19$\pm$2 & 55$\pm$3 & 10$\pm$1 & 31$\pm$4 & 
 15.4  & $\sim$190 & 3.4 (15-150)$^{C}$  \\
 & & & & & &  & & & 8.1 (0.3-10)$^{X}$ \\ 
 & & & & & &  & & &  \\
 
2PBC J0717.8$-$2156 & 30.6$\pm$1.4 & 5.20$\pm$0.15 & 12.7$\pm$0.4 & 1.9$\pm$0.1 & 8.3$\pm$0.2 & 1.3$\pm$0.1 & 
 16.1  & $\sim$260 & 4.9 (15-150)$^{C}$  \\
  & & & & & &  & & & 2.8 (0.3-10)$^{X}$\\ 
   & & & & & &  & & & 6.4 (14-195)$^{B}$\\ 
 & & & & & & &  & &  \\

2PBC J0800.5$-$4306 & 34$\pm$3 & 10.7$\pm$0.4 & 18.8$\pm$2.3 & 9.5$\pm$0.4 & 24$\pm$9 & 13$\pm$4 & 
 16.8  & $\sim$360 & 12.4 (15-150)$^{C}$  \\
  & & & & & &  & & & 9.6 (0.3-10)$^{X}$\\ 
    & & & & & &  & & & 5.8 (20-40)$^{I}$\\ 
      & & & & & &  & & & 10.1 (14-195)$^{B}$\\ 

\noalign{\smallskip} 
\hline
\noalign{\smallskip} 
\multicolumn{10}{l}{Note: EWs are expressed in \AA, line fluxes are
in units of 10$^{-15}$ \textit{erg cm$^{-2}$ s$^{-1}$}, X--ray luminosities
are in units of} \\
\multicolumn{10}{l}{10$^{31}$ erg s$^{-1}$, and the reference band (between 
round brackets) is expressed in keV.} \\
\multicolumn{10}{l}{In the last column, the upper case letter indicates the satellite and/or the 
instrument with which the} \\
\multicolumn{10}{l}{$^{C}$: from \it{Swift}/BAT catalogue (Cusumano et al. 2010), $^{X}$: from \it{Swift}/XRT, 
$^{B}$: from \it{Swift}/BAT (Baumgartner et al. 2013) } \\ 
\multicolumn{10}{l}{and $^{I}$: from \textit{INTEGRAL}/IBIS catalogue (Bird et al. 2016).} \\
\noalign{\smallskip} 
\hline
\hline
\noalign{\smallskip} 
\end{tabular} 
\end{center}
\end{table*}

\subsection{Active stars}


\begin{figure}[htbp]
\begin{center}
\includegraphics[height=6.0cm]{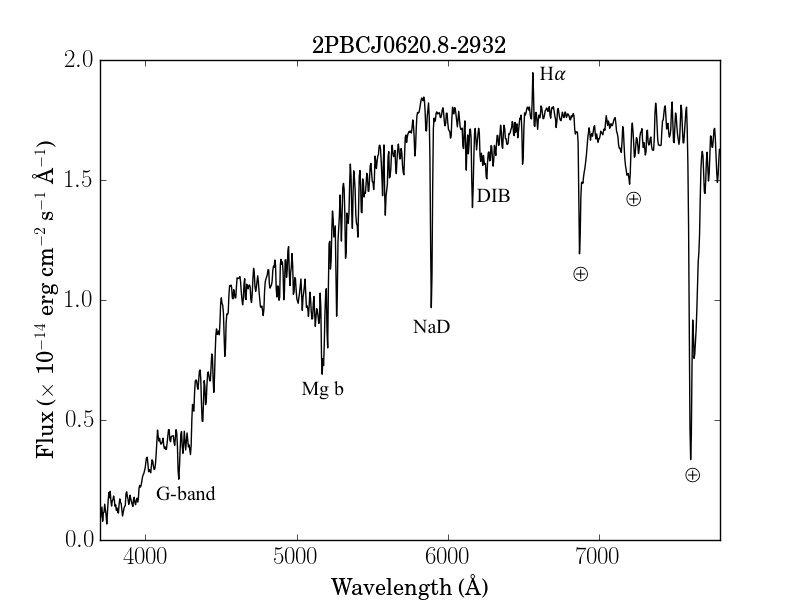}
\begin{spacing}{1}
\caption{Spectra (not corrected for the intervening Galactic absorption) of the optical spectra acquired for 2PBC J0620.8$-$2932, the optical counterpart of the active star belonging to the sample of BAT sources presented in this paper. For the spectrum, the main spectral features are labeled and $\oplus$ indicates telluric absorptions.}
\end{spacing}
\end{center}
\end{figure}

\begin{table*}
\caption[]{Main observational results for the active star as identified in 
the present sample of {\it 2PBC} sources.}
\scriptsize
\setlength{\tabcolsep}{3pt} 
\begin{center}
\begin{tabular}{lccccc}
\noalign{\smallskip}
\hline
\hline
\noalign{\smallskip}
\multicolumn{1}{c}{Object} & \multicolumn{2}{c}{H$_\alpha$} & 
$R$  & $d$ & \multicolumn{1}{c}{$L_{\rm X}$} \\
\cline{2-3}
\noalign{\smallskip} 
  & EW & Flux & mag  & (pc) & \\

\noalign{\smallskip}
\hline
\noalign{\smallskip}

2PBC J0620.8$-$2932 & 0.8$\pm$0.2  &   14.3$\pm$2.5 & $\sim$14.7  & $\sim$1500 & 108 (15-150)$^{C}$ \\
 &  &  &  &    & 6.81 (0.3-10)$^{X}$ \\ 

\noalign{\smallskip} 
\hline
\noalign{\smallskip}
\multicolumn{6}{l}{Note: EWs are expressed in \AA, line fluxes are
in units of 10$^{-15}$ \textit{erg cm$^{-2}$ s$^{-1}$}, } \\
\multicolumn{6}{l}{X--ray luminosities are in units of 10$^{31}$ erg s$^{-1}$, and the 
reference band (between round } \\
\multicolumn{6}{l}{brackets) is expressed in keV. In the last column, the upper case letter indicates the } \\
\multicolumn{6}{l}{satellite and/or the instrument with which the corresponding X--ray flux measurement } \\
\multicolumn{6}{l}{was obtained (see text). $^{C}$: from Cusumano et al. (2010).} \\
\noalign{\smallskip}
\hline
\hline
\end{tabular} 
\end{center} 
\end{table*}

The object 2PBC J0620.8$-$2932 shows the optical features of an active star in its spectrum (Fig. 3). We recall that this object is within the BAT error box at 99\% confidence level and does not display x-ray emission at above 3 keV in the XRT observations; nevertheless, we suggest that this source can be classified as a chromospherically active star, with a star-like continuum typical of late-G/early-K type stars, a faint emission of $H_{\alpha}$ and absorption bands such as G, NaD and Mg b at z = 0, and we think that the flux above 3 keV from this source is possibly transient and could be due to flares (Sguera et al. 2016; Krivonos et al. 2007; MVI; MIX;  MX). This conclusion is supported by the comparison between hard and soft fluxes of the source in Table 6, which is $F_{15-150 keV}$/$F_{0.3-10 keV}$ $\approx$ 16 indicating that the star could have been undetected by BAT during a flare. 

An estimate for the distance to this object can be evaluated, as suggested in Rodriguez et al. (2010), assuming similarity with the active star II Peg  (which has magnitude R $\sim$ 6.9 and distance 40 pc; Monet et al. 2003, van Leeuwen 2007). We considered that 2PBC J0620.8$-$2932 has a USNO-A2.0 magnitude R $\sim$ 14.7 and we consequently obtain the distance reported in Table 6, which is actually an upper limit since we considered $A_{V}$ $\sim$ 0 mag because we can not determine the reddening along the source line of sight with the presently available information.

\begin{table*}[th!]
\caption[]{Synoptic table containing the main results for the
QSO identified in the 
present sample of {\it 2PBC} sources.}
\scriptsize
\setlength{\tabcolsep}{6pt} 
\begin{center}
\begin{tabular}{lcccccc}
\noalign{\smallskip}
\hline
\hline
\noalign{\smallskip}
\multicolumn{1}{c}{Object} & $F_{\rm [OII]}$ & $F_{\rm MgII}$ & $z$ &
\multicolumn{1}{c}{$D_L$ (Mpc)} & $E(B-V)_{\rm Gal}$ & \multicolumn{1}{c}{$L_{\rm X}$} \\
\noalign{\smallskip}
\hline
\noalign{\smallskip}

2PBC J2010.3$-$2522 & 0.3$\pm$0.1 & 1.3$\pm$0.1  & 0.828 & 5234.4 & 0.164 & 42.6 (15-150)$^{C}$ \\ 
 & [0.6$\pm$0.2] & [1.8$\pm$0.3]  & & & & 4.03  (0.3-10)$^{X}$  \\ 
       & & & & & & 47.7 (14-195)$^{B}$\\ 

\noalign{\smallskip} 
\hline
\noalign{\smallskip} 
\multicolumn{7}{l}{Note: emission-line fluxes are reported both as 
observed and (between square brackets) corrected for the intervening} \\ 
\multicolumn{7}{l}{ Galactic absorption $E(B-V)_{\rm Gal}$ along the object line of sight 
(from Schlegel et al. 1998). Line fluxes are in units} \\
\multicolumn{7}{l}{ of 10$^{-15}$ \textit{erg cm$^{-2}$ s$^{-1}$}, X--ray luminosities are in units of 10$^{45}$ erg s$^{-1}$,
and the reference band (between round } \\
\multicolumn{7}{l}{brackets) is expressed in keV. In the last column, the upper case letter indicates the satellite and/or the 
instrument } \\
\multicolumn{7}{l}{with which the corresponding X--ray flux measurement was obtained. } \\
\multicolumn{7}{l}{$^{C}$: from Cusumano et al. (2010) and $^{X}$: from $\it{Swift}$/XRT. $^{B}$: from Baumgartner et al. (2013).} \\
\multicolumn{7}{l}{ The typical error of the redshift measurement is $\pm$0.001.} \\
\noalign{\smallskip} 
\hline
\hline
\end{tabular} 
\end{center} 
\end{table*}

\subsection{AGNs}


The results for the extragalactic sources are reported in Tables 7, 8 and 9, where for each source we list fluxes, the classification, the estimated redshift, the luminosity distance (in Mpc), the Galactic colour excess, the colour excess local to the AGN host. 

We found that among 45 extragalactic objects, 27 have strong redshifted broad and narrow emission lines that are typical of type 1 AGNs (Figs. 4, 5 and 6) and 18 have only strong and redshifted narrow emission lines that are typical of a type 2 AGN nature (see Figs. 7, 8, 9 and 10).

In our sample, we report the redshift value for most AGNs for the first time and most of them are in the range 0.004-0.294, that is, the AGNs of our sample are located in the local Universe (z $<$ 0.3) except 2PBC J2010.3$-$2522, the QSO reported in Table 5 which is at redshift 0.828.

Among our AGN sample, 20 objects are intermediate type Sy 1.2-1.5 and 5 are Sy 1.9, whereas of the 18 type 2 AGNs, 14 are Seyfert 2 galaxies, one is a starburst galaxy (2PBC J1555.0$-$6225, Fig. 8), one is a LINER (2PBC J2238.9+4050, Fig. 9) and one is a X-ray bright optically normal galaxy (XBONG, Comastri et al. 2002) which displays absorption features only (2PBC J1332.1$-$7751, Fig. 10).

For type 1 AGNs, black hole masses and Eddington luminosities are reported in Table 10, where we corrected the measurement of the observed FWHM of $H_{\beta}$ for the instrumental resolution corresponding to the relevant telescope setup (see Tables 2 and 3) in order to determine the BH masses.
We only report the $H_{\alpha}$ emission for narrow line sources (Table 9) as the $H_{\alpha}$ emission required more complex fits and sometimes had issues with blending of [NII] emission lines and atmospheric lines.

For three Sy 2 objects, 2PBC J0356.6$-$6252, 2PBC J0608.0$+$5749 and 2PBC J2029.4$-$6146, $ \left(\frac{F_{H_{\alpha}}}{F_{H_{\beta}}}\right) _{*}$ is less than 2.86, which implies that they can be type 2 AGN without local absorption (these are called $''$naked$''$ Sy 2; Hawkins 2004). But, in particular, the object 2PBC J0356.6$-$6252 has strong IR emission due to dust according to Yamada et al. 2013 and the analysis of its X-ray spectra according to Ricci et al. (2016) indicates a probably Compton thick nature, thus a more detailed study is needed to confirm the nature of this source.

As can be seen in Figure 6, the spectrum of 2PBC J1548.5$-$3208 displays a composition of an AGN type Sy 1.9 at redshift 0.048 and a foreground star due to the inability of our setup to disentangle the pair of sources.


\begin{figure}[htbp]
\begin{center}
\includegraphics[height=6.0cm]{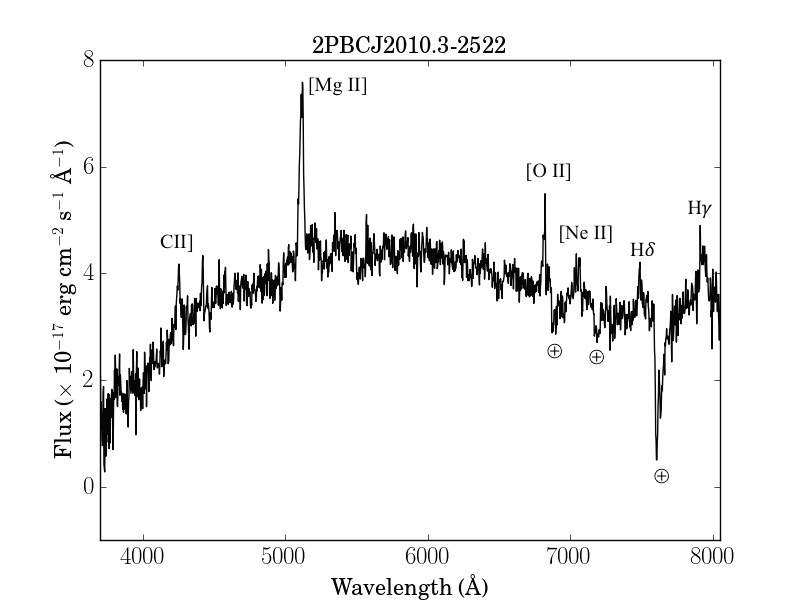}
\begin{spacing}{1}
\caption{Spectrum (not corrected for the intervening Galactic absorption) of the optical counterpart of the QSO belonging to the sample of BAT sources presented in this paper, 2PBC J2010.3-2522. For this spectrum, the main spectral features are labeled and $\oplus$ indicates telluric absorptions.} 
\end{spacing}
\end{center}
\end{figure}

We note that for the object 2PBC J0819.2$-$2508, Cusumano et al. (2010) proposed a ROSAT soft X-ray counterpart in the catalogue (1RXS J081915.0$-$251106) and thus we acquired and analysed the spectrum of the optical object associated with the ROSAT source. However, we subsequently found in the XRT pointings of this field that the source analysed here has no emission above 3 keV; rather, there is another source inside the corresponding BAT error box of 2PBC J0819.2$-$2508 with emission above 3 keV (labeled as "XRT object" in Fig. A.6 in Appendix A, left panel). This second source has coordinates (J2000) RA = 08:19:16.2, DEC = $-$25:07:06.9 with error radius of 6$''$ and is positionally associated with a R $\approx$ 19 optical source. We think that the proposed ROSAT soft X-ray counterpart in Cusumano et al. (2010) could actually not be the real soft X-ray counterpart of 2PBC J0819.2$-$2508 and instead this other source could be; at the very least, both can be contributing to the detected hard X-ray emission. The nature of this object and, in turn, an optical spectrum of this other source is required in order to confirm if this is a possible soft X-ray counterpart to 2PBC J0819.2$-$2508. 

Similarly, the object 2PBC J1520.2$-$0433 has a proposed ROSAT soft X-ray counterpart in the Cusumano et al. (2010) catalogue, 1RXS J152017.4-043621 (see Fig. A.6 in Appendix A right panel, in magenta). We, however, found in the XRT observation of this field that this ROSAT source has no X-ray emission inside its error circle. Rather, there is another source inside the 90\% BAT error circle of 2PBC J1520.2$-$0433, indicated with the black arrow in that figure, with emission above 3 keV. Thus, we considered this latter source as the actual soft X-ray counterpart for the optical follow-up. \\
 

As mentioned in Section 2, the object 2PBC J1020.5$-$0235 has 2 hard X-ray emitting sources just outside the 99\% BAT error circle (see Fig. A.5 in Appendix A), listed as SWIFT J1020.5$-$0237A and SWIFT J1020.5$-$0237B in the 70-month \textit{Swift}/BAT catalogue of Baumgartner et al. (2013); in this work, we analysed the object A and we identified it as a Sy 2 at z = 0.294 (see Table 9 and Figure 7). \\


For the object 2PBC J2010.3$-$2522 that we classified as type 1 QSO at z $\sim$ 0.8, we found in the literature that this object could be associated with a Fermi transient J2007$-$2518 that is also spatially coincident with a radio source (Kocevski et al. 2014). Strader et al. (2014) acquired an optical spectrum of this same source and obtained results equivalent to ours. We thus confirm the QSO nature of this object and its possible association to the Fermi transient J2007$-$2518.

We found that the object classified as Starburst galaxy in our sample, 2PBC J1555.0$-$6225, has WISE magnitudes and colors that indicate an AGN nature according to the criteria of Secrest et al. (2015): (W1 - W2) = 0.56, (W2 - W3) = 2.34 $>$ 2. This could be an elusive AGN as in Smith et al. (2014) and we are probably only seeing the starburst contribution in its optical spectrum, with the X-ray emission detected from the hidden nuclear activity.

For the object 2PBC J1332.1$-$7751, classsified as XBONG, Davies et al. (2015) reported a column density in X-ray of Log($N_{H}$) = 23.8 $cm^{-2}$, that is almost in the limit to be a Compton Thick object (with Log($N_{H}$) $\gtrsim$ 24 $cm^{-2}$) and they classified it as Seyfert galaxy, but no optical classification is considered for this object in that work. Thus, 2PBC J1332.1$-$7751 is a peculiar object that is absorbed in X-ray and optically obscured. Our classification as XBONG is then supported by the information from Davies et al. (2015): the AGN optical appearence is likely hidden due to dust obscuration revealed in X-ray spectrum.

We note that in the field of the object 2PBC J0838.7$+$2612, there is another object with emission above 3 keV inside the 99\% BAT error circle, indicated with tick marks (see Fig. A.2 in Appendix A). For the sake of completeness, in Appendix B we thus report the main optical spectroscopic information we gathered about this source (a galaxy pair). Furthermore, the soft X-ray source that we associated with the object 2PBC J2030.7$-$7530 is the hardest source with emission above 3 keV within the BAT error circle (its countrate is at least 20 times larger than the other sources within the BAT error box). In the same way, the object 2PBC J1419.2$+$6804 has a countrate 2 orders of magnitude stronger than the other objects inside the 90\% BAT error circle (see Appendix A).

We then looked for Compton-thick AGNs. In order to explore this nature, we apply the diagnostic T of Bassani et al. (1999), i.e. the ratio of the measured 2-10 keV X-ray flux to the unabsorbed flux of the [OIII]$\lambda$5007 forbidden emission line, to the objects which are classified as type 2 AGNs in our sample.

It is found that all objects show a T parameter $>$ 1 (from 2.5 to 430), so all of them apparently lie in the Compton thin regime, with $N_{H}$ $<$ $10^{24}$ $cm^{-2}$. We however stress that the T values should be considered as upper limits because the soft X-ray fluxes that we considered refer to band which are wider (0.3-10 keV) than the one for which this method should be applied (2-10 keV). 

Moreover, we support this result by applying the criterion of Malizia et al. (2007): we found that all our type 2 AGNs lie in the Compton thin regime with softness ratio Fx$_{soft}$/Fx$_{hard}$ from 0.02 to 0.41. Although we use a different range of soft and hard X-ray range (0.3-10 keV and 15-150 keV instead of 2-10 keV and 20-100 keV), the approach can be considered equivalent and the results comparable.

\subsection{Statistics}

We here give an update of the statistics concerning the identifications of new hard X--ray {\it Swift}/BAT sources including the results of our group in  this work and those in PI, PII and PIII.



\begin{figure*}[htbp]
\centering
\subfigure{\includegraphics[width=0.32\textwidth]{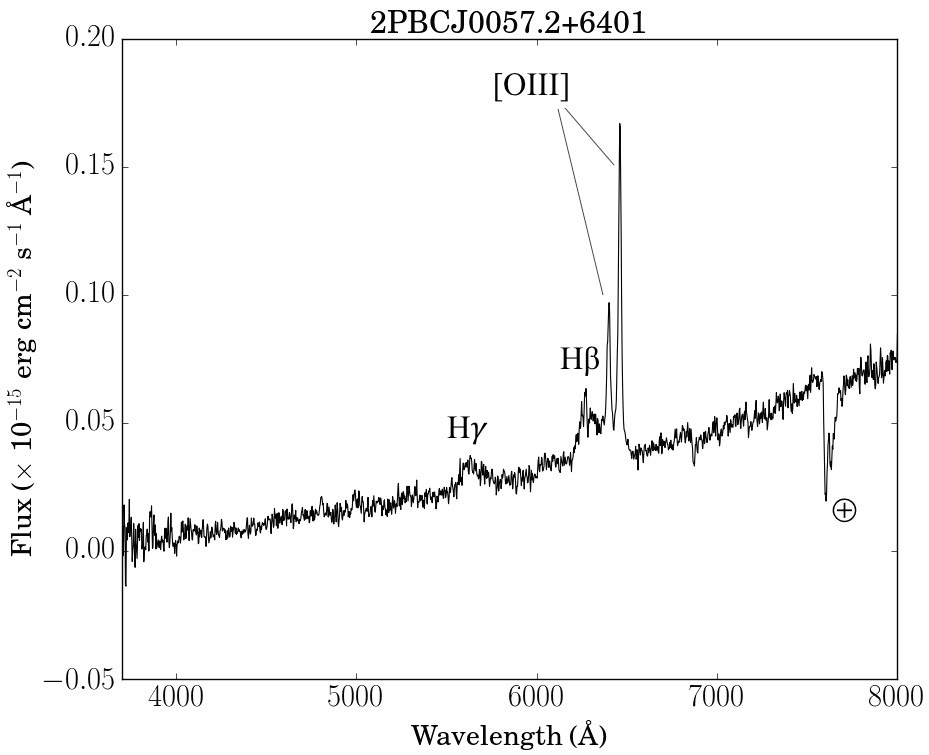}}
\subfigure{\includegraphics[width=0.32\textwidth]{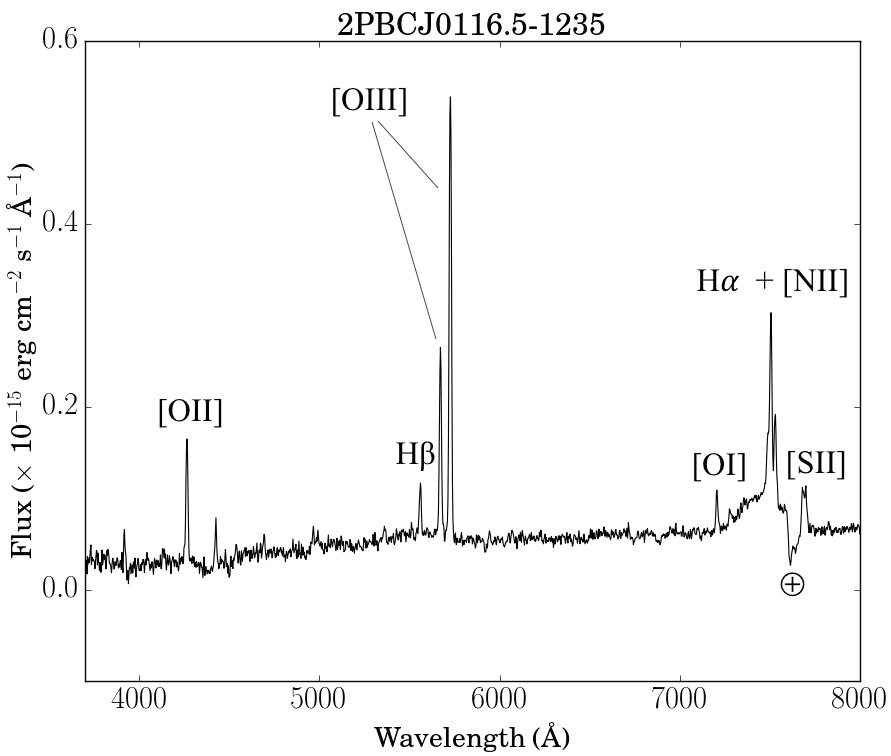}}
\subfigure{\includegraphics[width=0.32\textwidth]{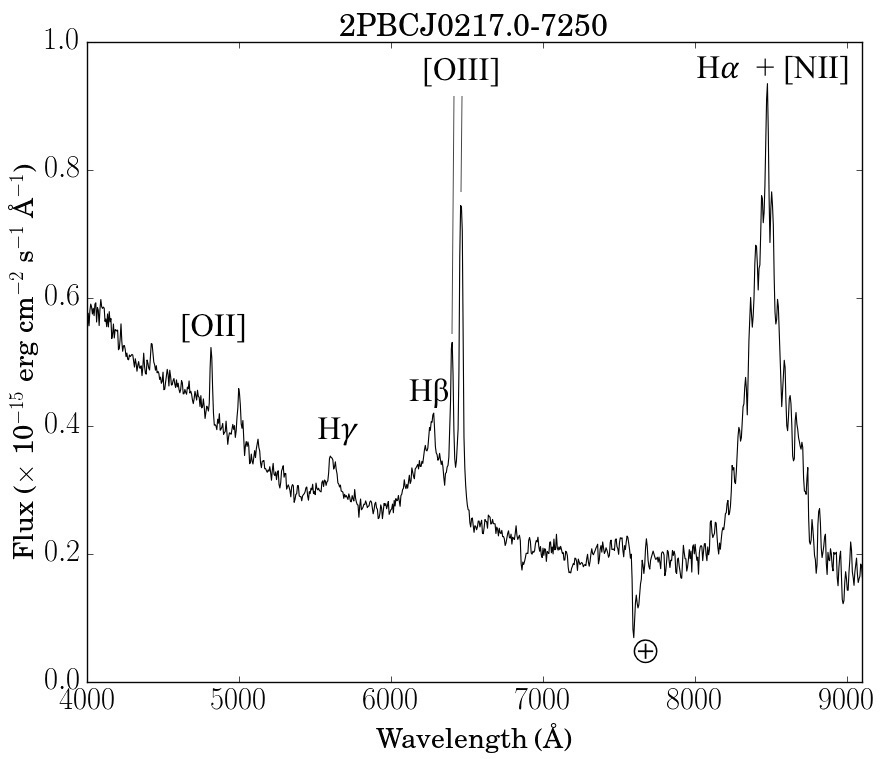}}
\subfigure{\includegraphics[width=0.32\textwidth]{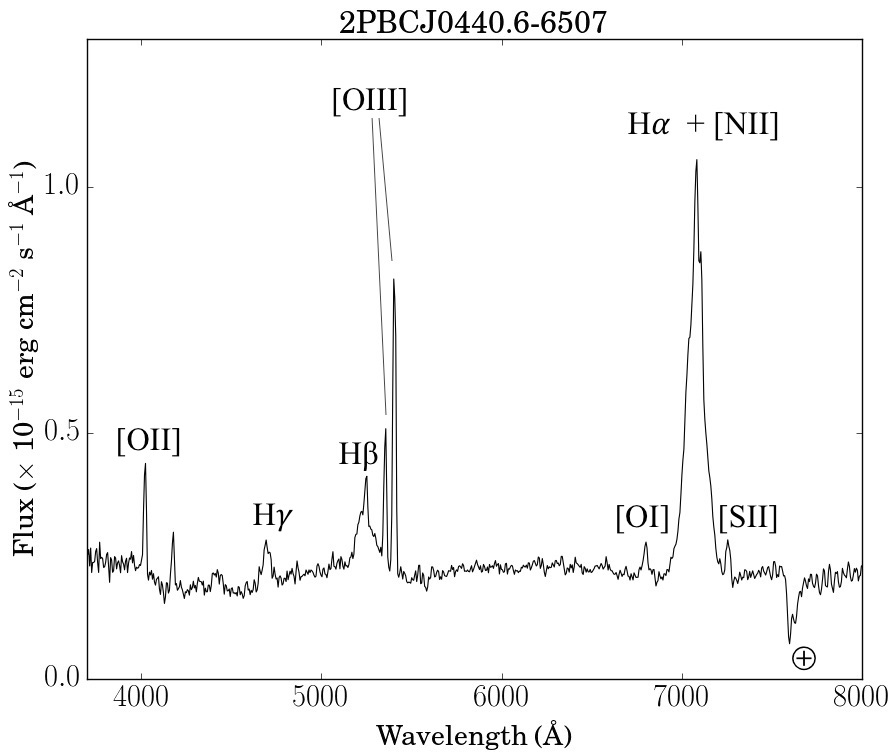}}
\subfigure{\includegraphics[width=0.32\textwidth]{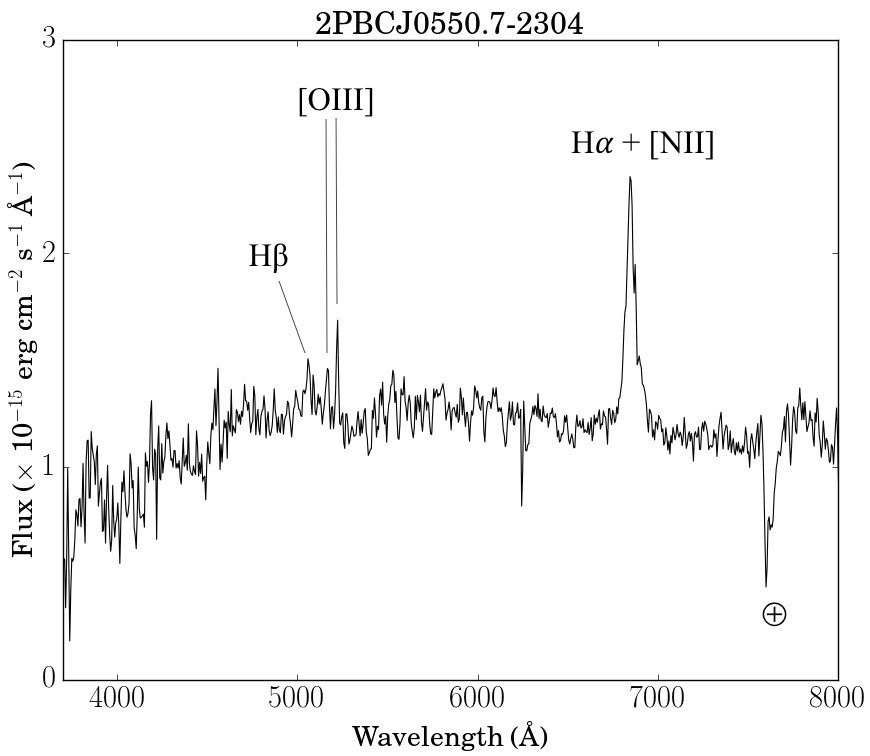}}
\subfigure{\includegraphics[width=0.32\textwidth]{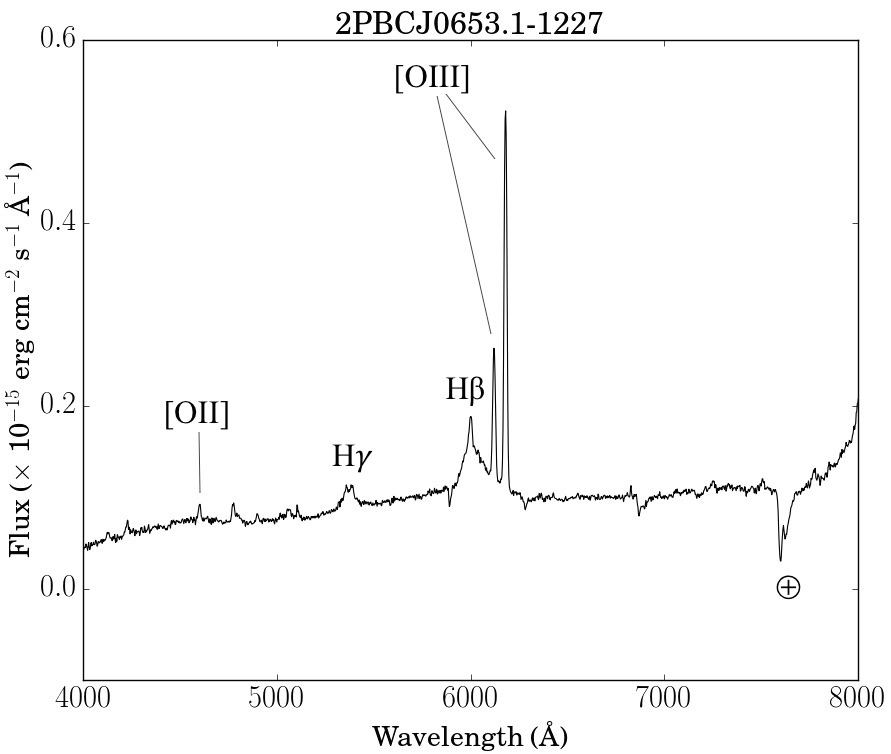}}
\subfigure{\includegraphics[width=0.32\textwidth]{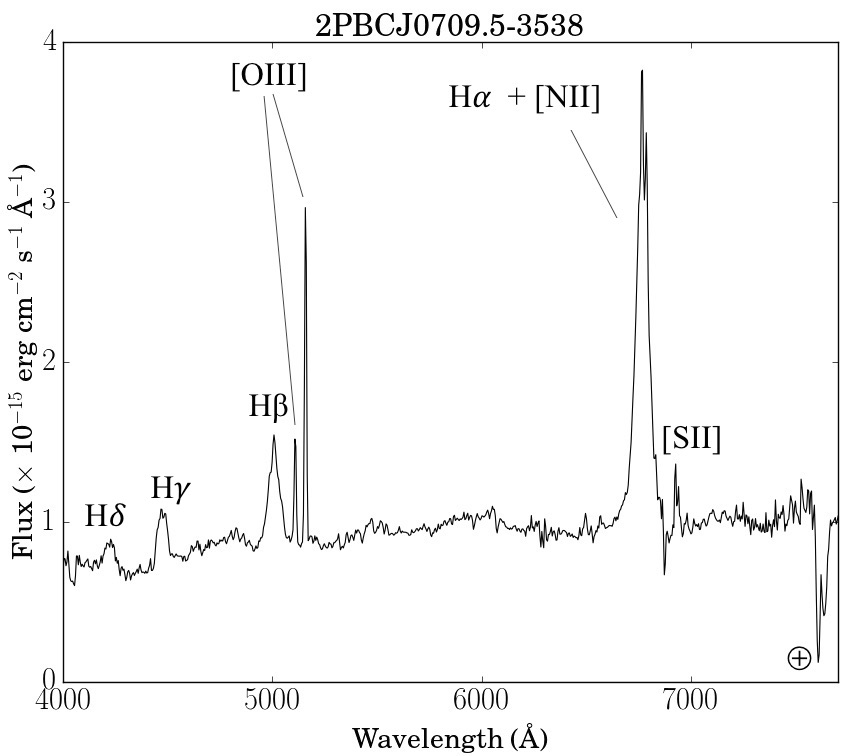}}
\subfigure{\includegraphics[width=0.32\textwidth]{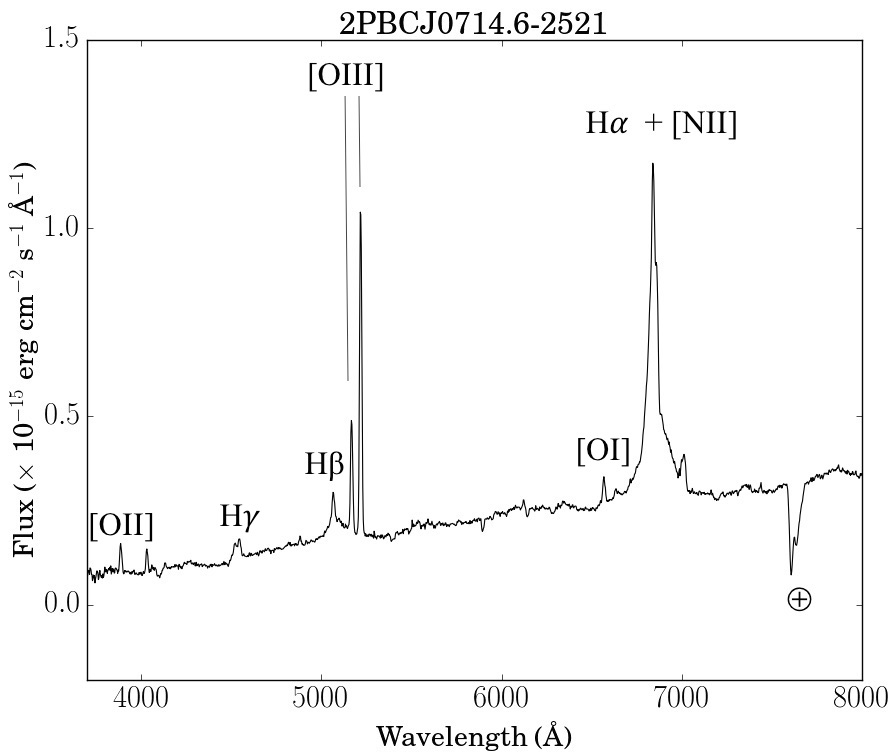}}
\subfigure{\includegraphics[width=0.32\textwidth]{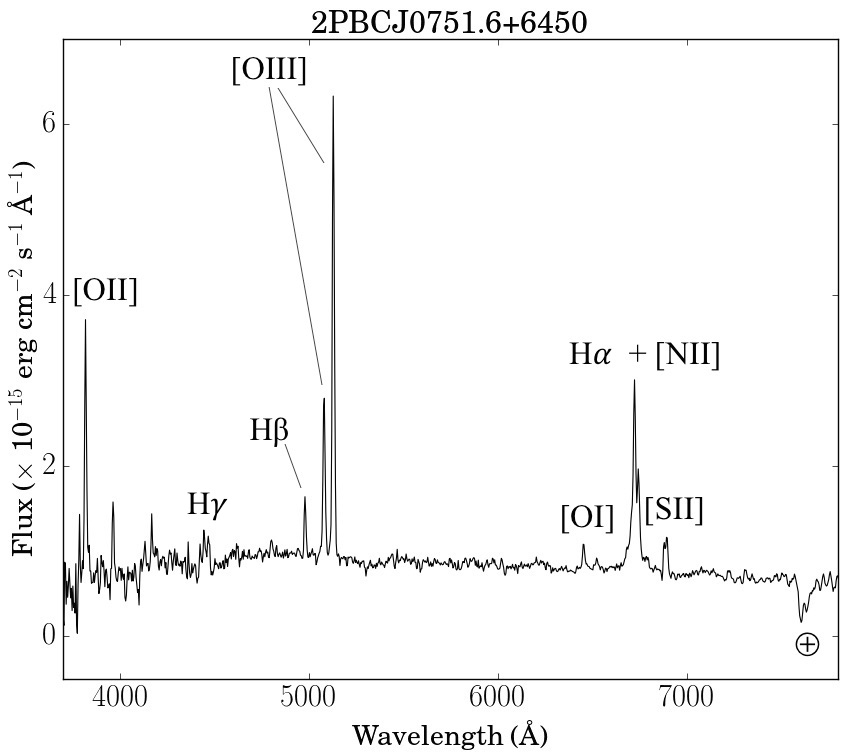}}
\subfigure{\includegraphics[width=0.32\textwidth]{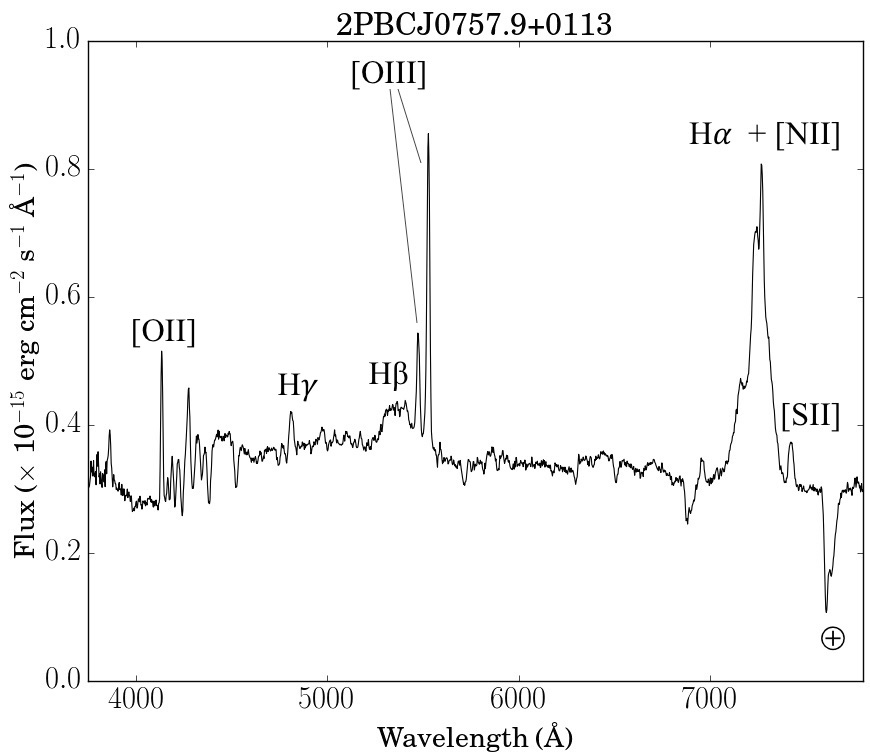}}
\subfigure{\includegraphics[width=0.32\textwidth]{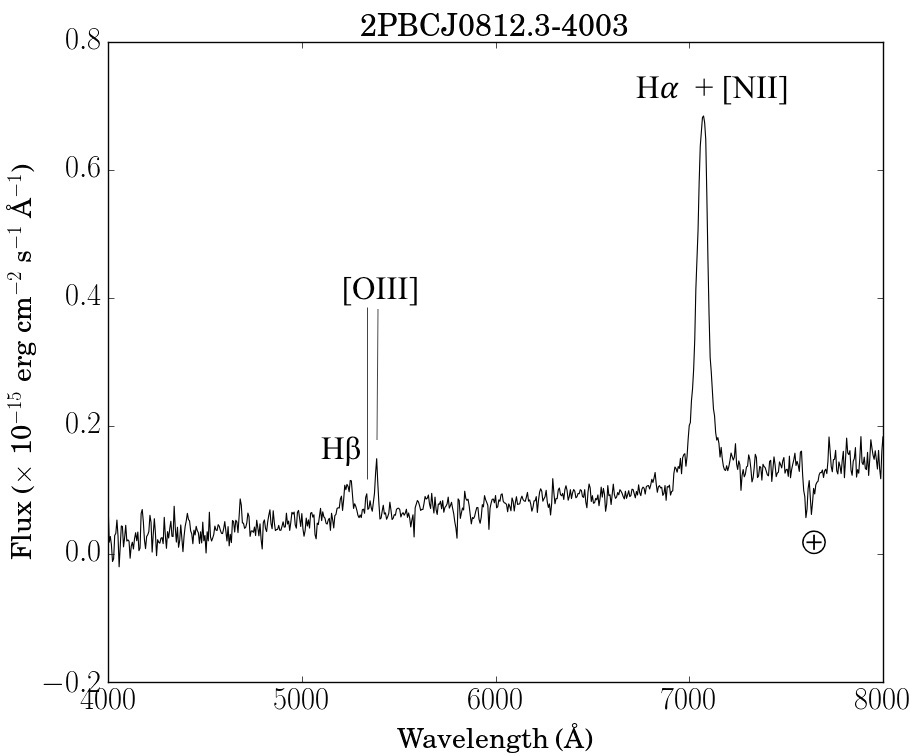}}
\subfigure{\includegraphics[width=0.32\textwidth]{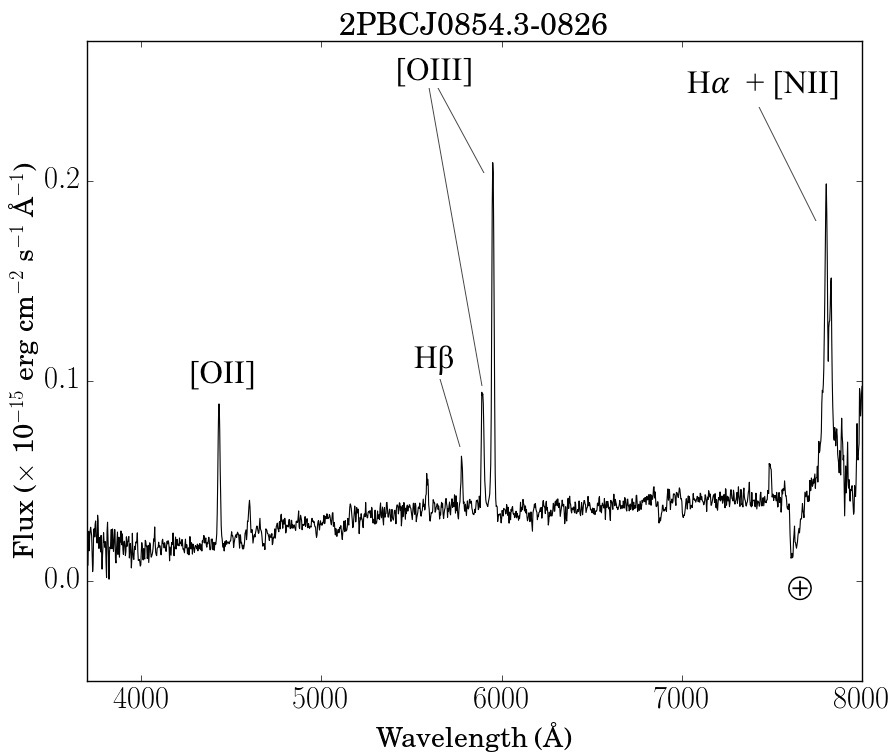}}
\caption{Optical spectra of type 1 AGNs presented in this work (not corrected for the intervening galactic absorption). For each spectrum, the main spectral features are labeled and $\oplus$ indicates telluric absorptions.}
\end{figure*}

\newpage

\begin{figure*}[htbp]
\centering
\subfigure{\includegraphics[width=0.3\textwidth]{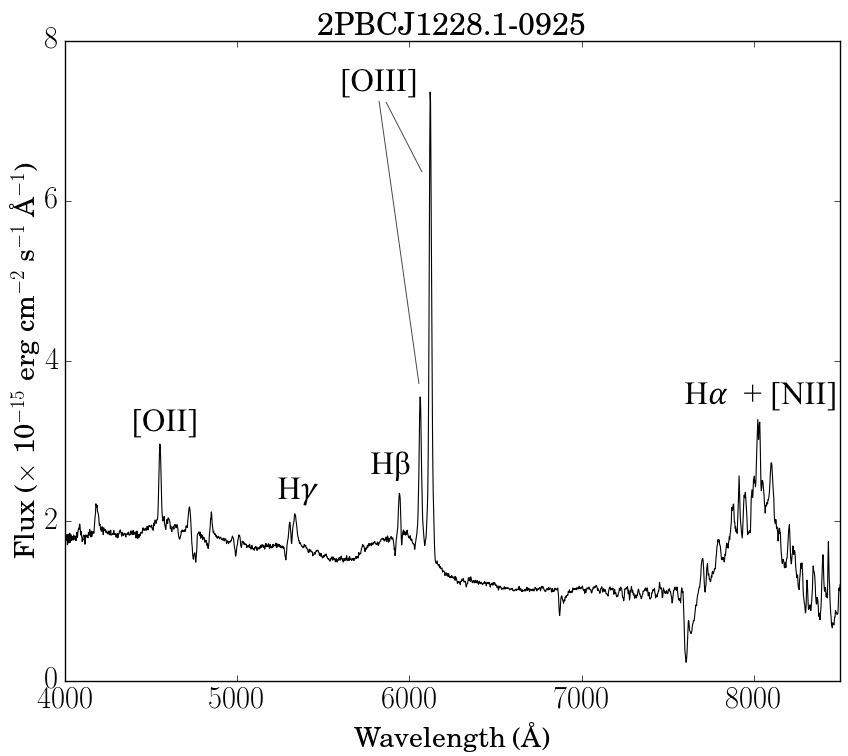}}
\subfigure{\includegraphics[width=0.3\textwidth]{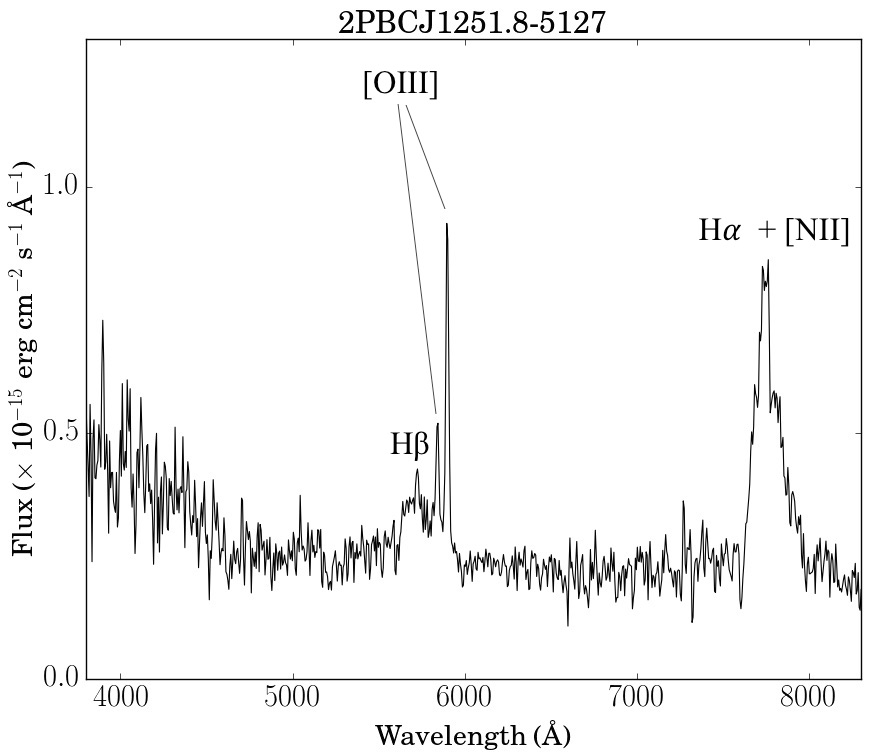}}
\subfigure{\includegraphics[width=0.3\textwidth]{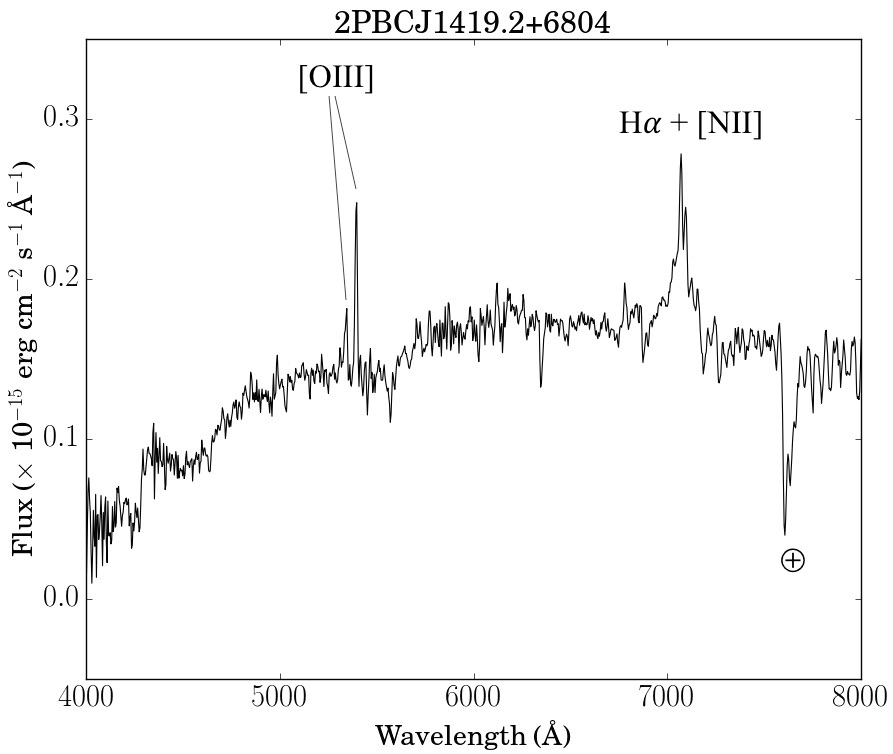}}
\subfigure{\includegraphics[width=0.3\textwidth]{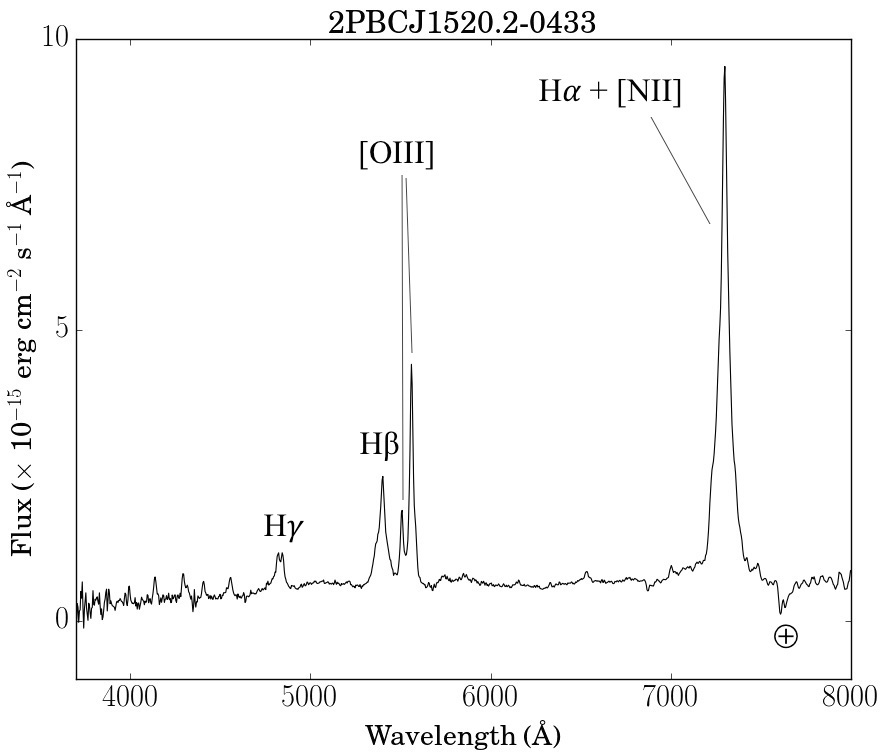}}
\subfigure{\includegraphics[width=0.3\textwidth]{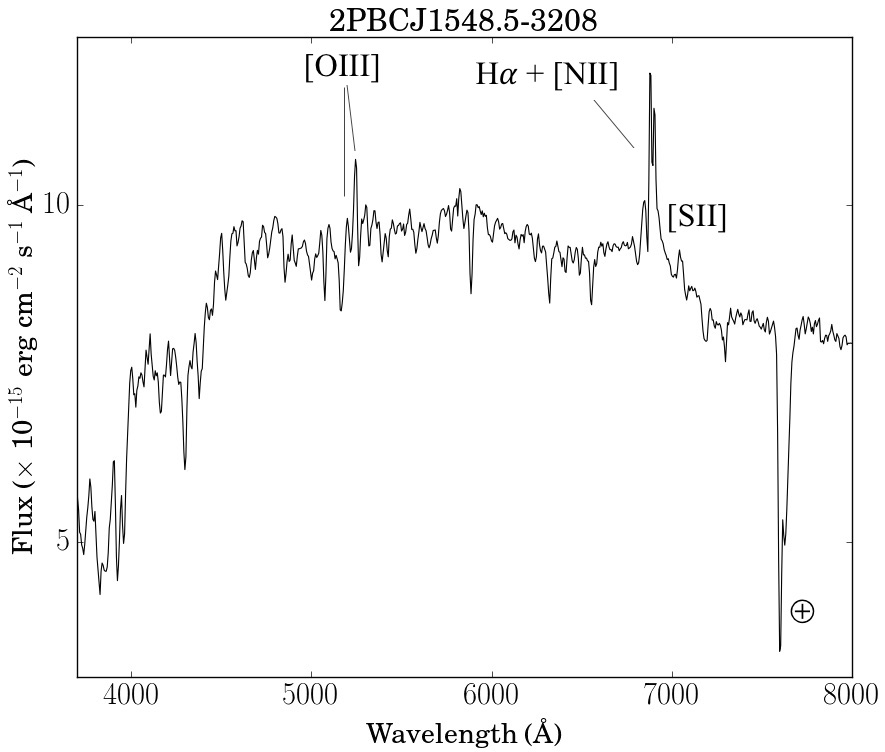}}
\subfigure{\includegraphics[width=0.3\textwidth]{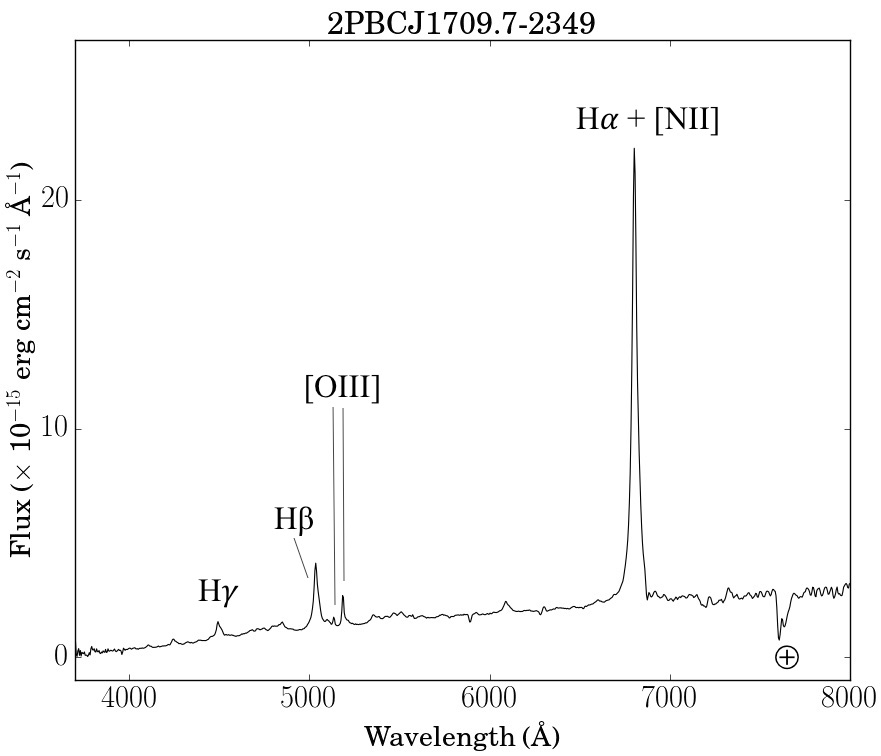}}
\subfigure{\includegraphics[width=0.3\textwidth]{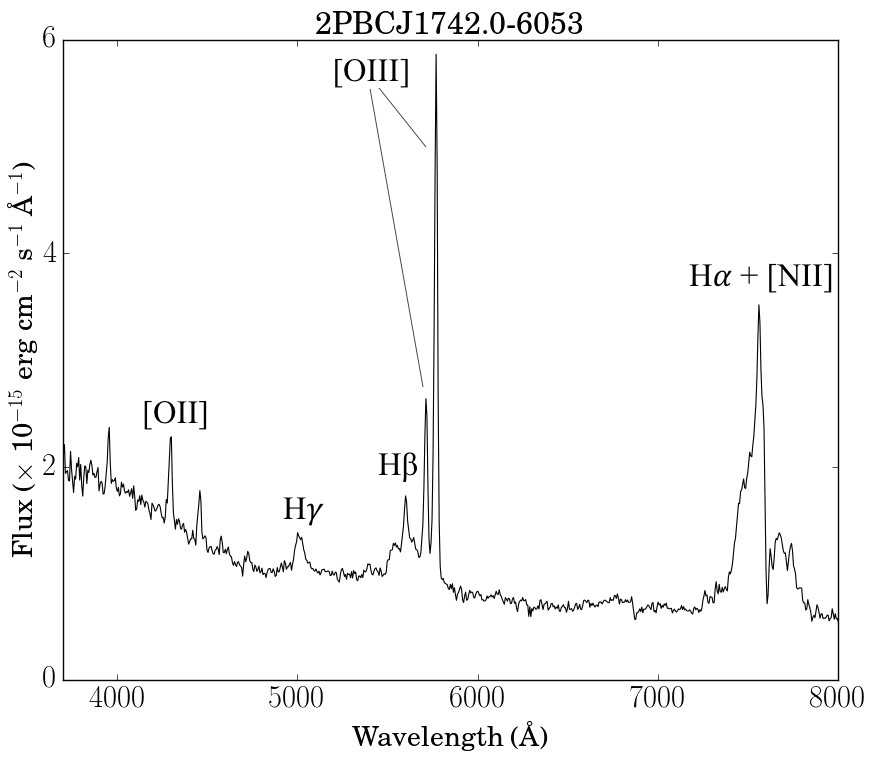}}
\subfigure{\includegraphics[width=0.3\textwidth]{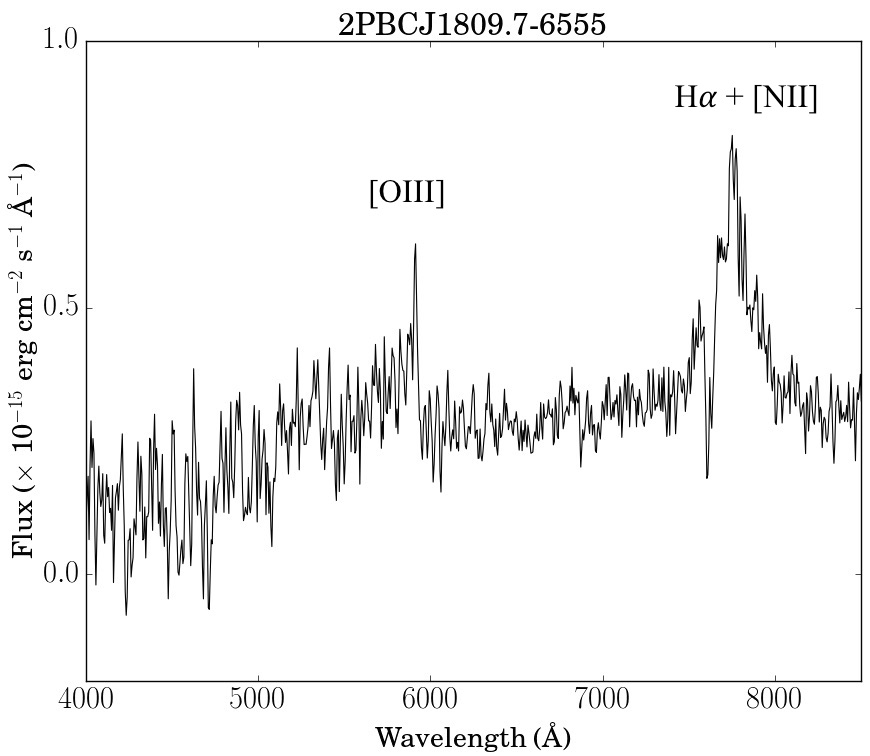}}
\subfigure{\includegraphics[width=0.3\textwidth]{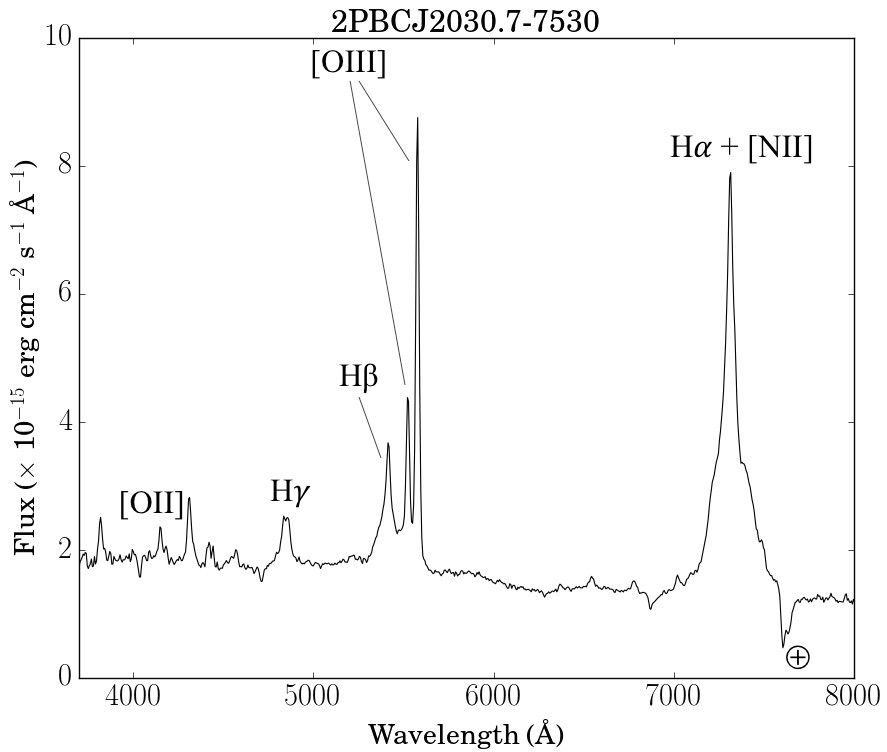}}
\subfigure{\includegraphics[width=0.3\textwidth]{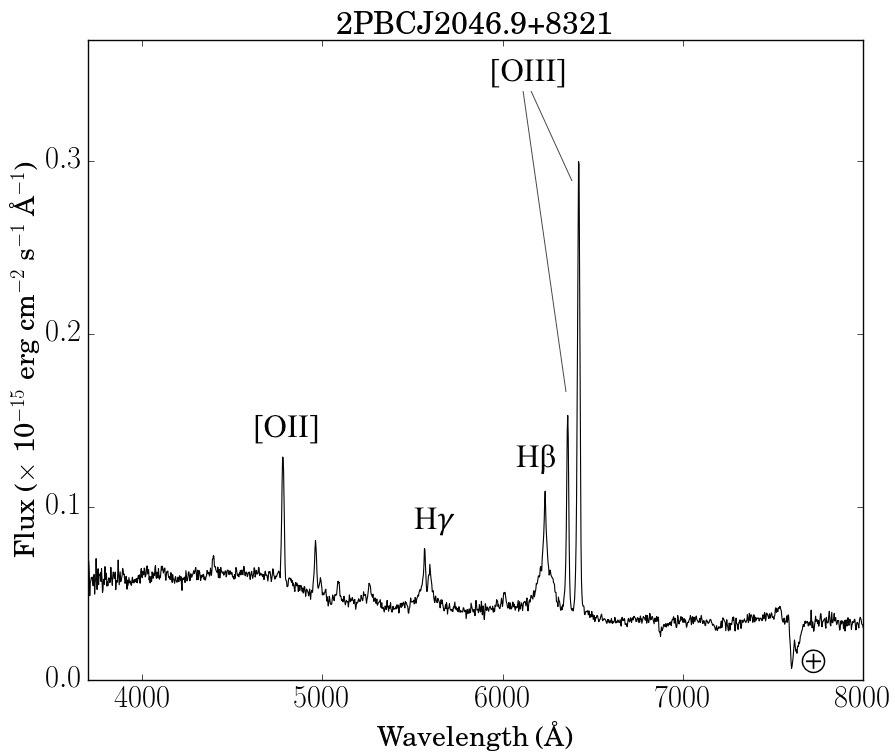}}
\subfigure{\includegraphics[width=0.3\textwidth]{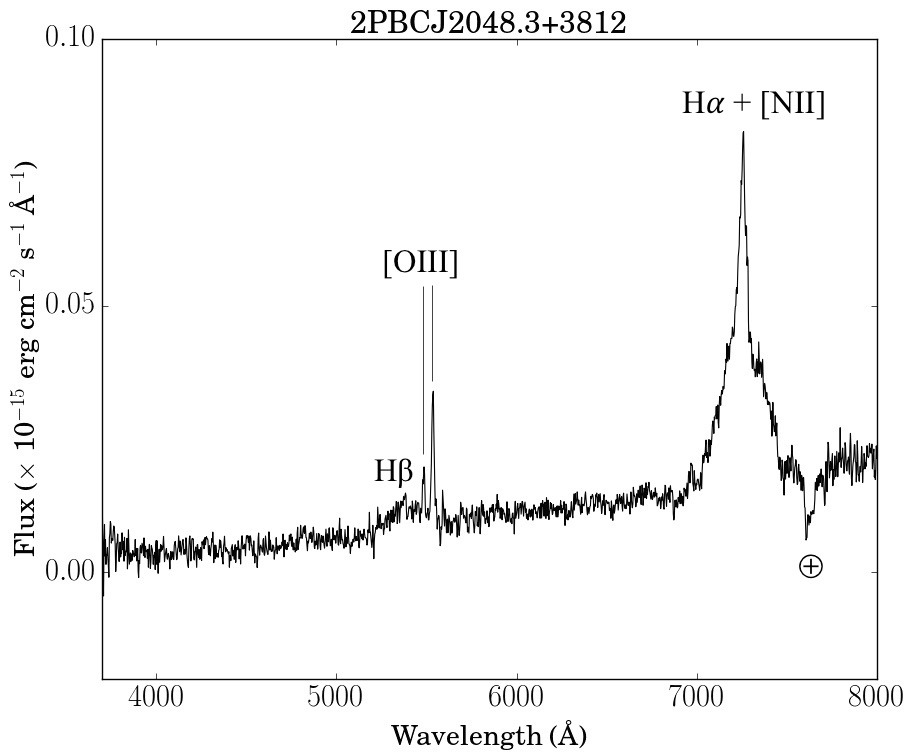}}
\subfigure{\includegraphics[width=0.3\textwidth]{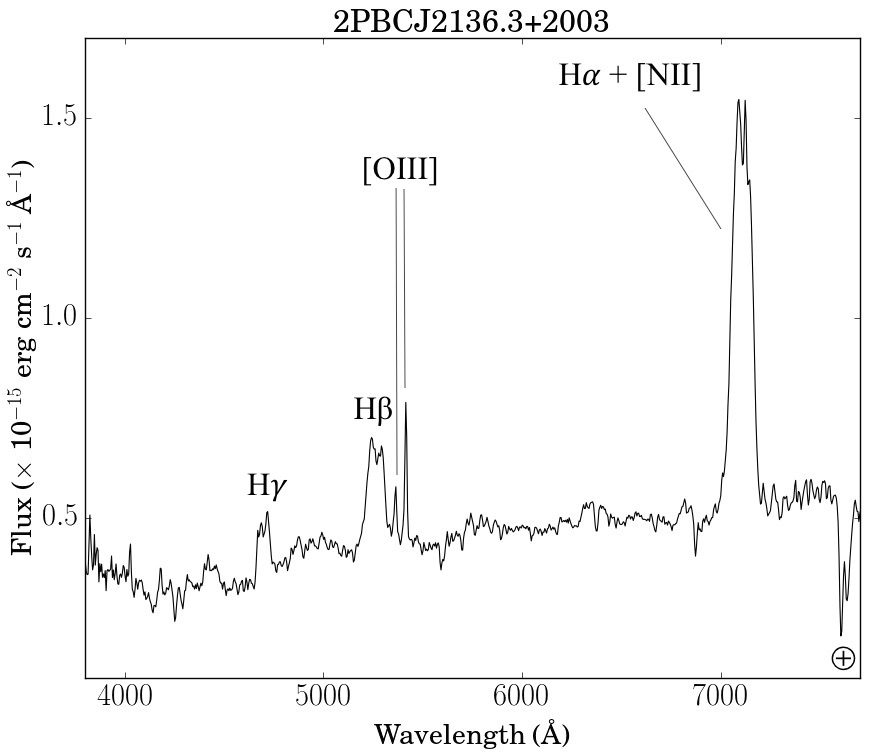}}
\subfigure{\includegraphics[width=0.3\textwidth]{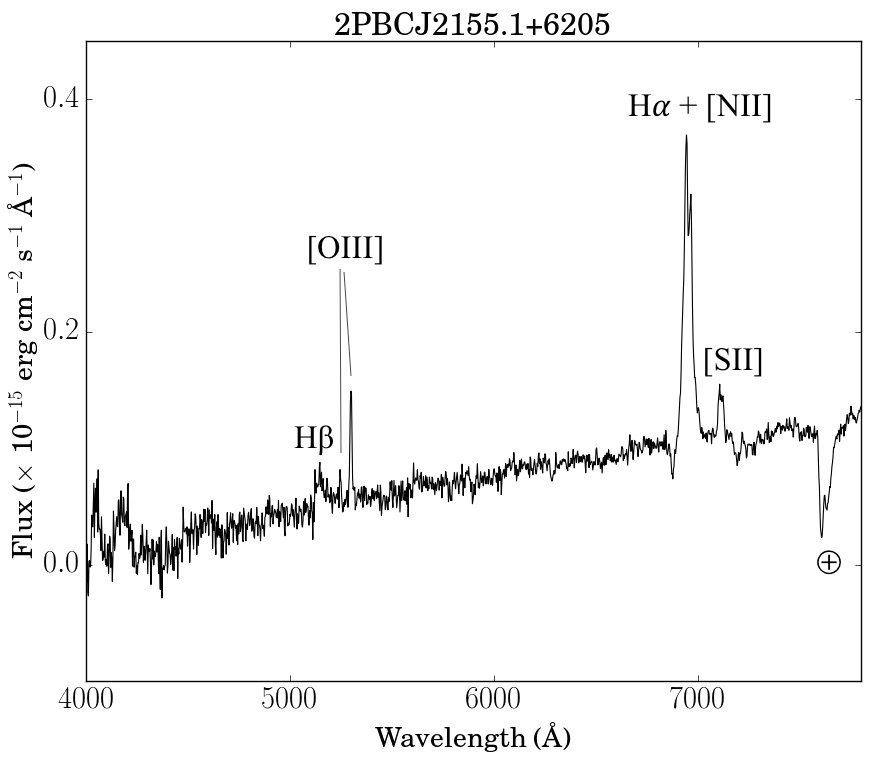}}
\subfigure{\includegraphics[width=0.3\textwidth]{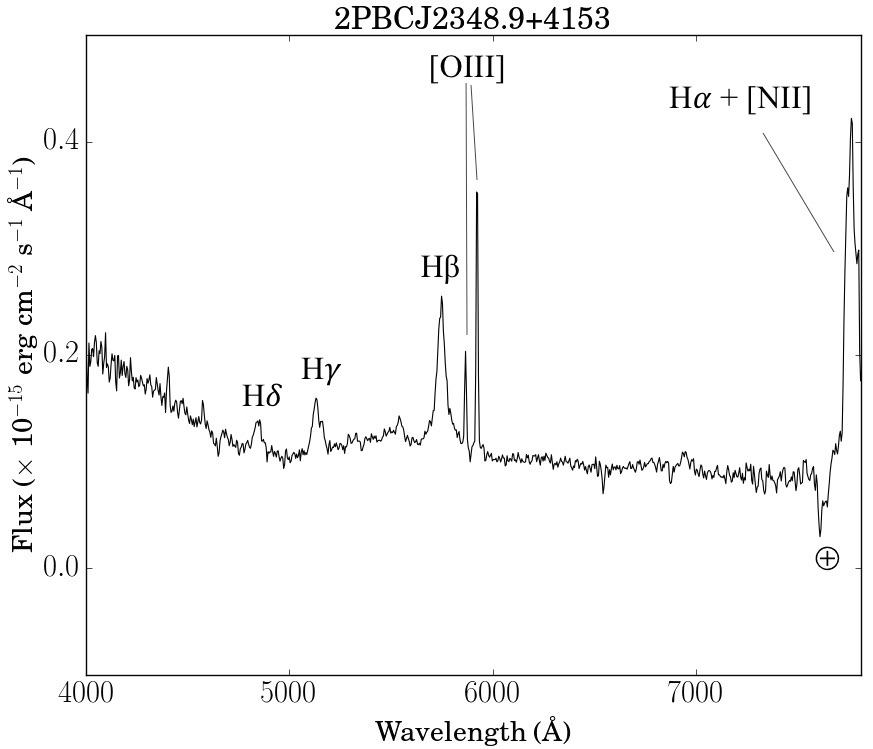}}
\caption{-- \emph{continued Figure 5}}
\end{figure*}

\newpage

\begin{figure*}[htbp]
\centering
\subfigure{\includegraphics[width=0.3\textwidth]{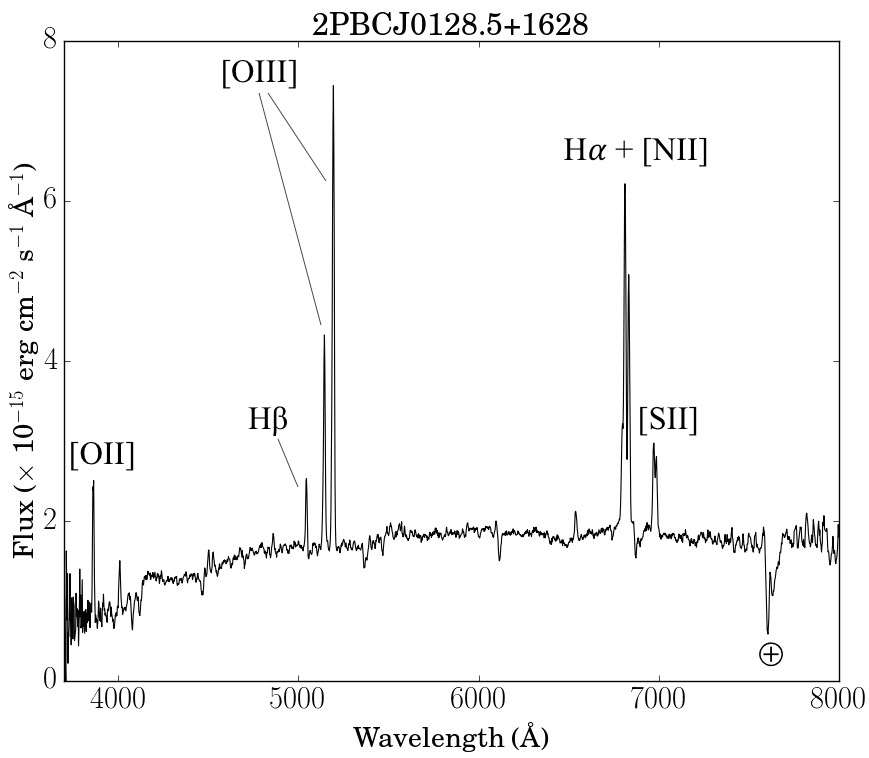}}
\subfigure{\includegraphics[width=0.3\textwidth]{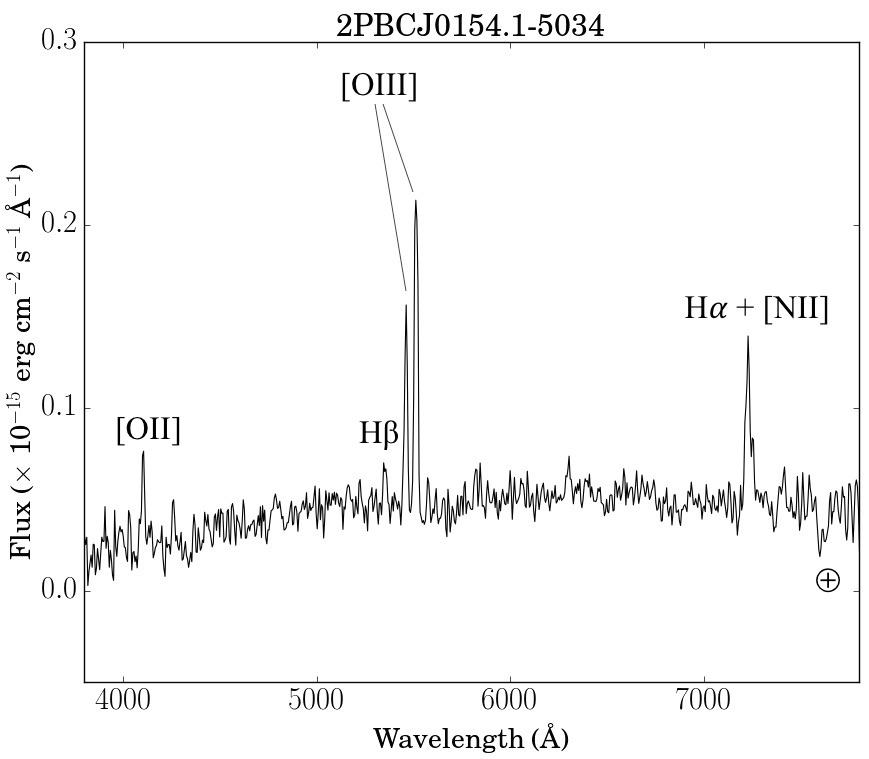}}
\subfigure{\includegraphics[width=0.3\textwidth]{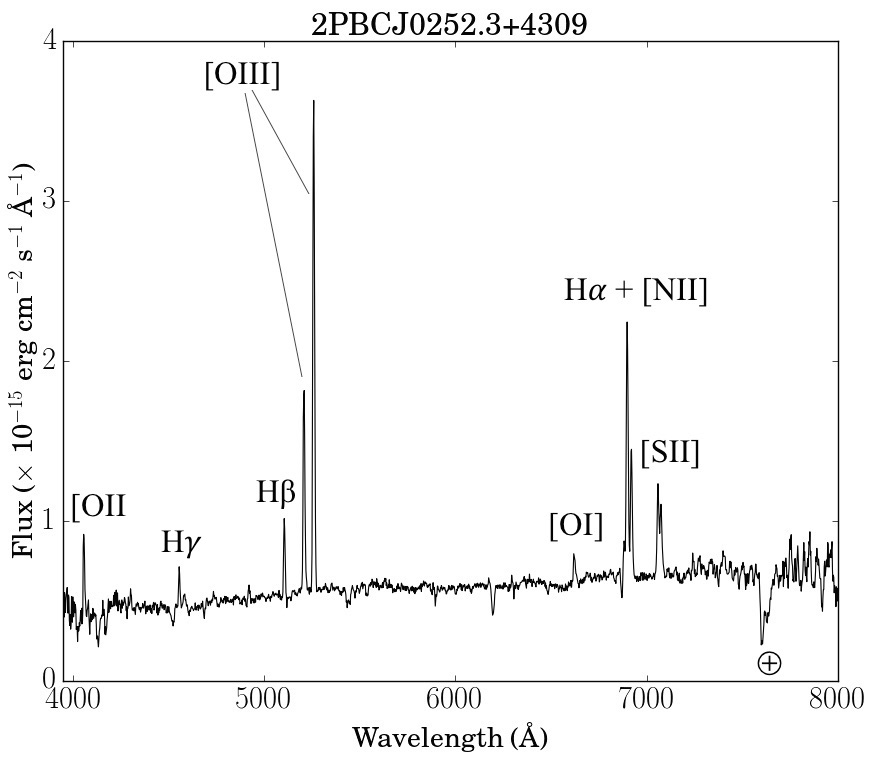}}
\subfigure{\includegraphics[width=0.3\textwidth]{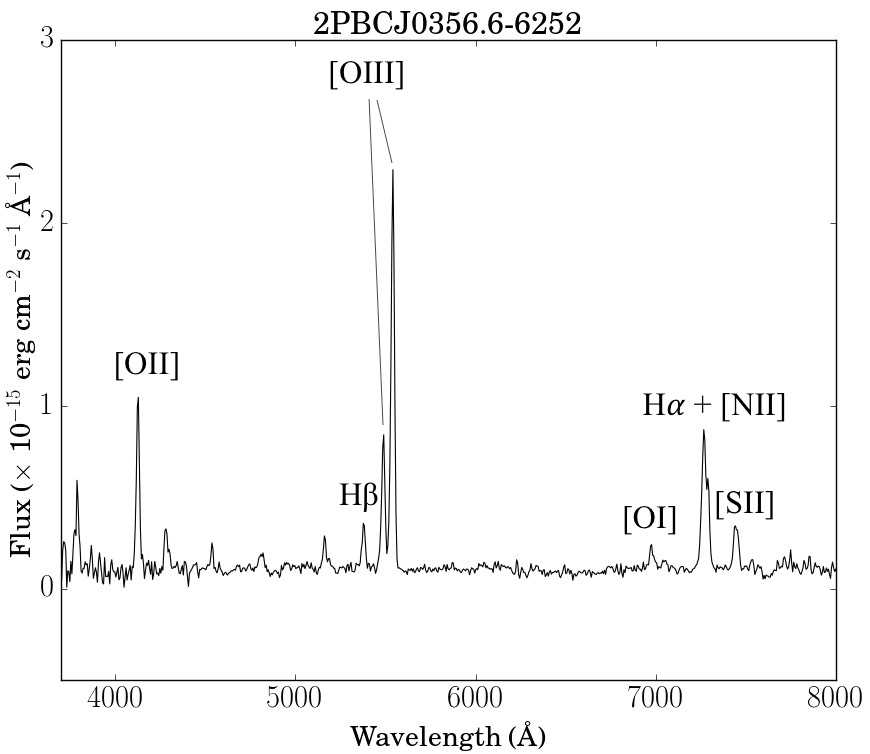}}
\subfigure{\includegraphics[width=0.3\textwidth]{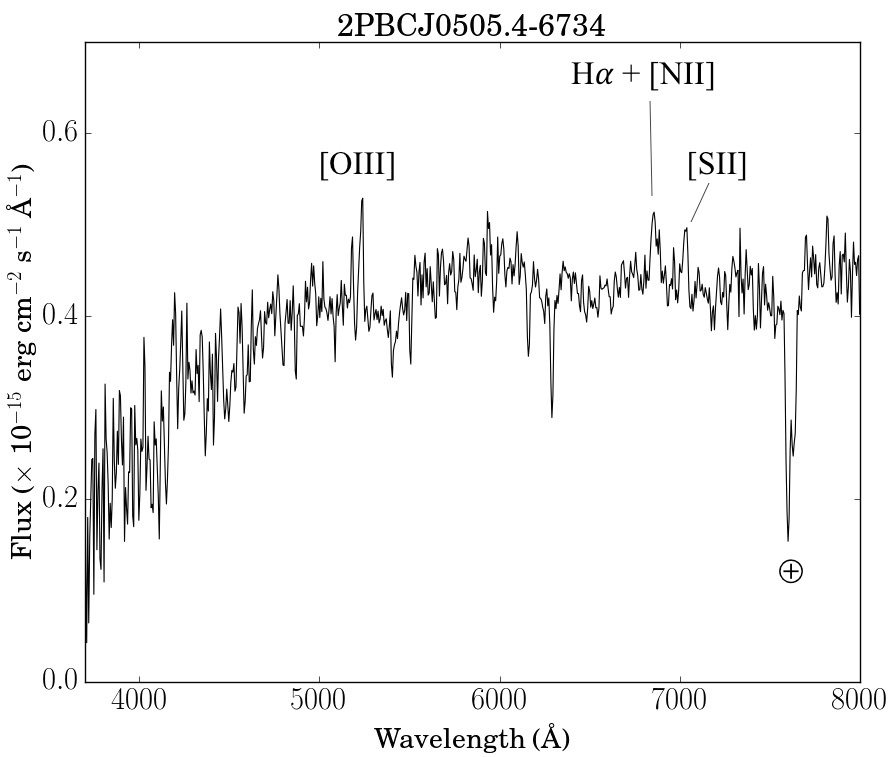}}
\subfigure{\includegraphics[width=0.3\textwidth]{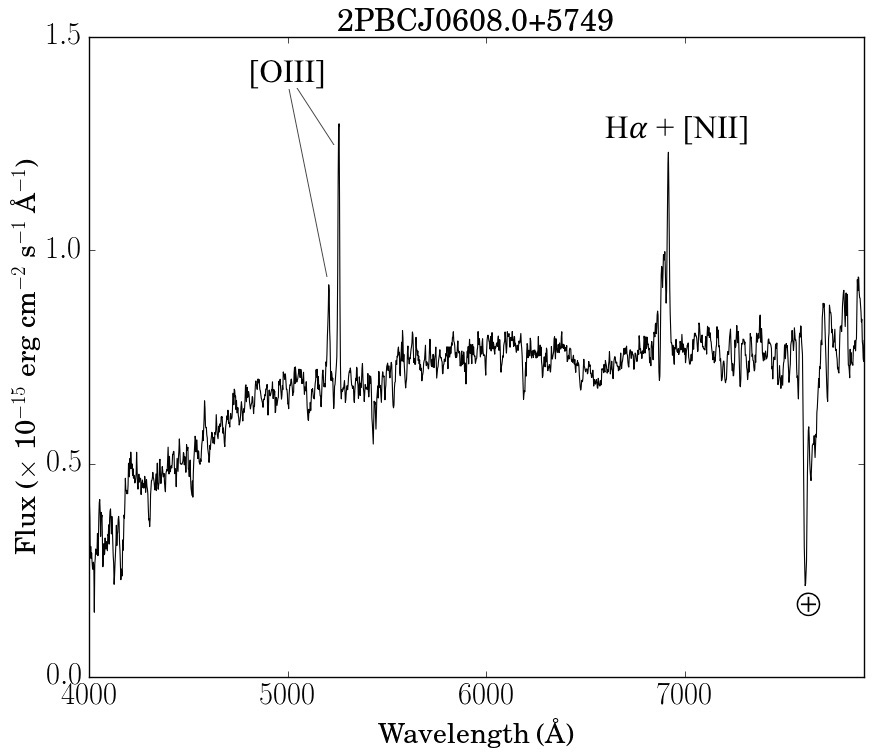}}
\subfigure{\includegraphics[width=0.3\textwidth]{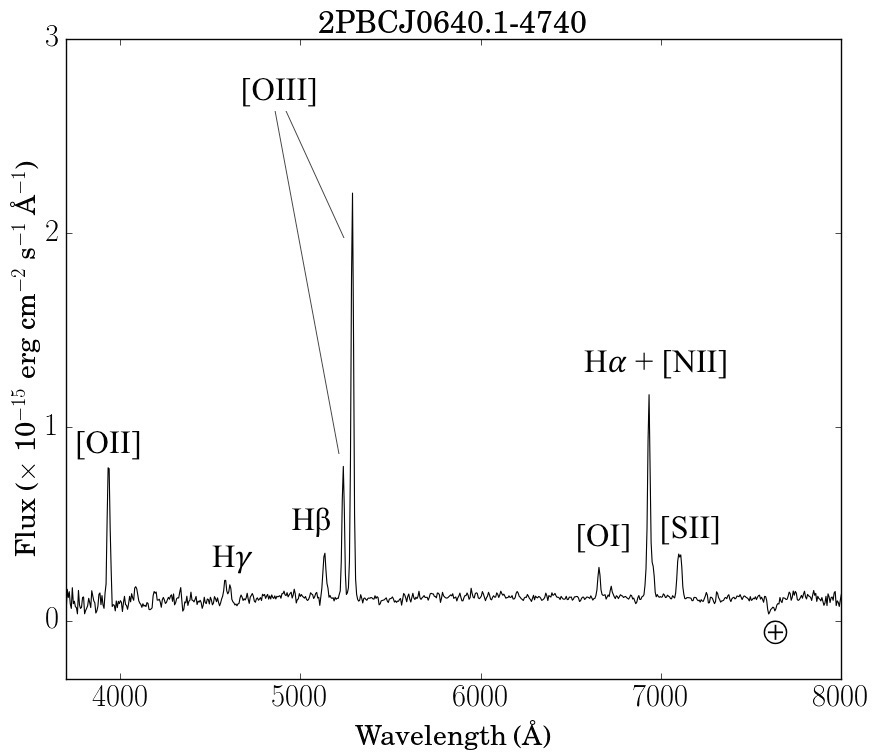}}
\subfigure{\includegraphics[width=0.3\textwidth]{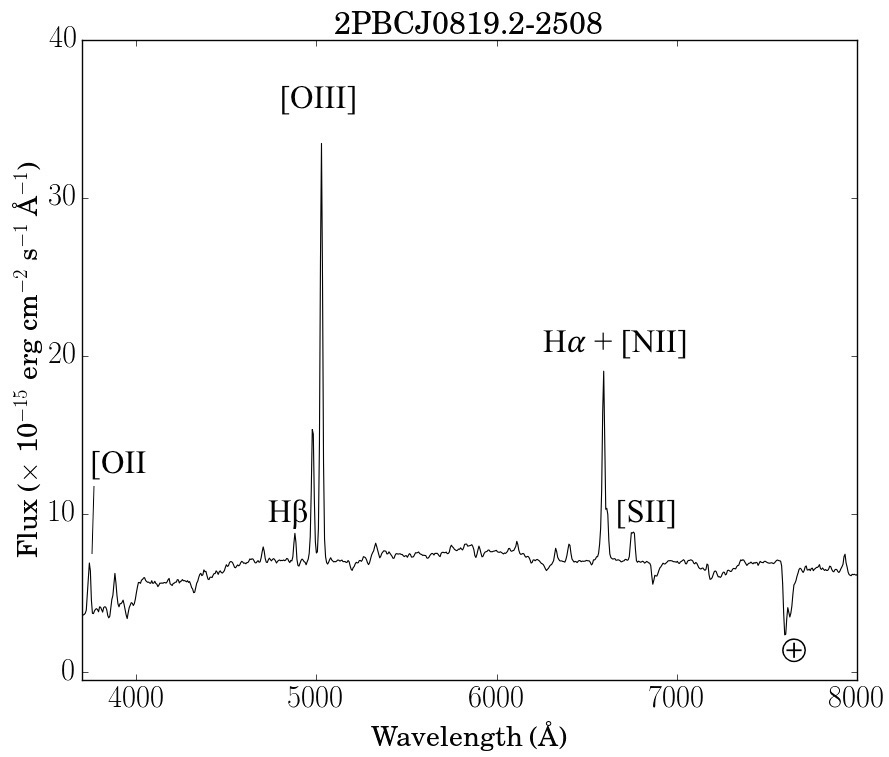}}
\subfigure{\includegraphics[width=0.3\textwidth]{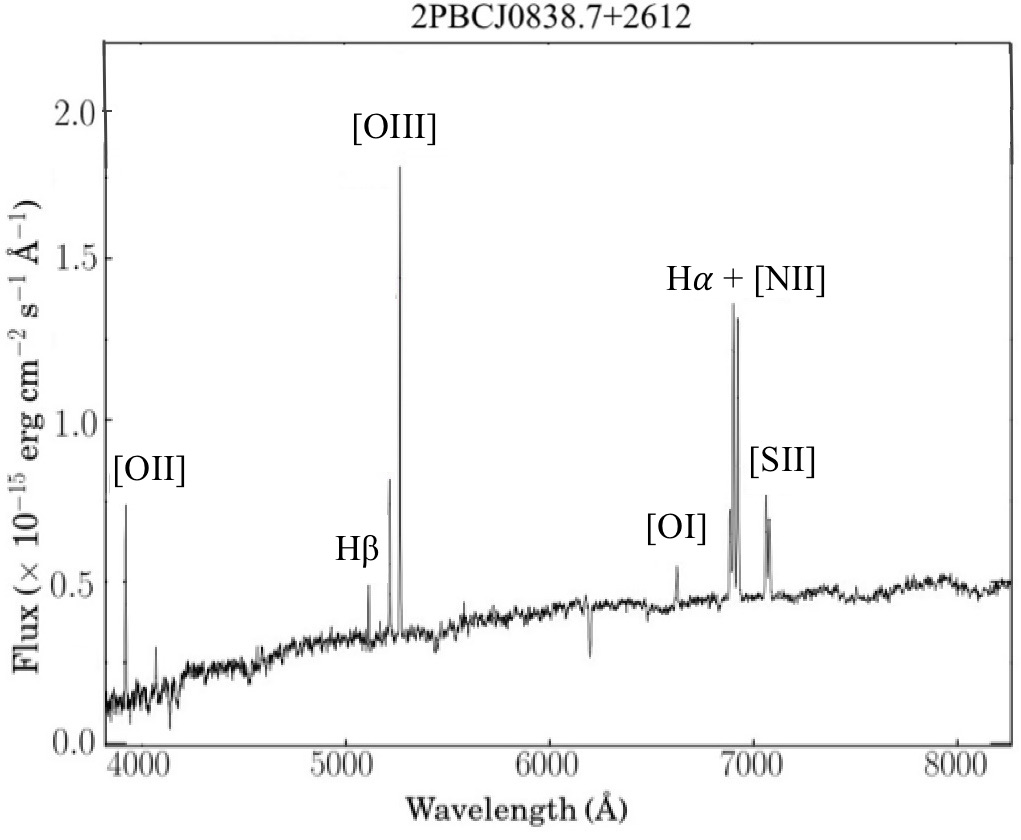}}
\subfigure{\includegraphics[width=0.3\textwidth]{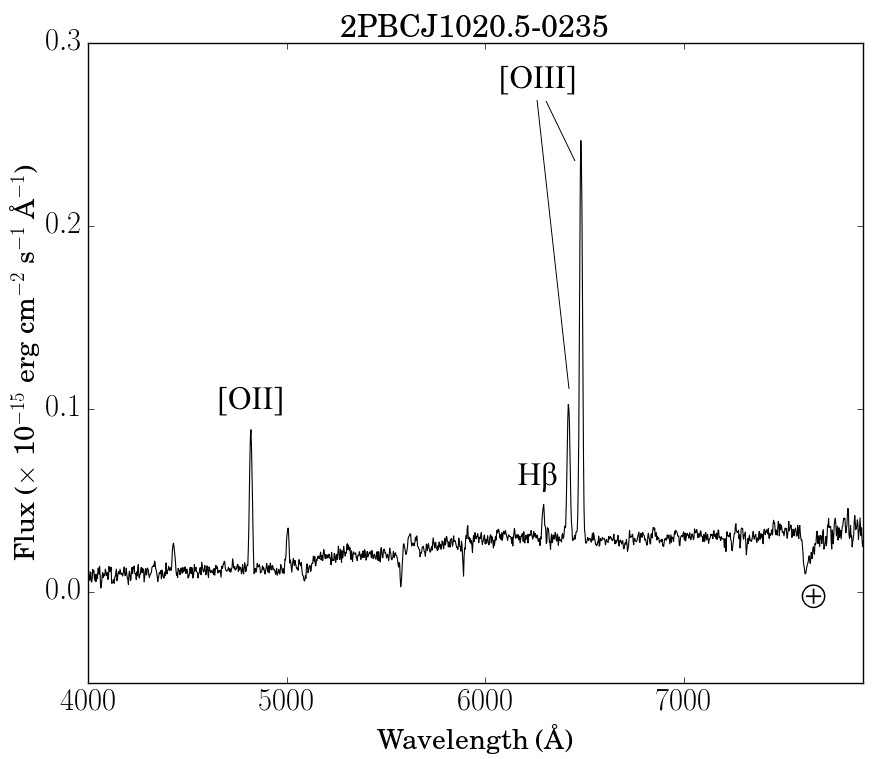}}
\subfigure{\includegraphics[width=0.3\textwidth]{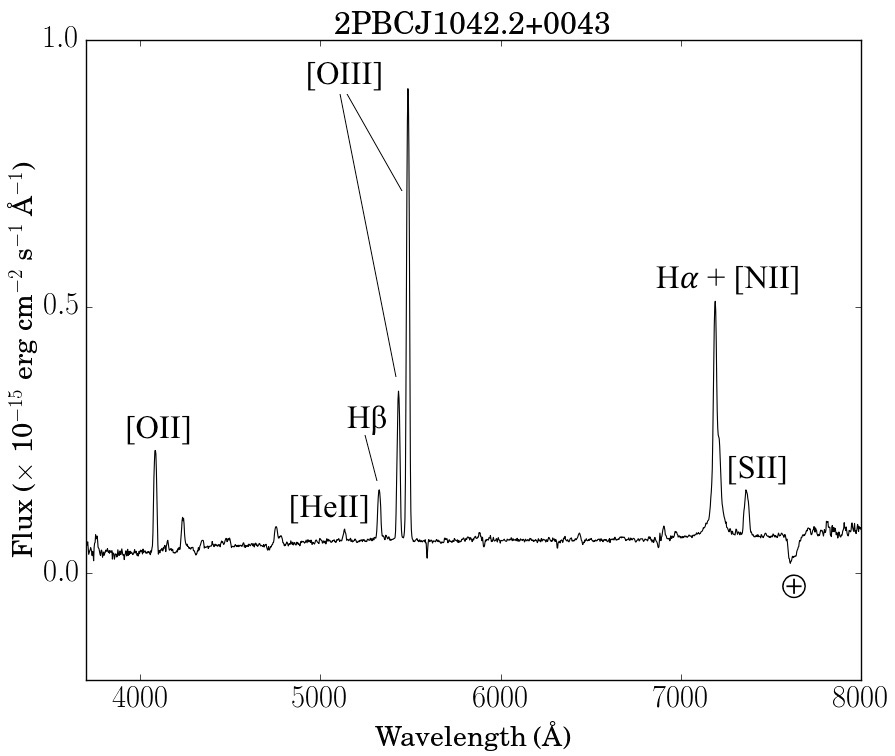}}
\subfigure{\includegraphics[width=0.3\textwidth]{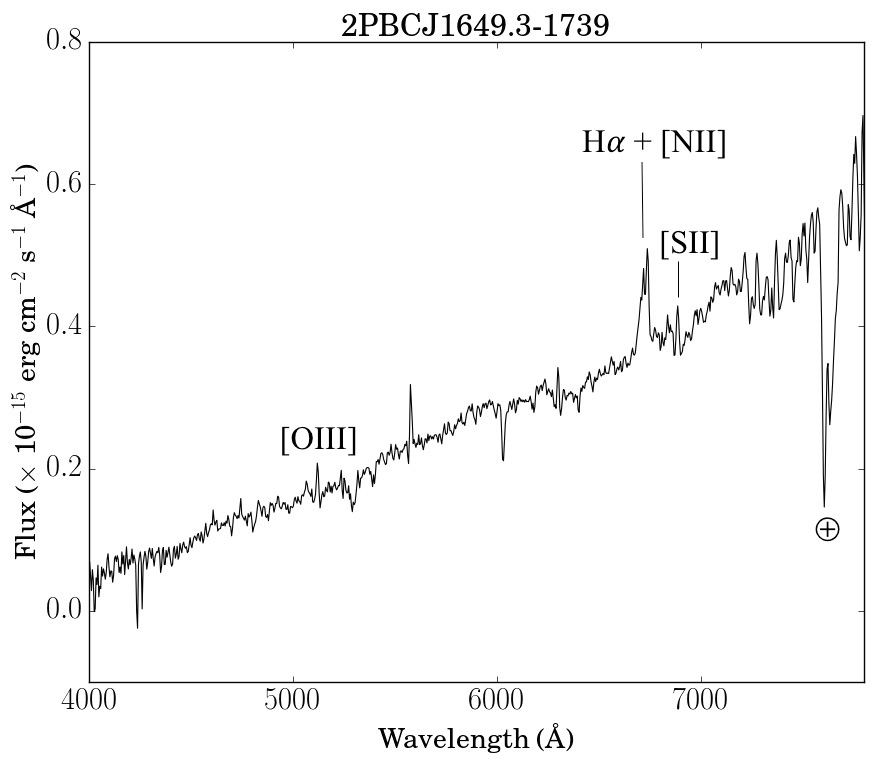}}
\subfigure{\includegraphics[width=0.3\textwidth]{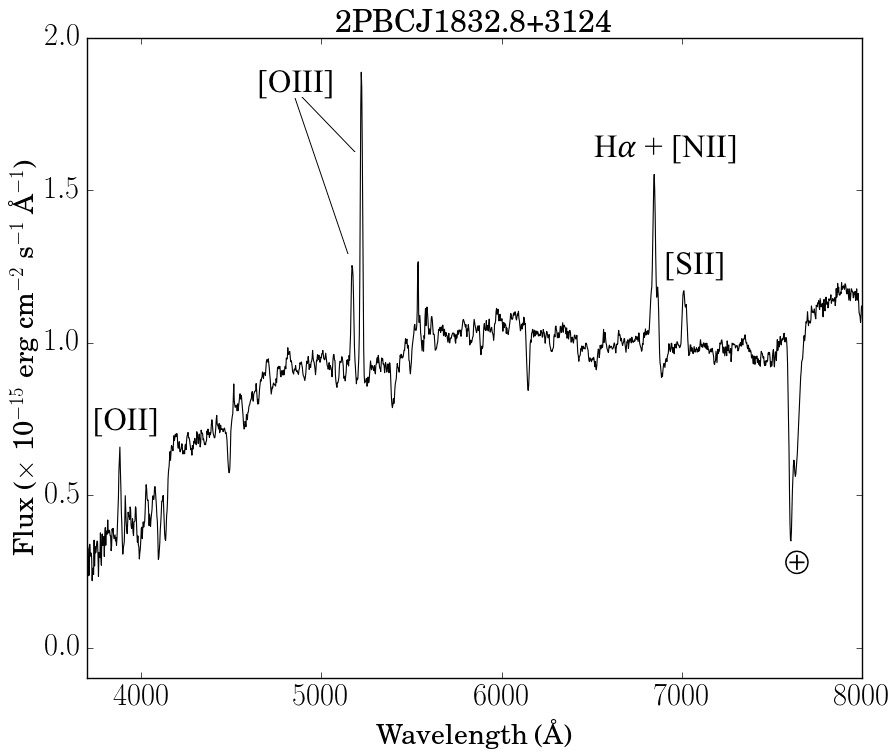}}
\subfigure{\includegraphics[width=0.3\textwidth]{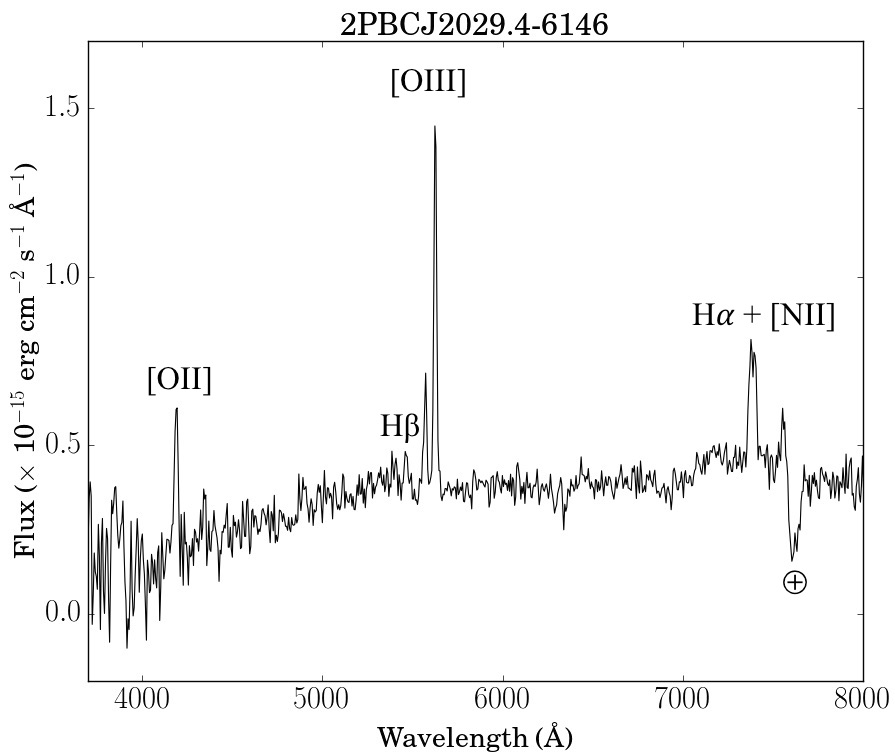}}
\subfigure{\includegraphics[width=0.3\textwidth]{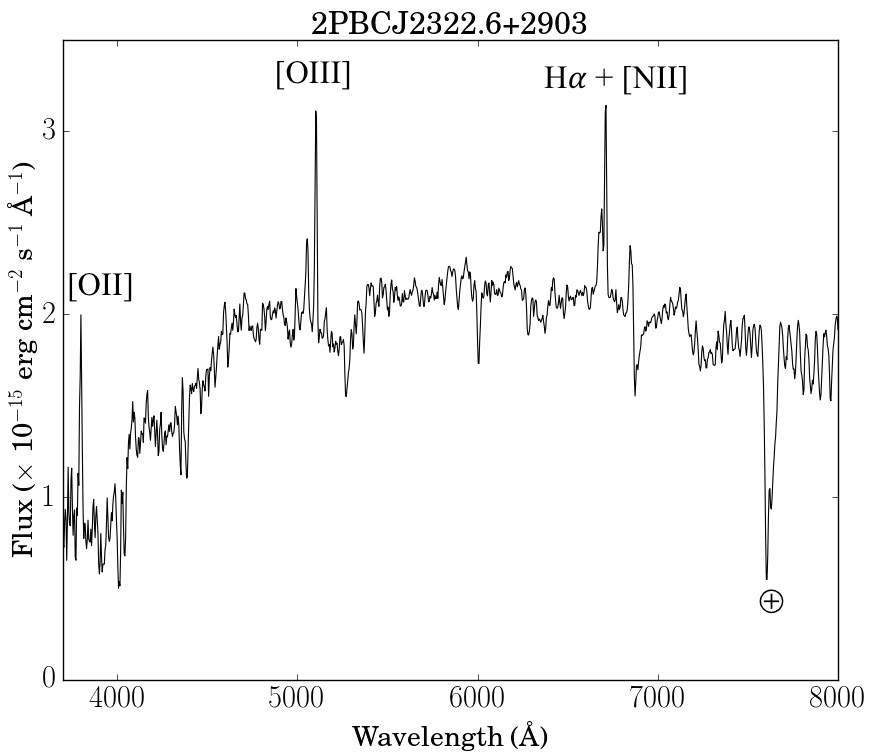}}
\caption{Optical spectra of type 2 AGNs presented in this work (not corrected for the intervening galactic absorption).} 
\end{figure*}

Considering all works, we identified 171 {\it Swift}/BAT objects through optical spectroscopy and their separation into the main classes
is the following: 157 (92\%) are AGNs, 2 are starburst galaxies (1\%), 1 (0.6\%) is probably a X--ray binary, 10 (5.8\%) are CVs, and 1 case (0.6\%) is likely identified as an active star. 
We were able to drastically decrease the percentage of unidentified objects in the BAT survey, as shown in Figure 11, where the unknown source type percentage decreased from more than 20\% to $\sim$ 10\% thanks to our systematic work (yellow bars). 

The green bars in Figure 11 represents the initial 54-month Palermo \textit{Swift}-BAT catalogue and the grey bars represents the classification of 20 AGNs and 18 galactic objects performed by other authors and available in the literature, as well as by our group using information already existing in the Simbad database (See Appendix C for the relevant information concerning these sources). We would like to stress that those other works have been able to reduce in $\sim$ 3\% the percentage of unclassified sources in the catalogue of Cusumano et al. (2010), whereas our enduring classification work has allowed us to decrease in $\sim$ 10\% the percentage of these objects.

All AGNs (Seyfert 1, Seyfert 2, starburst galaxy and LINERs) and XBONG in our sample and PI-PIII samples are at low redshifts ($z <$ 0.5) while only two QSOs are located at $z >$ 0.5.

We can see that the high-redshift QSOs found in our identification program are of Type 1. 
This is reasonably explained by the evidence that soft X--rays, needed for a precise localization of the 
source, are not easily detected from Type 2 QSOs due to local absorption in these sources, which adds 
to the faintness tied to their distance.


\begin{figure}[htbp]
\begin{center}
\includegraphics[height=6.0cm]{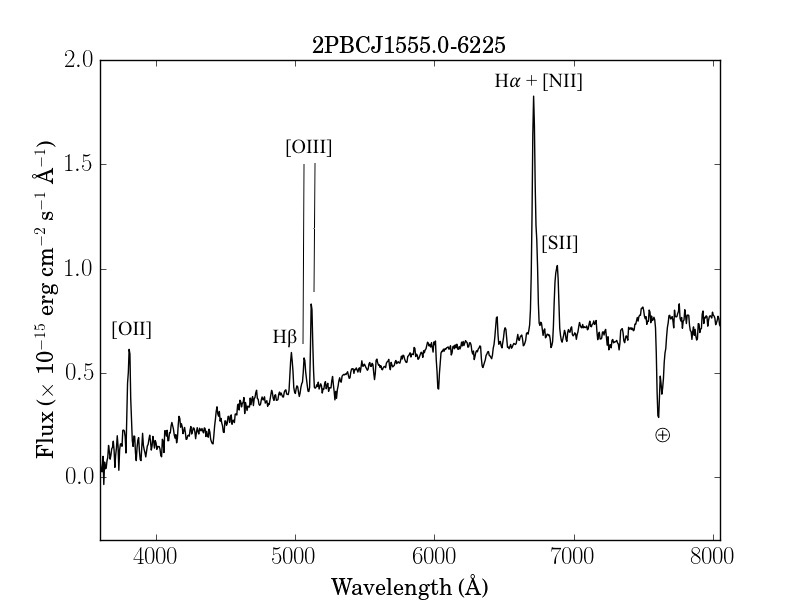}
\begin{spacing}{1}
\caption{Spectrum (not corrected for the intervening Galactic absorption) of the optical counterpart of the starburst galaxy belonging to the sample of BAT sources presented in this paper, 2PBC J1555.0-6225. For this spectrum, the main spectral features are labeled and $\oplus$ indicates telluric absorptions.}
\end{spacing}
\end{center}
\end{figure}


\begin{figure}[htbp]
\begin{center}
\includegraphics[height=6.0cm]{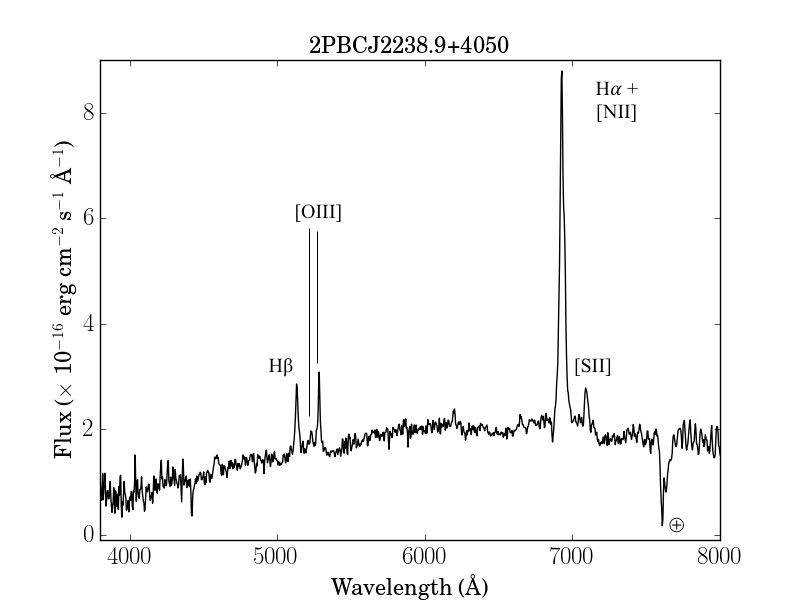}
\begin{spacing}{1}
\caption{Spectrum (not corrected for the intervening Galactic absorption) of the optical counterpart of the LINER belonging to the sample of BAT sources presented in this paper, 2PBC J2238.9+4050. For this spectrum, the main spectral features are labeled and $\oplus$ indicates telluric absorptions.}
\end{spacing}
\end{center}
\end{figure}


\begin{figure}[htbp]
\begin{center}
\includegraphics[height=6.0cm]{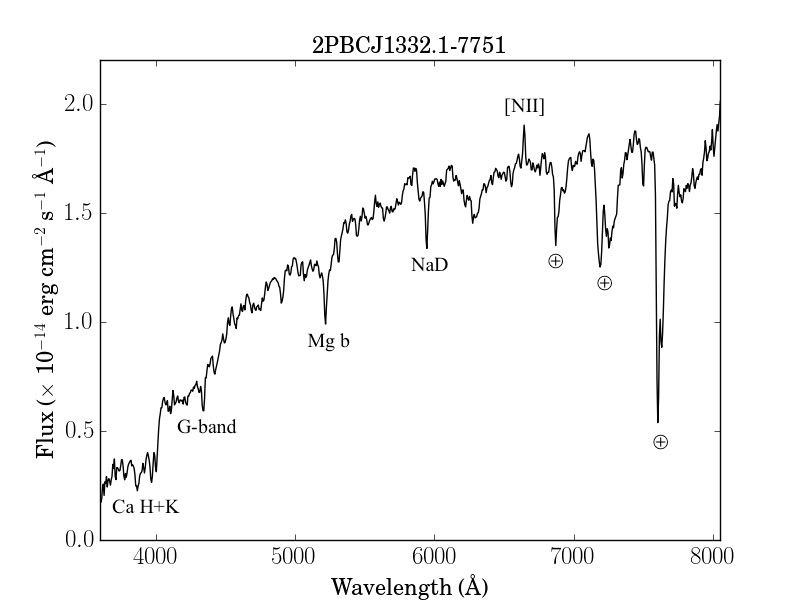}
\begin{spacing}{1}
\caption{Spectrum (not corrected for the intervening Galactic absorption) of the optical counterpart of the XBONG belonging to the sample of BAT sources presented in this paper, 2PBC J1332.1-7751. For this spectrum, the main spectral features are labeled and $\oplus$ indicates telluric absorptions.}
\end{spacing}
\end{center}
\end{figure}

\begin{figure}[htbp]
\begin{center}
\includegraphics[height=5.8cm]{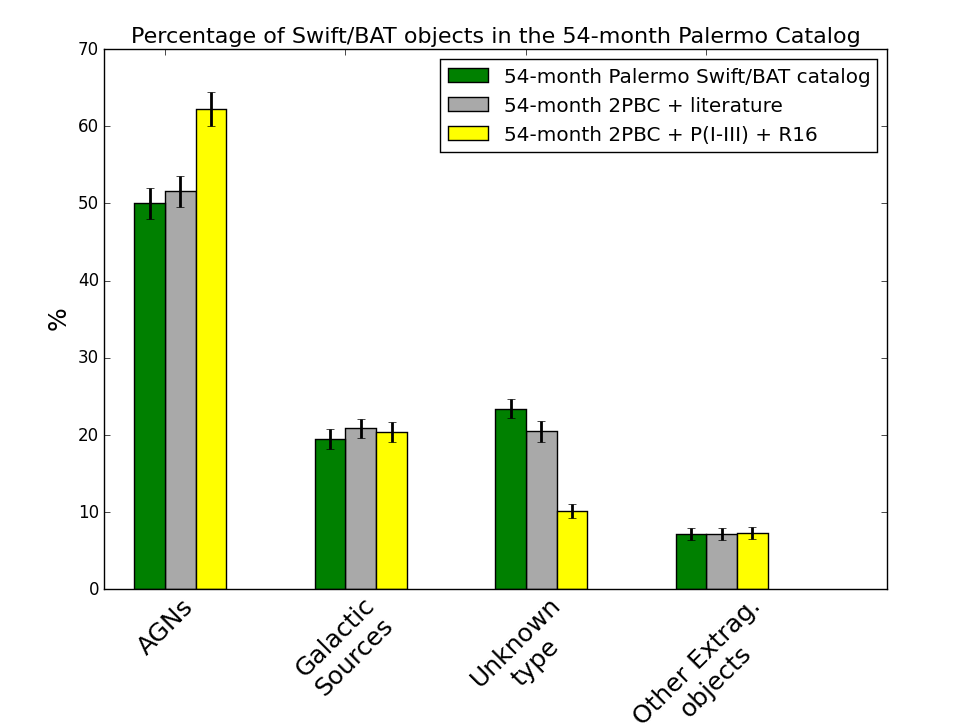}
\begin{spacing}{1}
\caption{Histogram, subdivided into source types, showing how the initial percentage of objects in the 54 months Palermo \textit{Swift}/BAT catalogue (in green) has changed by means of the identifications performed by other authors (in grey) and by our group in PI-PIII and in this work (R16) (in yellow).} 
\end{spacing}
\end{center}
\end{figure}

Figure 12 shows the redshift distribution of AGNs analysed in this work and the previous works of PI, PII and PIII, represented by a histogram binned according to rules of Knuth (2006), which is based on the optimization of a Bayesian fitness function across fixed-width bins. 

The redshift distributions for the present sample and that of PI-III merged sample were studied through a two sample Kolmogorov-Smirnov (KS) test. When we compare our redshift sample of Sy 1 with that of PI-III, we find that the KS test gives a p-value $\sim$ 0.001, whereas the comparison between the Sy 2 samples gives a p-value $\sim$ 0.178. This suggests that our Sy 1 sample sample is drawn from a different redshift distribution compared with the Sy 1 sample of PI-III. There is no univocal explanation to this: the difference can be due for instance to the fact that we on average used larger telescopes than those used in PI-III works, thus we are able to observe fainter (and likely more distant) objects. On the other hand, when we compare the distributions of Sy 2 galaxies in the two samples, we see that they are consistent with belonging to the same parent population; therefore the above interpretation cannot be applied to this class of AGNs.

\begin{figure}[htbp]
\begin{center}
\includegraphics[height=5.8cm]{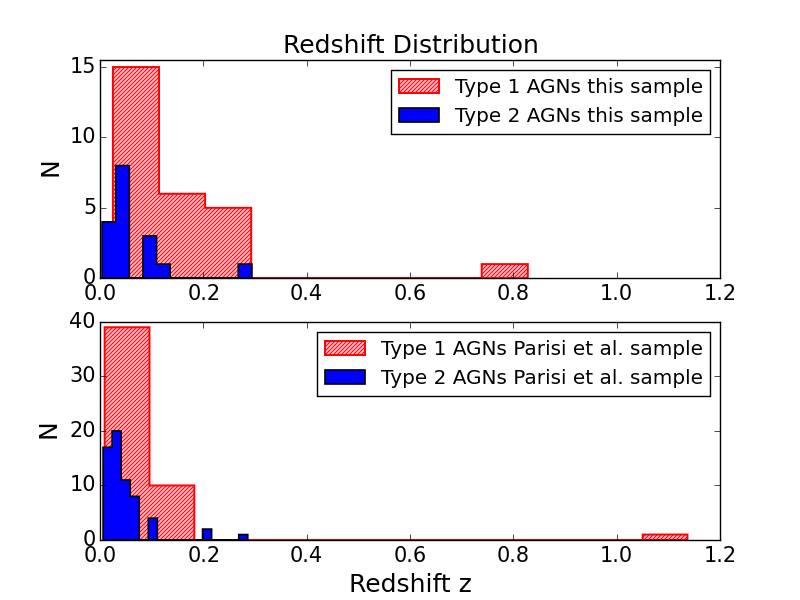}
\begin{spacing}{1}
\caption{Histograms showing the redshift distribution of {\it Swift/BAT} objects analysed and classified as AGNs 
in this sample (above) and those in the previous works of PI-III (bellow).}
\end{spacing}
\end{center}
\end{figure}

Moreover, we noted that Sy 1 sample has on average a higher z ($\sim$ 0.130) than the Sy 2 sample (at z $\sim$ 0.066). Interestingly the objects classified as naked Sy 2 are those few at higher z. This can explain why we see a difference with PI-III samples: indeed, in our selection of sources we are observing more distant Sy 1 galaxies, while this done not occur in the case of Sy 2 AGNs (but the naked ones) given that they are expected to be more absorbed and thus even fainter with distance (which means, difficult to observe with medium-sized telescopes).

Considering this work and PI-III, we found 12 Galactic objects: 10 of which are CVs (i.e. 84\% of the galactic identifications) and most of them are of magnetic type, whereas one object is an X-ray Binary, most probably a LXRB (8\%) and the remaining one is an active star (8\%). 
We can confirm that the \textit{Swift} satellite is efficient in detecting magnetic CVs because these objects are hard X-ray emitting sources due to their magnetic nature, and because the BAT surveys allow the study of objects outside the Galactic plane and, among them, hard X-ray emitting CVs.


We additionally tested the distribution of our AGNs by comparing the [OIII]$\lambda$5007 unabsorbed flux with the soft X-ray flux following Berney et al. (2015). Specifically, we constructed a plot of the relative strength of the unabsorbed [OIII]$\lambda$5007 emission line and the soft X-ray flux. It is important to stress that the soft X-ray fluxes considered refer to a wider band (0.3-10 keV) than that in Berney et al. (2015; 2-10 keV). So, these fluxes should conservatively be considered as upper limits. 
We found that when we compare the soft X-ray flux with that of the [OIII]$\lambda$5007 emission line in our type 2 AGN sample, the relation between them shows a Pearson correlation coefficient of R$_{Pear}$ = 0.39 and p-value = 0.13, with a standard deviation of $\sim$0.50 dex. 
The plot of the relation is shown in Fig 13 left panel, with the best fit model (slope 0.49 $\pm$ 0.09, orange dashed line) and the locus for which the parameter T is equal to 1 (red continuous line). In this panel, we also show the best fit model for our type 1 AGNs, as a dash-dot line (with slope of 0.60 $\pm$ 0.09, in green). We found that for this sample the Pearson correlation coefficient is R$_{Pear}$ = 0.37 with a p-value = 0.06 and a standard deviation of $\sim$0.55 dex. 
The dotted line (in pink) in both panels indicates the Berney et al. (2015) relation for the ratio between the soft X-ray flux and the [OIII]$\lambda$5007 unabsorbed flux. Finally, in the right panel we compare the soft X-ray flux with that of the [OIII]$\lambda$5007 emission line in our entire AGN sample (type 1 and type 2), with the best fit as a dashed line (with slope 0.52 $\pm$ 0.03, in grey), finding a Pearson correlation coefficient of R$_{Pear}$ = 0.41 with a p-value = 0.007 and a standard deviation of $\sim$0.53 dex. We can see that the Berney et al. (2015) relation has a steeper slope than our fits. This is possibly due to the range of the soft X-ray band we considered. Nevertheless, the results agree with those of Berney et al. (2015), suggest to us that any correlation between these quantities, if present, is marginal at best.

\begin{figure*}[htbp!]
\centering
\includegraphics[height=6.5cm]{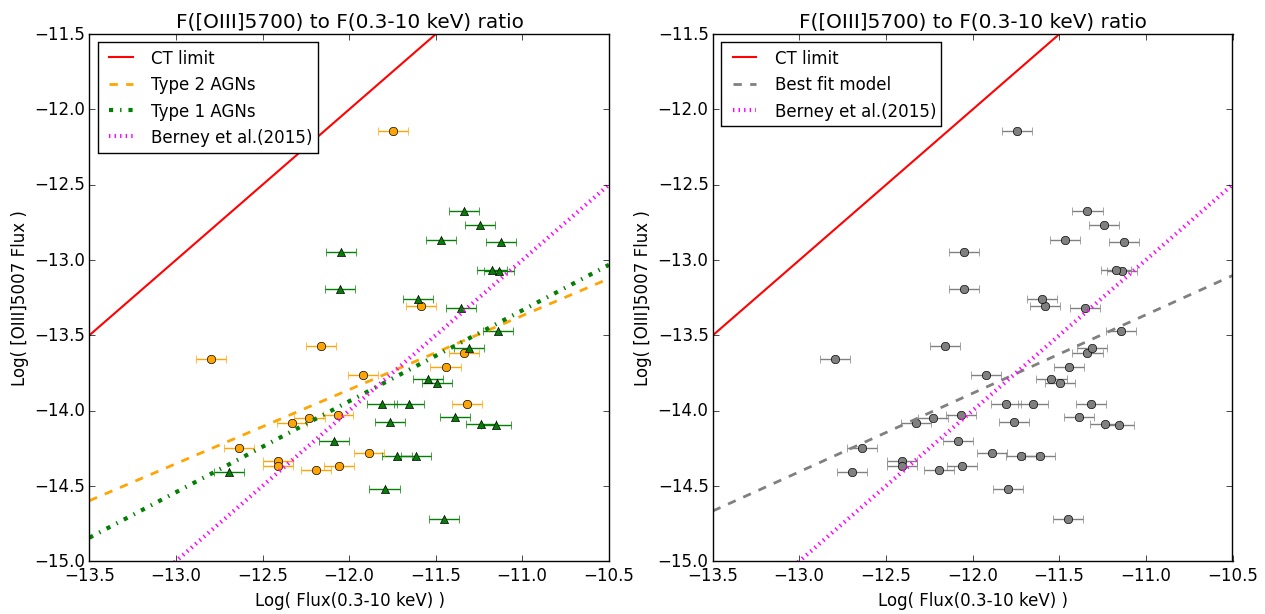}
\caption{Distribution of the [OIII]$\lambda$5007 to 0.3-10 keV ratio for the AGNs of our sample and corresponding best-fit correlations (see text for details).} 
\end{figure*}

\section{Conclusions}

In the present work, we identified and characterized 50 X-ray emitting objects with unknown or poorly explored nature, and which are listed in the 2PBC \textit{Swift} survey. This was accomplished with the use of seven telescopes of different apertures, from 1.5 to 3.58 metres, and of archival data from one spectroscopic survey. 

Our main results are as follows.

\begin{itemize}

\item The majority of identifications is made up of extragalactic objects. Most of them are AGNs (26 are Seyferts of Type 1, 15 cases are Seyferts of Type 2, one is a LINER, one is a XBONG and one is a QSO of type 1) and one is a starburst galaxy. This confirms the trend of our previous findings that optical spectroscopy eminently allows the identification of hard X-ray emitting AGNs.

\item In this work, we reported 30 new redshifts, 13 confirmations and 2 more accurate redshift values.
All AGNs lie at redshift z$\lesssim$ 0.3 except one of them, 2PBC J2010.3$-$2522, which is a QSO at z $\sim$ 0.8.
For most type 1 AGNs we estimated the BLR size, velocity, and the central black-hole mass as well as their Eddington ratio, and for type 2 AGNs we studied the local absorption.

\item Our sample is thus dominated by extragalactic objects; among them we highlighted some peculiar objects, such as one galaxy displaying LINER features, one starburst galaxy, one XBONG and two "naked" Seyfert 2 AGNs in the Compton thin regime. 

\item Five objects in our sample are galactic sources: one is probably a distant LXRB, 3 are CVs of magnetic nature and one is an active star. 

\end{itemize}

This confirms the importance of this identification work on catalogued but unidentified high-energy sources, also because peculiar objects can be found within the considered samples (see MI-X and PI-III).

Surveys at optical and NIR wavelengths, such as the Vista Variables in the V\'{i}a L\'actea (VVV: Minniti et al. 2010; Saito et al. 2012) public NIR survey, will permit the identification of variable sources in the fields of the objects detected in published and forthcoming hard X-ray catalogues. This will facilitate the detection of putative NIR counterparts for these high-energy sources by means of accurate positional information and/or variability studies. In turns this is facilitating the detection of transient and persistent hard X--ray sources, especially in crowded fields such as those along the Galactic Plane and Bulge (Rojas et al. 2012ab, 2013; Masetti et al. 2016; Saito et al. 2016). 

Furthermore, the NuSTAR satellite (Harrison et al. 2013) with its high spatial resolution can allow the identification of blended sources such as the case of the object 2PBC J1548.5-3208 in our sample. Furthermore, its high energy high energy resolution, NuSTAR can allow us to identify the different hard X-ray contributions of two sources in a same field, for example in the case of 2PBC J0819.2$-$2508, such as the work done by Koss et al. (2016b) with SWIFT J2028.5+2543.


 \begin{acknowledgements}
We thank Silvia Galleti for Service Mode observations at the Loiano telescope, and Roberto Gualandi, Ivan Bruni and Antonio De Blasi for night assistance; Aldo Fiorenzano, Vania Lorenzi and Walter Boschin for Service Mode observations at the TNG; Manuel Hern\'andez, Rodrigo Hern\'andez and Jacqueline Seron for Service Mode observations at the CTIO telescope and Fred Walter for coordinating them. 
We thank the anonymous referee for useful comments and suggestions. 
We also acknowledge the use of public data from the {\it Swift} data archive. 
This work made use of data supplied by the UK \textit{Swift} Science Data Centre at the University of Leicester, 
the ASI Science Data Center Multimission 
Archive, of the NASA Astrophysics Data System Abstract Service, 
the NASA/IPAC Extragalactic Database (NED), of the NASA/IPAC Infrared 
Science Archive, which are operated by the Jet Propulsion Laboratory, 
California Institute of Technology, under contract with the National 
Aeronautics and Space Administration, 
and of data obtained from the High Energy Astrophysics Science Archive Research Center (HEASARC), 
provided by NASA's GSFC.
This publication made use of data products from the Two Micron All 
Sky Survey (2MASS), which is a joint project of the University of 
Massachusetts and the Infrared Processing and Analysis Center/California 
Institute of Technology, funded by the National Aeronautics and Space 
Administration and the National Science Foundation.
This research has also made use of data extracted from the 
Sloan Digitized Sky Survey archive;
the SIMBAD database operated at CDS, Strasbourg, 
France, and of the HyperLeda catalogue operated at the Observatoire de 
Lyon, France.
Alejandra Rojas acknowledges financial support from Universidad Andr\'es Bello.
Lorenzo Morelli acknowledges financial support from Padova University grant CPS0204.
The IASF and IAPS coauthors acknoweledge support from ASI/INAF agreement 2013.025.R0.
Gaspar Galaz acknowledges the support of Basal center for Astrophysics and Technologies (CATA), PFB-06.
\end{acknowledgements}


\newpage


\begin{table*}[H!]
\caption[]{Synoptic table containing the main results for the 26
low-$z$ broad emission-line AGNs identified or observed in the 
present sample of {\it 2PBC} sources.}
\scriptsize
\setlength{\tabcolsep}{6pt} 
\begin{center}
\begin{tabular}{lccccrcl}
\noalign{\smallskip}
\hline
\hline
\noalign{\smallskip}
\multicolumn{1}{c}{Object} & $F_{\rm H_\beta}$ & $F_{\rm [OIII]}$ & Class & $z$ &
\multicolumn{1}{c}{$D_L$ (Mpc)} & $E(B-V)_{\rm Gal}$ & \multicolumn{1}{c}{$L_{\rm X}$} \\
\noalign{\smallskip}
\hline
\noalign{\smallskip}

2PBC J0057.2$+$6401 & 5$\pm$1 & 2.4$\pm$0.1 & Sy1.2 & 0.291 & 1499.1 & 1.252 & 2.15 (15-150)$^{C}$  \\
 & [62.3$\pm$5.2] & [54.7$\pm$4.3] & & & & & 0.67  (0.3-10)$^{X}$  \\ 
 &  & & & & & & 2.7  (14-195)$^{B}$  \\ 

2PBC J0116.5$-$1235 & 0.8$\pm$0.1 & 7.7$\pm$0.1 & Sy1.9 & 0.144 & 682.1 & 0.020 & 0.45 (15-150)$^{C}$  \\
 & [0.7$\pm$0.1] & [8.1$\pm$0.2]  & & & & & 0.32 (0.3-10)$^{X}$  \\ 
  &  & & & & & &  0.5 (14-195)$^{B}$  \\ 

 2PBC J0217.0$-$7250 & 33$\pm$1 & 14.8$\pm$0.1 & Sy1.2 & 0.291 & 1499.1 & 0.035 & 1.88 (15-150)$^{C}$  \\ 
 & [35$\pm$1] & [16.1$\pm$0.1]  & & & & & 0.77 (0.3-10)$^{X}$  \\ 

  2PBC J0440.6$-$6507 & 16$\pm$1 & 13.0$\pm$0.1 & Sy1.5 & 0.080 & 363.3 & 0.045 & 0.11 (15-150)$^{C}$  \\ 
 & [18$\pm$1] & [15.2$\pm$0.4]  & & & & & 0.05 (0.3-10)$^{X}$  \\

2PBC J0550.7$-$2304 & 12$\pm$3 & 7.0$\pm$0.7 & Sy1.5 & 0.043 & 190.2 & 0.037 & 0.05 (15-150)$^{C}$ \\
 & [18$\pm$6] & [8.0$\pm$1.5] & & & & & 0.03 (0.3-10)$^{X}$ \\ 

2PBC J0653.1$-$1227 & 7.7$\pm$0.3 & 7.80$\pm$0.02 & Sy1.5 & 0.234 & 1169.1 & 0.861 & 1.31 (15-150)$^{C}$ \\
 & [45$\pm$6] & [48.0$\pm$1.5] & & & & & 0.72  (0.3-10)$^{X}$  \\ 

2PBC J0709.5$-$3538 & 38.6$\pm$2.5 & 24.6$\pm$0.4 & Sy1.5 & 0.030 & 131.4 & 0.401 & 0.02 (15-150)$^{C}$ \\
 & [141$\pm$15] & [85$\pm$1] & & & & & 0.015 (0.3-10)$^{X}$  \\ 

2PBC J0714.6$-$2521 & 6.3$\pm$2.5 & 13.5$\pm$0.2 & Sy1.5 & 0.042 & 185.6 & 0.512 & 0.04 (15-150)$^{C}$ \\
 & [34.3$\pm$3.4] & [64.1$\pm$0.7] & & & & &  0.004 (0.3-10)$^{X}$   \\ 
  &  & & & & & & 0.08  (14-195)$^{B}$  \\ 

2PBC J0751.6$+$6450 & 7.5$\pm$0.5 & 73.5$\pm$0.8 & Sy1.9 & 0.025 & 109.1 & 0.045 & 0.009 (15-150)$^{C}$  \\
 & [8.6$\pm$0.5] & [84$\pm$1] & & & & & 0.002 (0.3-10)$^{X}$  \\ 

2PBC J0757.9$+$0113 & 14$\pm$1 & 9.3$\pm$0.7 & Sy1.5 & 0.104 & 480.0 & 0.044 & 0.3 (15-150)$^{C}$ \\
 & [17$\pm$1] & [11$\pm$1] & & & & &  0.061 (0.3-10)$^{X}$  \\ 

2PBC J0812.3$-$4003 & 3.6$\pm$0.3 & 1.6$\pm$0.2 & Sy1.2 & 0.076 & 344.1 & 1.632 & 0.1 (15-150)$^{C}$ \\
  & [461$\pm$78] & [172$\pm$8] & & & & & 0.08 (0.3-10)$^{X}$  \\ 
   &  & & & & & &  0.2 (14-195)$^{B}$  \\ 

2PBC J0854.3$-$0826 & 0.3$\pm$0.1 & 2.80$\pm$0.03 & Sy1.9 & 0.189 & 920.2 & 0.032 & 0.8 (15-150)$^{C}$ \\
  & [0.3$\pm$0.2] & [3.0$\pm$0.1] & & & & &  0.16 (0.3-10)$^{X}$  \\ 
   &  & & & & & &  1.3 (14-195)$^{B}$  \\ 

2PBC J1228.1$-$0925 & 197$\pm$37 & 104$\pm$6 & Sy1.5 & 0.223 & 1107.3 & 0.036 & 1.47 (15-150)$^{C}$  \\
  & [183$\pm$56] & [113$\pm$10] & & & & &  0.13 (0.3-10)$^{X}$ \\ 

2PBC J1251.8$-$5127 & 21.8$\pm$1.2 & 15.0$\pm$0.1 & Sy1.5 & 0.178 & 861.0 & 0.201 & 0.9 (15-150)$^{C}$ \\
 & [34$\pm$5] & [26$\pm$5] & & & & & 0.43  (0.3-10)$^{X}$  \\ 

2PBC J1419.2$+$6804 & --- & 1.7$\pm$0.2 & likely Sy1.9 & 0.077 & 348.9 & 0.035 & 0.1 (15-150)$^{C}$ \\
 &  & [1.9$\pm$0.2] & & & & & 0.05 (0.2-12)$^{N}$ \\ 

2PBC J1520.2$-$0433 & 96$\pm$1  & 98$\pm$1 & Sy1.5 & 0.111 & 514.7 & 0.113 & 0.2 (15-150)$^{C}$ \\
 & [134$\pm$1] & [135$\pm$1] & & & & & 0.11 (0.3-10)$^{X}$ \\ 

2PBC J1548.5$-$3208 & --- & 22.5$\pm$3.4 & likely Sy1.9 & 0.048 & 213.1 & 0.109 & 0.04 (15-150)$^{C}$ \\
 &  & [34$\pm$5] & & & & & 0.04 (0.3-10)$^{X}$ \\ 

2PBC J1709.7$-$2349 & 63$\pm$13  & 18$\pm$2 & Sy1.2 & 0.036 & 158.4 & 0.536 & 0.03 (15-150)$^{C}$ \\
 & [392$\pm$37] & [86$\pm$10] & & & & & 0.02  (0.2-12)$^{N}$ \\ 

2PBC J1742.0$-$6053 & 93.4$\pm$8.4  & 109$\pm$10 & Sy1.5 & 0.152 & 723.6 & 0.077 & 0.6 (15-150)$^{C}$ \\
 & [110$\pm$10] & [132$\pm$10] & & & & & 0.47  (0.3-10)$^{X}$ \\ 
  &  & & & & & &  0.7 (14-195)$^{B}$  \\ 

2PBC J1809.7$-$6555 & ---  & 6.8$\pm$2.5 & Sy1.5 & 0.181 & 877.0 & 0.067 & 0.7 (15-150)$^{C}$  \\
 &  & [9$\pm$3] & & & & & 0.38 (0.3-10)$^{X}$ \\ 

2PBC J2030.7$-$7530 & 138$\pm$11  & 168$\pm$10 & Sy1.5 & 0.114 & 529.6 & 0.088 & 0.3 (15-150)$^{C}$ \\
 & [161$\pm$17] & [213$\pm$10] & & & & & 0.15 (0.3-10)$^{X}$  \\ 
  &  & & & & & &  0.4 (14-195)$^{B}$  \\ 

2PBC J2045.9$+$8321 & 3.4$\pm$0.6  & 4.5$\pm$0.1 & Sy1.5 & 0.283 & 1451.9 & 0.142 & 1.26 (15-150)$^{C}$ \\
 & [3.8$\pm$0.6] & [6.3$\pm$0.2] & & & & & 0.21 (0.3-10)$^{X}$ \\ 

2PBC J2048.3$+$3812 & 1.1$\pm$0.5  & 0.4$\pm$0.1 & Sy1.2 & 0.106 & 489.9 & 0.890 & 0.2 (15-150)$^{C}$ \\
 & [16.3$\pm$0.5] & [5.0$\pm$0.1] & & & & & 0.05  (0.3-10)$^{X}$ \\ 
  &  & & & & & &  0.2 (14-195)$^{B}$  \\ 

2PBC J2136.3$+$2003 & 27.0$\pm$0.2  & 2.0$\pm$0.4 & Sy1 & 0.081 & 368.1 & 0.113 & 0.1 (15-150)$^{C}$ \\
 & [37$\pm$1] & [5$\pm$2] & & & & &  0.04 (0.3-10)$^{X}$ \\ 

2PBC J2155.1$+$6205 & 1.4$\pm$0.1  & 1.2$\pm$0.1 & Sy1.5 & 0.058 & 259.3 & 0.754 & 0.05 (15-150)$^{C}$ \\
 & [14$\pm$1] & [11$\pm$1] & & & & &  0.012 (0.3-10)$^{X}$ \\ 

2PBC J2348.9$+$4153 & 7.2$\pm$0.3  & 2.9$\pm$0.1 & Sy1.2 & 0.183 & 887.8 & 0.112 & 0.4 (15-150)$^{C}$ \\
 & [9.6$\pm$0.3] & [3.9$\pm$0.1] & & & & & 0.02 (0.3-10)$^{X}$ \\ 

\noalign{\smallskip} 
\hline
\noalign{\smallskip} 
\multicolumn{8}{l}{Note: emission-line fluxes are reported both as 
observed and (between square brackets) corrected for the intervening } \\
\multicolumn{8}{l}{Galactic absorption $E(B-V)_{\rm Gal}$ along the object line of sight 
(from Schlegel et al. 1998). Line fluxes are in units of } \\ 
\multicolumn{8}{l}{10$^{-15}$ \textit{erg cm$^{-2}$ s$^{-1}$}, X--ray luminosities are in units of 10$^{45}$ erg s$^{-1}$, and the reference band (between round brackets) } \\ 
\multicolumn{8}{l}{is expressed in keV. In the last column, the upper case letter indicates the satellite 
and/or the instrument with which the } \\
\multicolumn{8}{l}{corresponding X--ray flux measurement was obtained. } \\
\multicolumn{8}{l}{$^{C}$: from $\it{Swift}$/BAT, Cusumano et al. (2010). $^{B}$: from $\it{Swift}$/BAT, Baumgartner et al. (2013).} \\
\multicolumn{8}{l}{$^{X}$: from $\it{Swift}$/XRT. $^{N}$: from XMM-Newton. The typical error of the  redshift measurement is $\pm$0.001.} \\
\noalign{\smallskip} 
\hline
\hline
\end{tabular} 
\end{center} 
\end{table*} 

\newpage


\begin{table*}[htbp!]
\caption[]{Synoptic table containing the main results for the 
18 narrow emission-line AGNs observed in the present sample of {\it 2PBC} 
sources.}
\scriptsize
\setlength{\tabcolsep}{6pt} 
\begin{center}
\begin{tabular}{lcccccrccl}
\noalign{\smallskip}
\hline
\hline
\noalign{\smallskip}
\multicolumn{1}{c}{Object} & $F_{\rm H_\alpha}$ & $F_{\rm H_\beta}$ & $F_{\rm [OIII]}$ & Class & $z$ &
\multicolumn{1}{c}{$D_L$} & \multicolumn{2}{c}{$E(B-V)$} &
\multicolumn{1}{c}{$L_{\rm X}$} \\
\cline{8-9}
\noalign{\smallskip}
 & & & & & & (Mpc) & Gal. & AGN & \\
\noalign{\smallskip}
\hline
\noalign{\smallskip}

2PBC J0128.5$+$1628 & 59$\pm$1 & 9.0$\pm$0.7 & 78$\pm$1 & Sy2 & 0.038 & 167.5 & 0.062 & 0.82 & 4.03(15-150)$^{C}$ \\
 & [68$\pm$2] & [11$\pm$1] & [93$\pm$1] & & & & & & 0.29 (0.3-10)$^{X}$   \\ 
  &  & &  & & & & & & 4.9 (14-195)$^{B}$   \\ 
 & & & & & & & & &  \\

2PBC J0154.1$-$5034 & 1.7$\pm$0.1 & 0.6$\pm$0.2 & 4.0$\pm$0.1 & Sy 2  & 0.101 & 465.2 & 0.017 & $\sim$0.04  & 20.70 (15-150)$^{C}$  \\ 
  & [1.7$\pm$0.1] & [0.6$\pm$0.2] & [4.3$\pm$0.1] & & & & & & 1.0 (0.3-10)$^{X}$ \\ 
 & & & & & & & & &  \\

2PBC J0252.3$+$4309 &19.3$\pm$0.3 &4.9$\pm$0.3 & 34.8$\pm$0.1 & Sy2 & 0.051 & 226.9 & 0.093 & 0.27 & 7.39 (15-150)$^{C}$ \\
 & [24$\pm$1] & [6.3$\pm$0.3] & [46$\pm$1] & & & & & & 0.24 (0.3-10)$^{X}$  \\ 
   &  & &  & & & & & & 8.04 (14-195)$^{B}$   \\ 
 & & & & & & & & &  \\

2PBC J0356.6$-$6252 & 15.7$\pm$0.3 & 5.7$\pm$0.4 & 49.8$\pm$0.3 & $"$Naked$"$ Sy2 & 0.107 & 494.8 & 0.039 & $\sim$0.0 & 29.3 (15-150)$^{C}$  \\
 & [17$\pm$5] & [6.1$\pm$0.6] & [56$\pm$1] & & & & & & 0.67  (0.3-10)$^{X}$  \\ 
   &  & &  & & & & & & 36.1 (14-195)$^{B}$   \\ 
 & & & & & & & & &  \\

2PBC J0505.4$-$6734 & --- & --- & 2.5$\pm$0.3 & likely Sy2 & 0.046 & 203.9 & 0.229 & --- & 3.48 (15-150)$^{C}$  \\
 &  &  & [5.2$\pm$0.6] & & & & & & 0.65 (0.2-12)$^{N}$  \\ 
   &  & &  & & & & & & 5.9 (14-195)$^{B}$   \\ 
 & & & & & & & & &  \\

 2PBC J0608.0$+$5749 & 1.5$\pm$0.4 & $<$3 & 6.7$\pm$0.3 & $"$Naked$"$ Sy2 & 0.050 & 222.3 & 0.099 & $\sim$0.0  & 5.91  (15-150)$^{C}$ \\
 & [4.9$\pm$0.4] & [$<$5] & [8.9$\pm$0.3] & & & & & &  0.35 (0.3-10)$^{X}$  \\ 
 & & & & & & & & & \\

2PBC J0640.1$-$4740 &17.7$\pm$0.1 &5.5$\pm$0.6 & 36.5$\pm$0.3 &Sy2 & 0.056 & 250.0 & 0.099 & 0.13 & 6.73 (15-150)$^{C}$ \\
 & [21.9$\pm$0.1] & [7.1$\pm$0.6] & [50$\pm$1] & & & & & & 1.92  (0.3-10)$^{X}$ \\ 
   &  & &  & & & & & & 10.3 (14-195)$^{B}$   \\ 
 & & & & & & & & &  \\

2PBC J0819.2$-$2508 & 193$\pm$2 & 25$\pm$4 & 453$\pm$12 & Sy2 & 0.004 & 17.2 & 0.145 & 0.876 &  0.02  (15-150)$^{C}$ \\
& [272$\pm$10] & [40$\pm$10] & [719$\pm$12] & & & & & & 0.006  (0.3-10)$^{X}$ \\ 
 & & & & & & & & & \\

2PBC J0838.7$+$2612 & 7.8$\pm$0.1 & 0.8$\pm$0.1 & 7.2$\pm$0.1 & Sy2 & 0.052 & 231.5 & 0.048 &1.220 & 6.41 (15-150)$^{C}$ \\
 & [9$\pm$1]& [0.9$\pm$0.2]   &  [8.3$\pm$0.1]  & & & & & &  0.30 (0.3-10)$^{X}$  \\ 
  & & & & & & & & &  \\

2PBC J1020.5$-$0235 & --- & 0.2$\pm$0.1 & 3.9$\pm$0.1 & Sy2 & 0.294 & 1517.0 & 0.042 & --- & 303.0 (15-150)$^{C}$  \\
 &  &  [0.3$\pm$0.1]  &  [4.3$\pm$0.1] & & & & & & 23.8 (0.3-10)$^{X}$ \\ 
   &  & &  & & & & & & 194.7 (14-195)$^{B}$   \\ 
 & & & & & & & & &  \\

2PBC J1042.2$+$0043 & 8.2$\pm$0.2 & 1.9$\pm$0.1 & 16.6$\pm$0.2 & Sy2 & 0.096 & 440.7 & 0.056 & 0.350  & 23.2 (15-150)$^{C}$  \\
 & [9.3$\pm$0.2] &  [2.3$\pm$0.1]  &  [19.5$\pm$0.2] & & & & & &  8.36 (0.3-10)$^{X}$ \\ 
   &  & &  & & & & & & 28.2 (14-195)$^{B}$   \\ 
 & & & & & & & & &   \\

2PBC J1332.1$-$7751 & --- & --- & --- & XBONG & 0.009 & 38.8 & 0.210 & --- & 0.25 (15-150)$^{C}$ \\
 &  &   & & & & & & &  0.02 (0.3-10)$^{X}$  \\ 
   &  & &  & & & & & & 0.3 (14-195)$^{B}$   \\ 
 & & & & & & & & &  \\

2PBC J1555.0$-$6225 & 27.8$\pm$4.1 & 4.1$\pm$0.1 & 7.3$\pm$0.3 & likely Starburst Galaxy & 0.022 & 95.8 & 0.271 & 0.599 & 1.1  (15-150)$^{C}$ \\
 & [52$\pm$5] &  [10.0$\pm$0.3]  &  [17.1$\pm$0.4] & & & & & & 0.13 (0.3-10)$^{X}$   \\ 
 & & & & & & & & & \\

2PBC J1649.3$-$1739 & 1.3$\pm$0.1 & --- & 0.6$\pm$0.1 & type 2 AGN & 0.023 & 100.2 & 0.938 & --- & 1.68  (15-150)$^{C}$ \\
 & [11$\pm$3] &    &  [11$\pm$2] & & & & & & 0.58 (0.3-10)$^{X}$  \\ 
 & & & & & & & & &  \\

2PBC J1832.8$+$3124 & 10.1$\pm$0.6  & ---  & 15.9$\pm$0.7 & Sy2 & 0.043 & 190.2 & 0.116 & --- &  2.60 (15-150)$^{C}$ \\
 & [13$\pm$1] &   &  [22$\pm$1] & & & & & &  0.07 (0.3-10)$^{X}$ \\ 
  & & & & & & & & &  \\
 
2PBC J2029.4$-$6146 & 6.5$\pm$1.5 & 1.9$\pm$0.1  & 21.0$\pm$0.1 & $"$Naked$"$ Sy2 & 0.124 & 579.9 & 0.052 & $\sim$0.0  & 44.3 (15-150)$^{C}$  \\
 & [8$\pm$3] &  [3.4$\pm$0.2]  &  [24.0$\pm$0.1] & & & & & &  18.5  (0.3-10)$^{X}$ \\ 
   &  & &  & & & & & & 63.6 (14-195)$^{B}$   \\ 
 & & & & & & & & &  \\

2PBC J2238.9$+$4050 & 14.9$\pm$0.2  & 3.4$\pm$0.2  & 2.0$\pm$0.1 & LINER & 0.055 & 245.4 &0.242  & 0.296 & 5.04 (15-150)$^{C}$  \\
 & [28$\pm$7] &  [7.2$\pm$0.6]  &  [4.0$\pm$0.4] & & & & & & 0.46 (0.3-10)$^{X}$\\ 
 & & & & & & & & &  \\

2PBC J2322.6$+$2903 & 6.5$\pm$0.2  & ---  & 16.8$\pm$0.4 & Sy2 & 0.019 & 82.6 & 0.154 & --- &  0.74 (15-150)$^{C}$ \\
 & [8.9$\pm$0.2] &    &  [26.7$\pm$0.4] & & & & & & 0.06 (0.3-10)$^{X}$  \\ 
 & & & & & & & & &  \\

\noalign{\smallskip} 
\hline
\noalign{\smallskip} 
\multicolumn{10}{l}{Note: emission-line fluxes are reported both as 
observed and (between square brackets) corrected for the intervening Galactic} \\ 
\multicolumn{10}{l}{absorption $E(B-V)_{\rm Gal}$ along the object line of sight 
(from Schlegel et al. 1998). Line fluxes are in units of 10$^{-15}$ \textit{erg cm$^{-2}$ s$^{-1}$},} \\
\multicolumn{10}{l}{X--ray luminosities are in units of 10$^{43}$ erg s$^{-1}$,
and the reference band (between round brackets) is expressed in keV.} \\ 
\multicolumn{10}{l}{In the last column, the upper case letter indicates the satellite and/or the 
instrument with which the corresponding X--ray flux } \\
\multicolumn{10}{l}{measurement was obtained. $^{C}$: from $\it{Swift}$/BAT, Cusumano et al. (2010). $^{B}$: from $\it{Swift}$/BAT, Baumgartner et al. (2013).} \\
\multicolumn{10}{l}{ $^{X}$: from $\it{Swift}$/XRT. $^{N}$: from XMM-Newton. The typical error of the redshift measurement is $\pm$0.001.} \\
\noalign{\smallskip} 
\hline
\hline
\end{tabular}
\end{center}
\end{table*}

\newpage
\onecolumn

\begin{table}[htbp!]
\caption{BLR gas velocities (in km s$^{-1}$), central black 
hole masses (in units of 10$^8$ $M_\odot$), Eddington Luminosity 
and apparent Eddington ratios 
for broad line AGNs discussed in this paper.}
\setlength{\tabcolsep}{3pt} 
\begin{center}
\begin{tabular}{lrccc}
\noalign{\smallskip}
\hline
\hline
\noalign{\smallskip}
\multicolumn{1}{c}{Object} & \multicolumn{1}{c}{$v_{\rm BLR}$} & $M_{\rm BH}$ & $L_{Edd}$ & $L_{BOL}/L_{\rm Edd}$\\
\noalign{\smallskip}
\hline
\noalign{\smallskip}
2PBCJ0057.2$+$6401  &	5803	  &		11	  &		136	  &		0.09	\\
2PBCJ0217.0$-$7250  &	10622	  &		25	  &		308	  &		0.04	\\
2PBCJ0440.6$-$6507  &	5046	  &		0.5	  &		6.4	  &		0.10	\\
2PBCJ0550.7$-$2304  &	2550	  &		0.1	  &		0.7	  &		0.44	\\
2PBCJ0653.1$-$1227  &	5184	  &		5	  &		62	  &		0.12	\\
2PBCJ0709.5$-$3538  &	2986	  &		0.2	  &		2.3	  &		0.05	\\
2PBCJ0714.6$-$2521  &	4086	  &		0.2	  &		2.6	  &		0.09	\\
2PBCJ0757.9$+$0113  &	9241	  &		2.4	  &		30	  &		0.06	\\
2PBCJ0812.3$-$4003  &	3229	  &		1.8	  &		22	  &		0.03	\\
2PBCJ1228.1$-$0925  &	16600	  &		123	  &		1542  &		0.01	\\
2PBCJ1251.8$-$5127  &	6732	  &		4.5	  &		57	  &		0.09	\\
2PBCJ1520.2$-$0433  &	2937	  &		1.1	  &		14	  &		0.09	\\
2PBCJ1709.7$-$2349  &	1180	  &		0.1	  &		1.0	  &		0.19	\\
2PBCJ1742.0$-$6053  &	7895	  &		11	  &		138	  &		0.03	\\
2PBCJ2030.7$-$7530  &	4076	  &		2.5	  &		31	  &		0.06	\\
2PBCJ2045.9$+$8321  &	2325	  &		0.2	  &		3.1	  &		2.41	\\
2PBCJ2048.3$+$3812  &	11448	  &		3.7	  &		46	  &		0.03	\\
2PBCJ2136.3$+$2003  &	4501	  &		0.7	  &		8.4	  &		0.07	\\
2PBCJ2155.1$+$6205  &	2475	  &		0.1	  &		0.8	  &		0.36	\\
2PBCJ2348.9$+$4153  &	2373	  &		0.2	  &		3.1	  &		0.76	\\
\noalign{\smallskip}
\hline 
\noalign{\smallskip}
2PBC J2010.3$-$2522    & 1575  & 1.3  &  16   &  15.4   \\ 
\noalign{\smallskip}
\hline
\noalign{\smallskip}
\multicolumn{5}{l}{Note: the final uncertainties on the black hole mass} \\
\multicolumn{5}{l}{estimates are about 50\% of their values. The velocities} \\
\multicolumn{5}{l}{ were determined using H$_\beta$ emission or Mg {\sc ii} (upper} \\
\multicolumn{5}{l}{ and lower part of the table, respectively), whereas the} \\
\multicolumn{5}{l}{ apparent Eddington Luminosity (in units of $10^{45}$ erg $s^{-1}$)} \\
\multicolumn{5}{l}{ and ratios were computed using the bolometric luminosity} \\
\multicolumn{5}{l}{ obtained from the (observed or rescaled, see text)} \\
\multicolumn{5}{l}{ 15--150 keV luminosities.} \\
\noalign{\smallskip}
\hline
\hline
\noalign{\smallskip}
\end{tabular}
\end{center}
\end{table}

\newpage
\appendix
\section*{Appendix A: Finding charts of the new soft X-ray identifications}

\setcounter{figure}{0}
\renewcommand{\thefigure}{A.\arabic{figure}}


\begin{figure*}[htbp]
\centering
\subfigure{\includegraphics[width=0.32\textwidth]{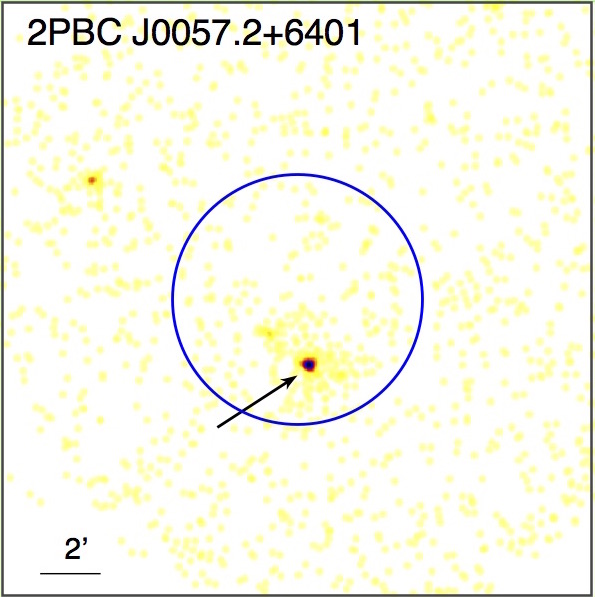}}
\subfigure{\includegraphics[width=0.32\textwidth]{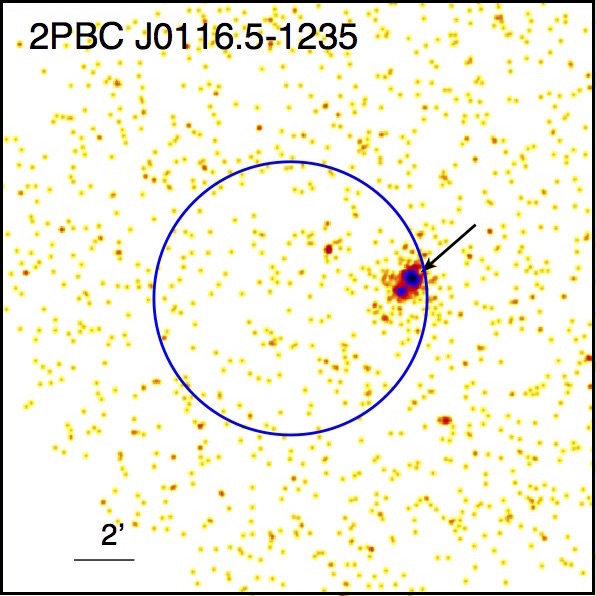}}
\subfigure{\includegraphics[width=0.32\textwidth]{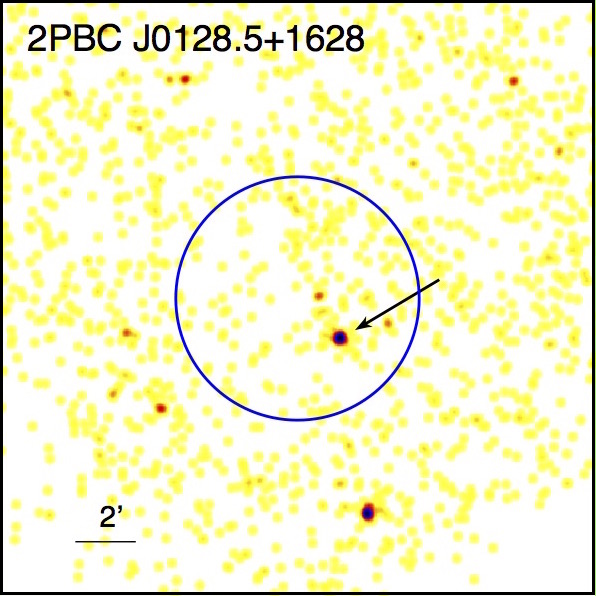}}
\subfigure{\includegraphics[width=0.32\textwidth]{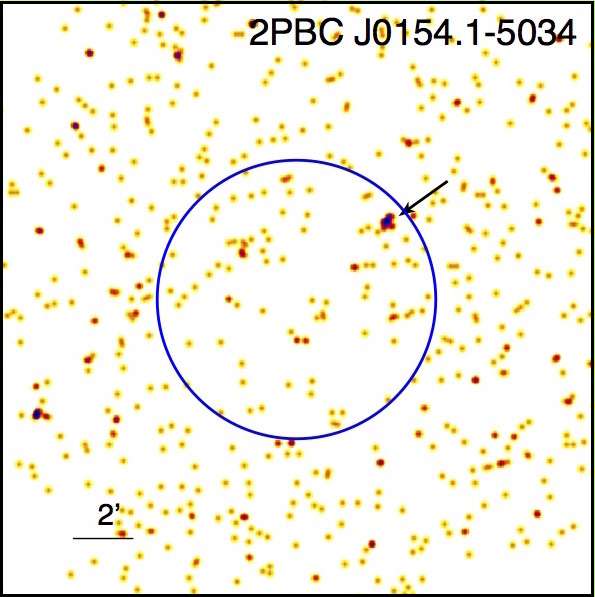}}
\subfigure{\includegraphics[width=0.32\textwidth]{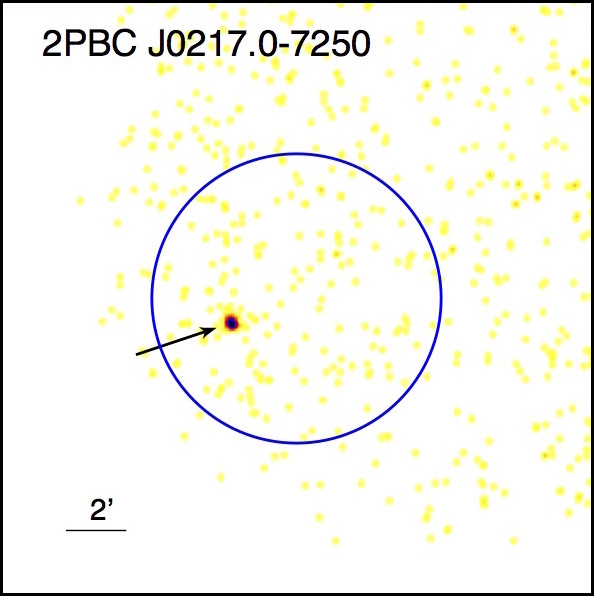}}
\subfigure{\includegraphics[width=0.32\textwidth]{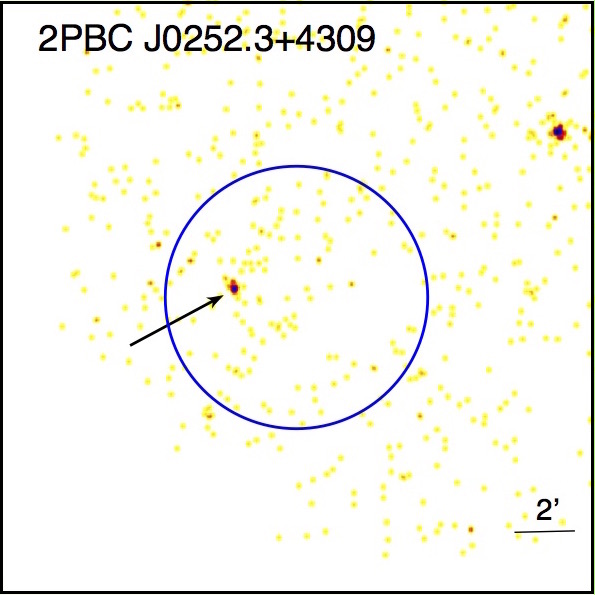}}
\subfigure{\includegraphics[width=0.32\textwidth]{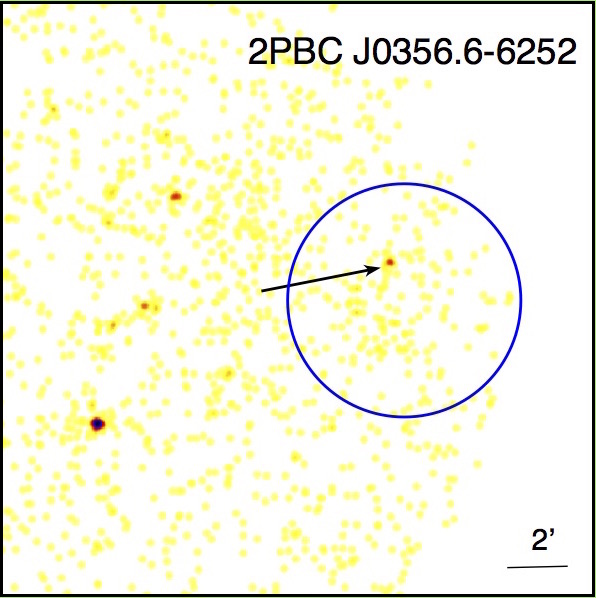}}
\subfigure{\includegraphics[width=0.32\textwidth]{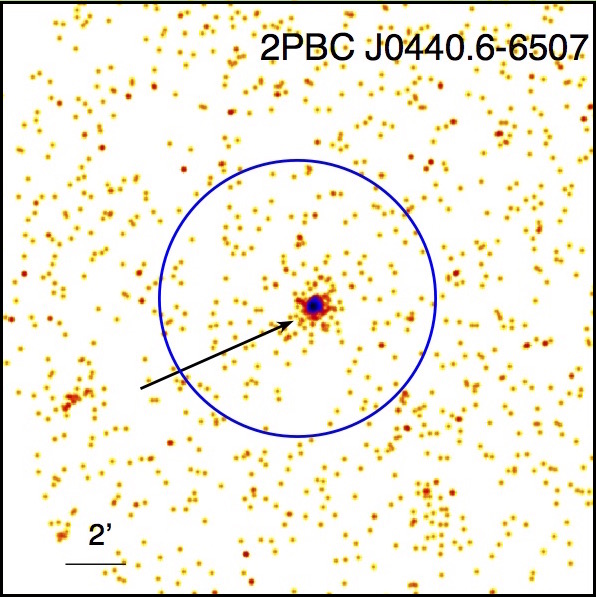}}
\subfigure{\includegraphics[width=0.32\textwidth]{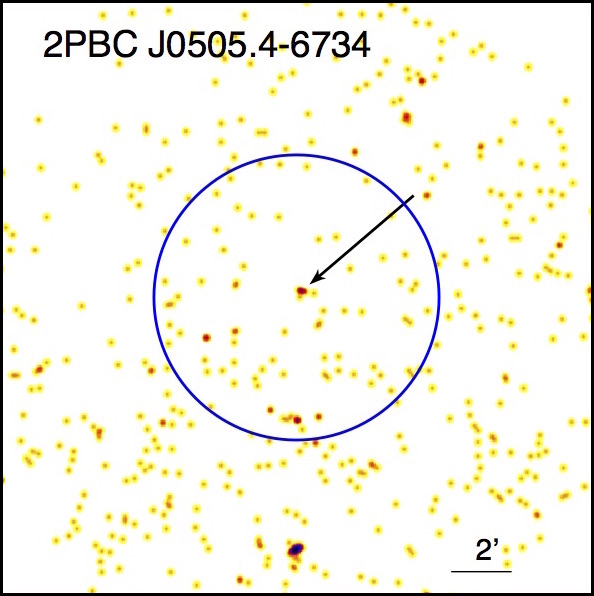}}
\caption{Soft X-ray images (0.3 - 10 keV) of the fields of \textit{2PBC} hard X-ray sources selected in this paper for optical spectroscopic follow-up. The object name is indicated in each panel. The BAT error circle is shown in blue and the new soft X-ray counterpart identifications for which the optical spectra were acquired are indicated with black arrows. In all cases, the objects selected as soft X-ray counterparts are the only ones with emission above 3 keV or with emission at 3$\sigma$ considering the entire band of 0.3-10 keV. Field sizes are 20$'$ x 20$'$. Nearly all images were extracted from the ASI-SDC XRT database. In all cases, north is up and east to the left.}
\end{figure*}


\begin{figure*}[htbp]
\centering
\subfigure{\includegraphics[width=0.32\textwidth]{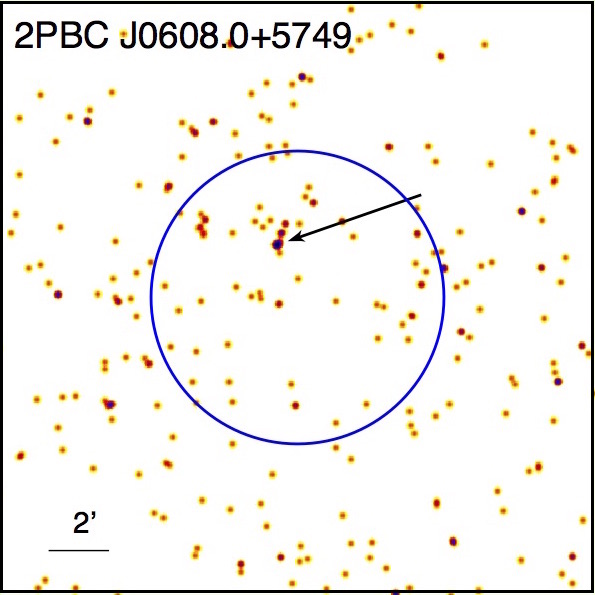}}
\subfigure{\includegraphics[width=0.32\textwidth]{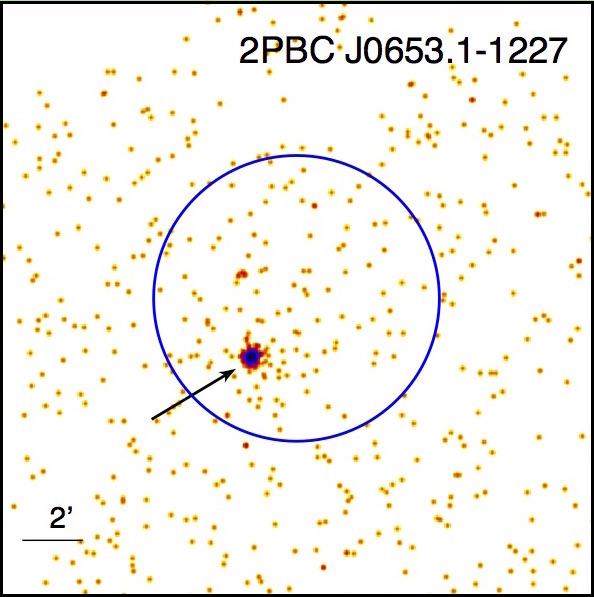}}
\subfigure{\includegraphics[width=0.32\textwidth]{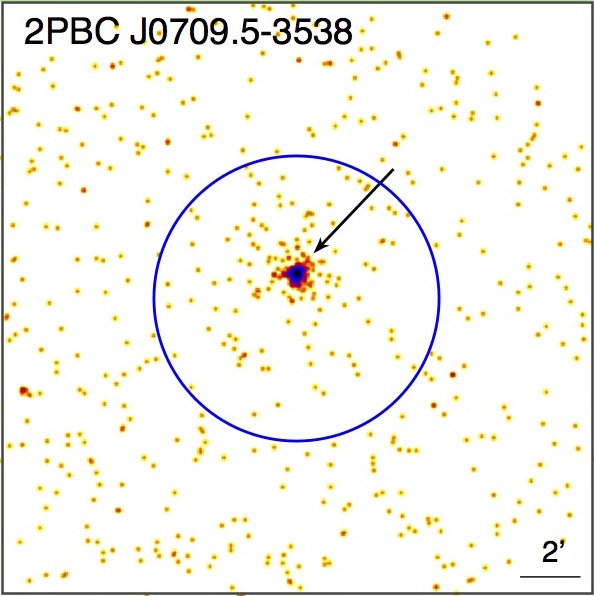}}
\subfigure{\includegraphics[width=0.32\textwidth]{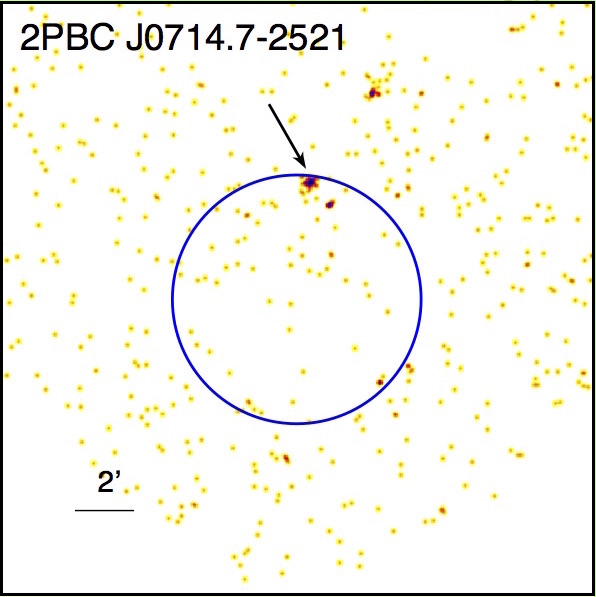}}
\subfigure{\includegraphics[width=0.32\textwidth]{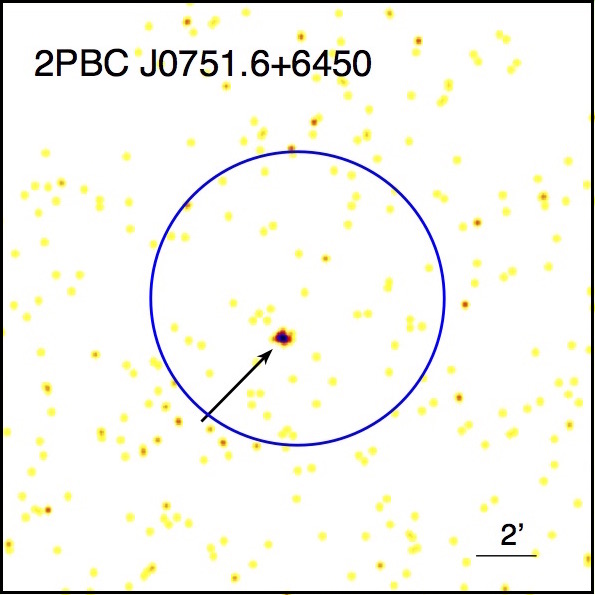}}
\subfigure{\includegraphics[width=0.32\textwidth]{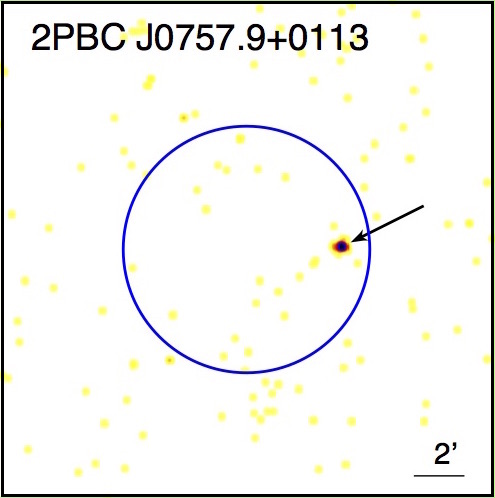}}
\subfigure{\includegraphics[width=0.32\textwidth]{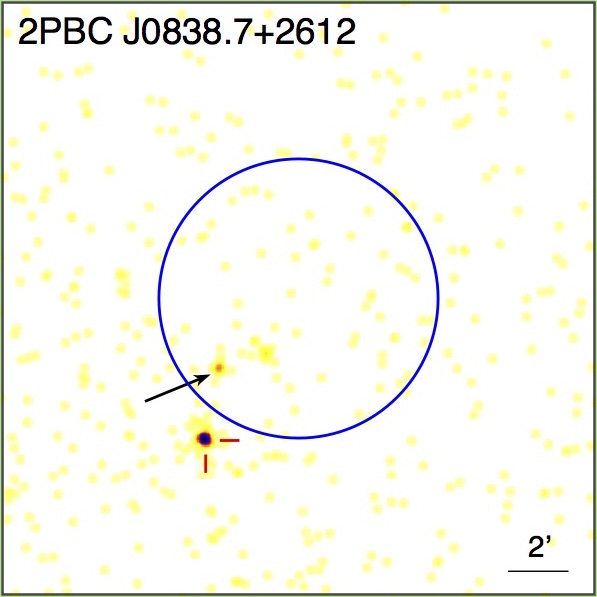}}
\subfigure{\includegraphics[width=0.32\textwidth]{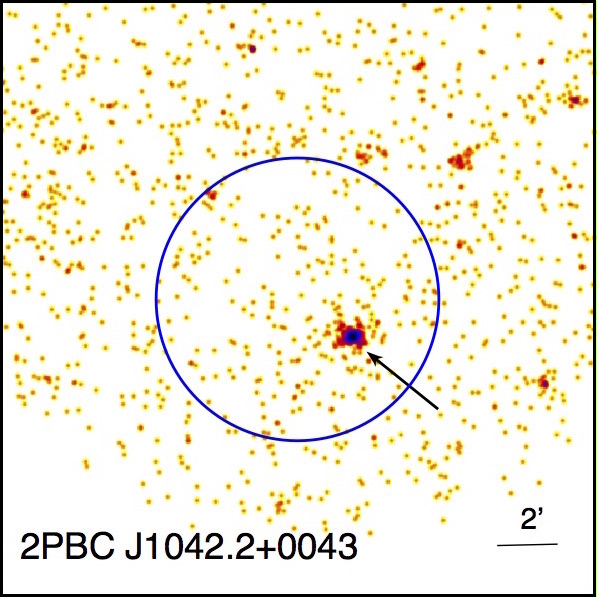}}
\subfigure{\includegraphics[width=0.32\textwidth]{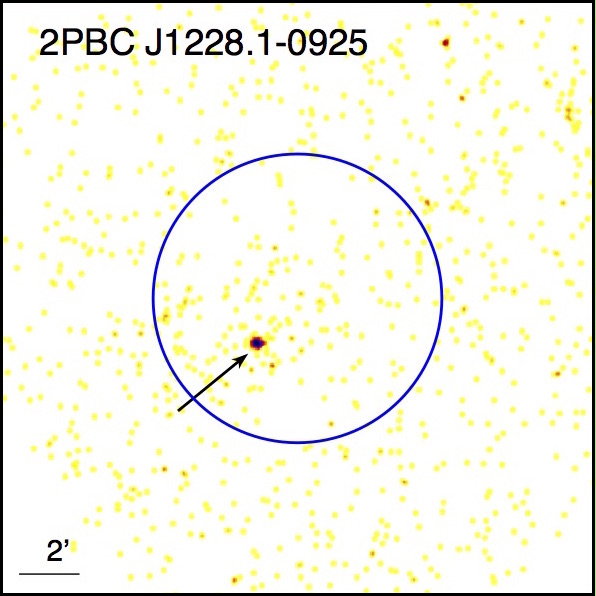}}
\subfigure{\includegraphics[width=0.32\textwidth]{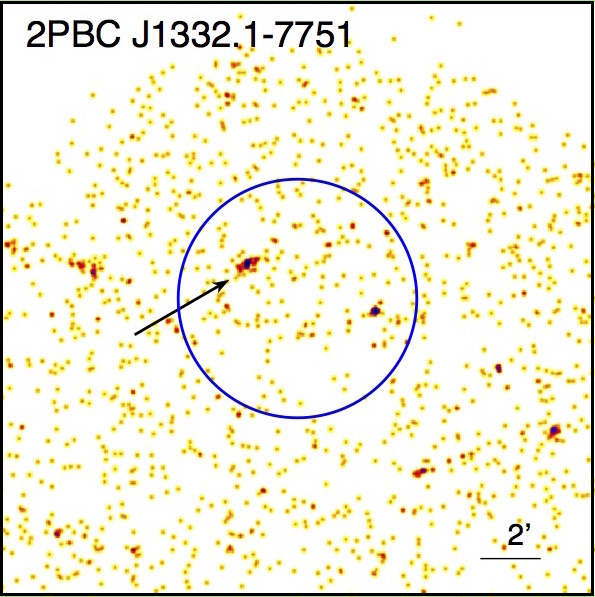}}
\subfigure{\includegraphics[width=0.32\textwidth]{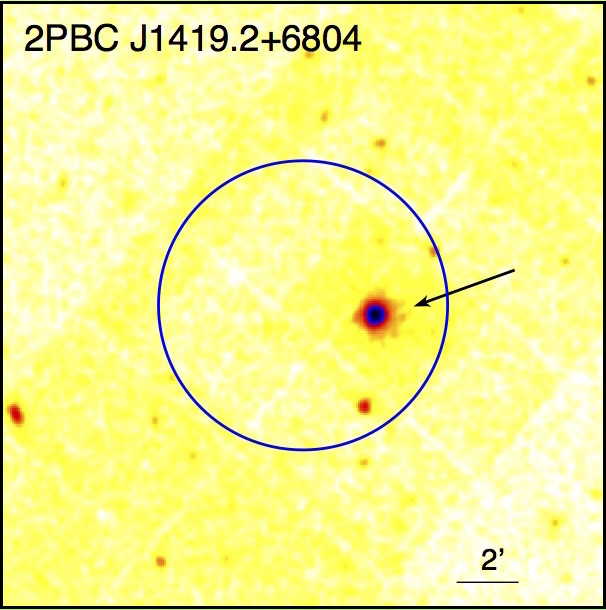}}
\subfigure{\includegraphics[width=0.32\textwidth]{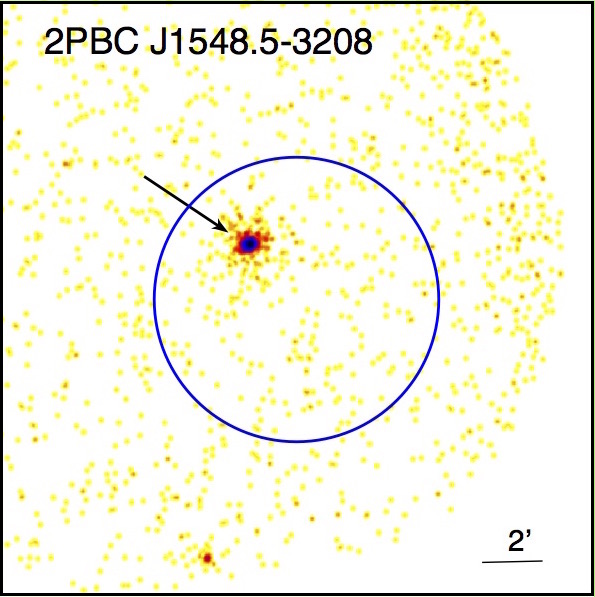}}
\caption{Continued Fig. A.1. The soft X-ray image of the field of 2PBC J1419.2$+$6804 was obtained from an XMM-Newton observation downloaded from the HEASARC database.}
\end{figure*}


\begin{figure*}[htbp]
\centering
\subfigure{\includegraphics[width=0.32\textwidth]{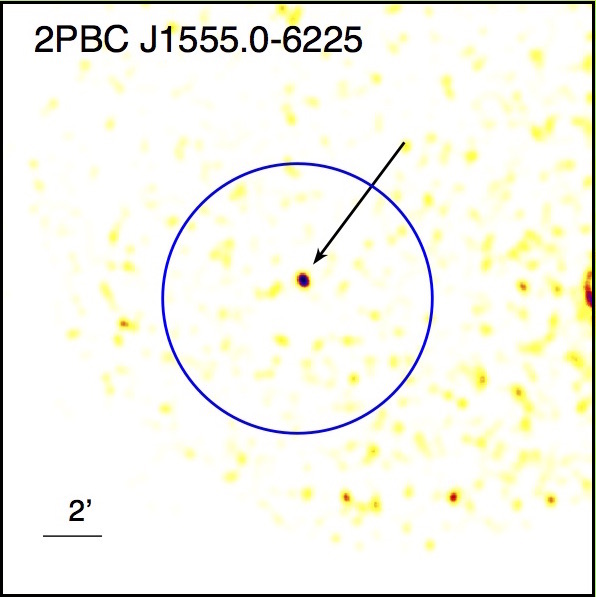}}
\subfigure{\includegraphics[width=0.32\textwidth]{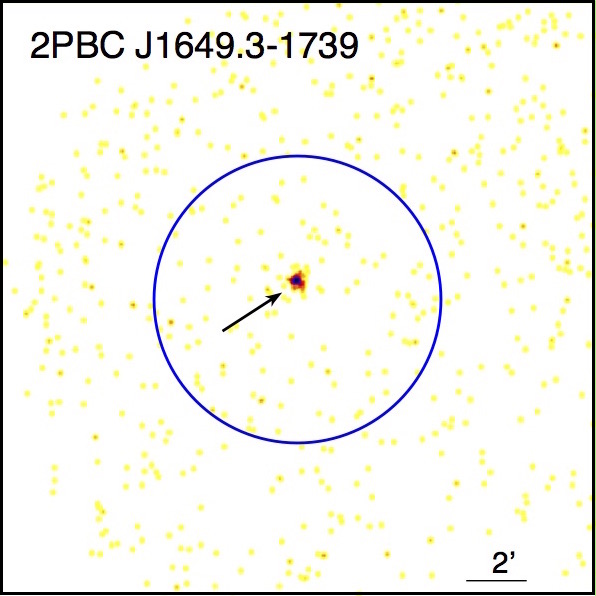}}
\subfigure{\includegraphics[width=0.32\textwidth]{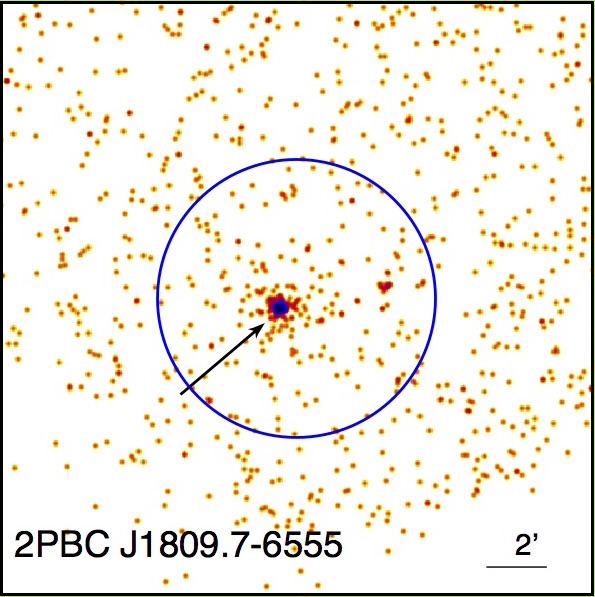}}
\subfigure{\includegraphics[width=0.32\textwidth]{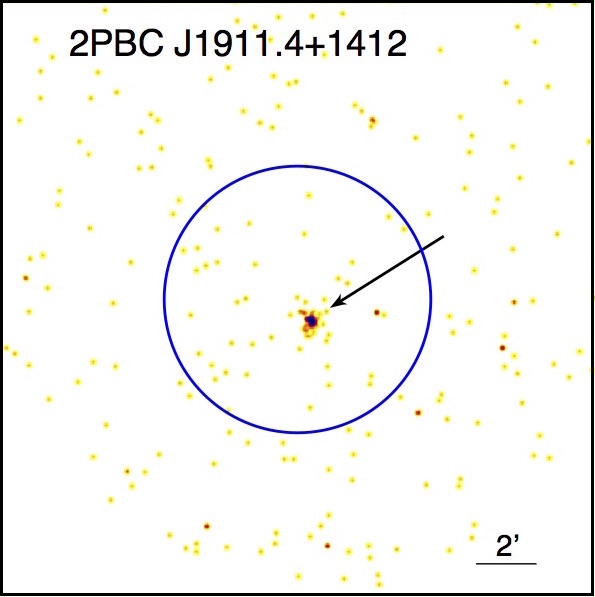}}
\subfigure{\includegraphics[width=0.32\textwidth]{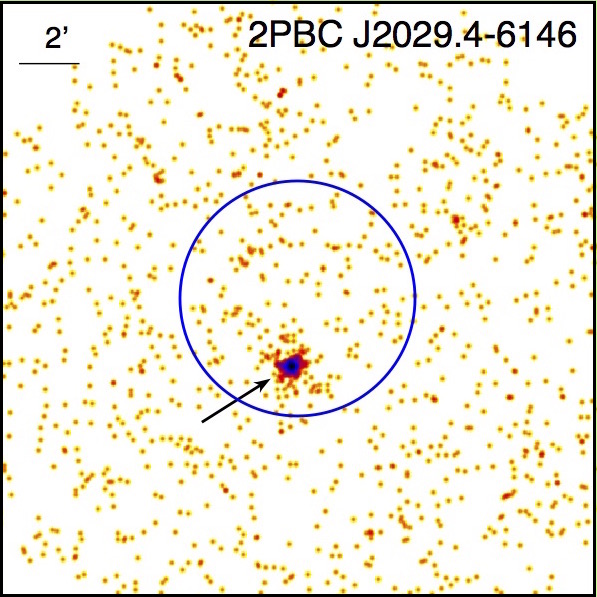}}
\subfigure{\includegraphics[width=0.32\textwidth]{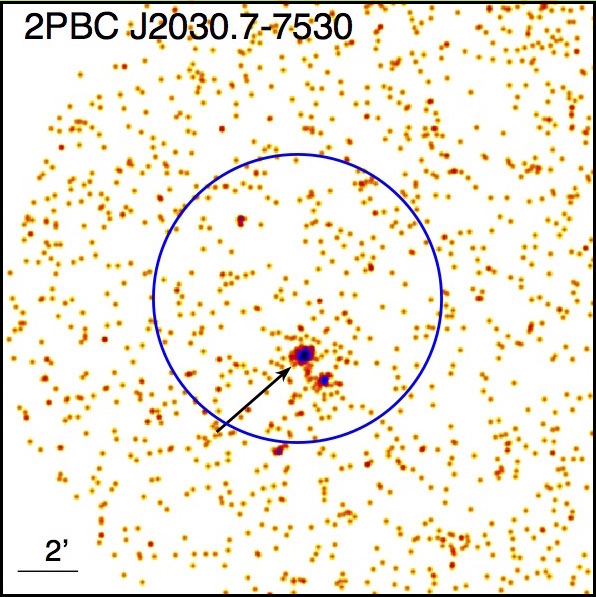}}
\subfigure{\includegraphics[width=0.32\textwidth]{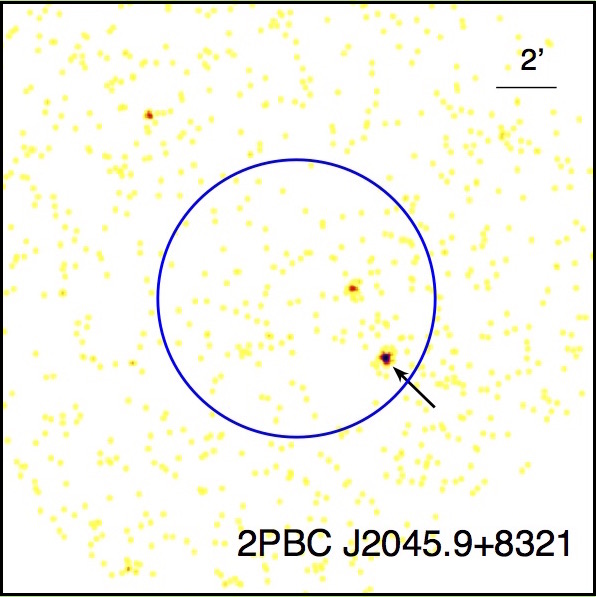}}
\subfigure{\includegraphics[width=0.32\textwidth]{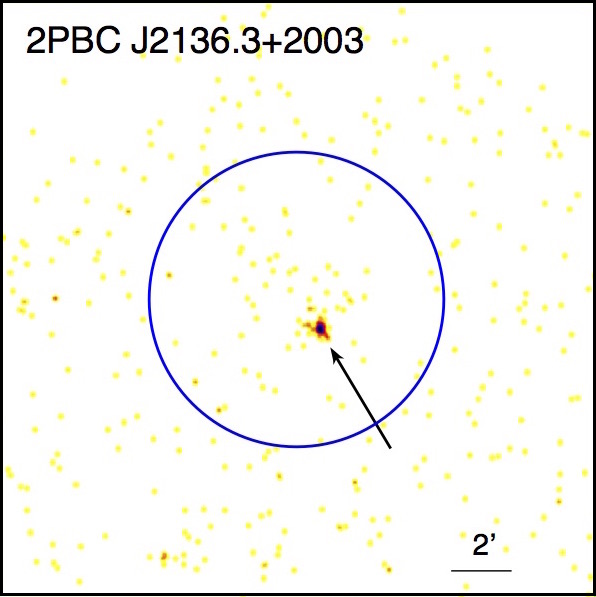}}
\subfigure{\includegraphics[width=0.32\textwidth]{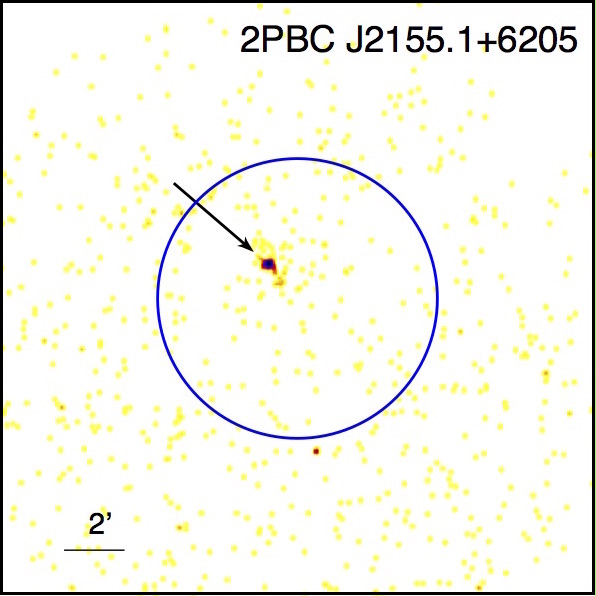}}
\subfigure{\includegraphics[width=0.32\textwidth]{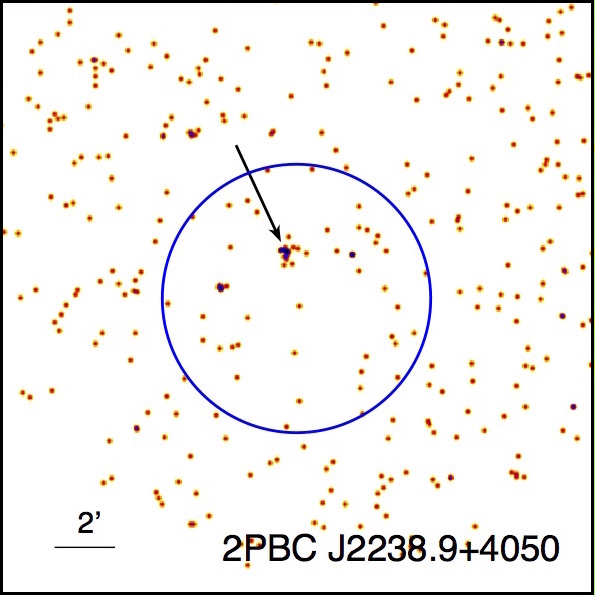}}
\subfigure{\includegraphics[width=0.32\textwidth]{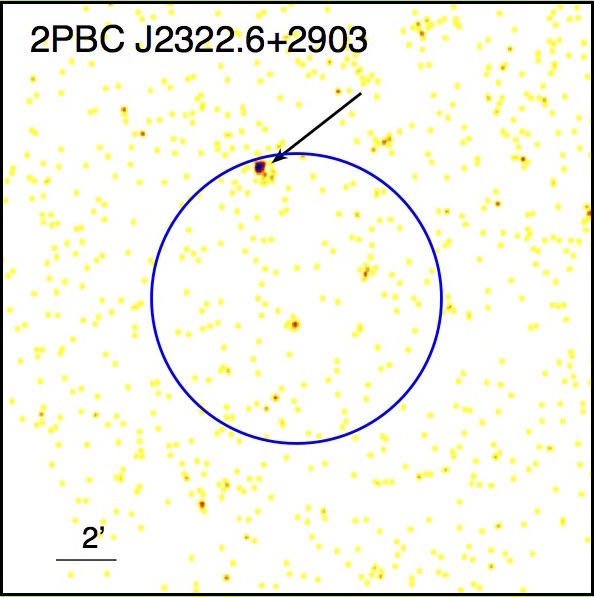}}
\subfigure{\includegraphics[width=0.32\textwidth]{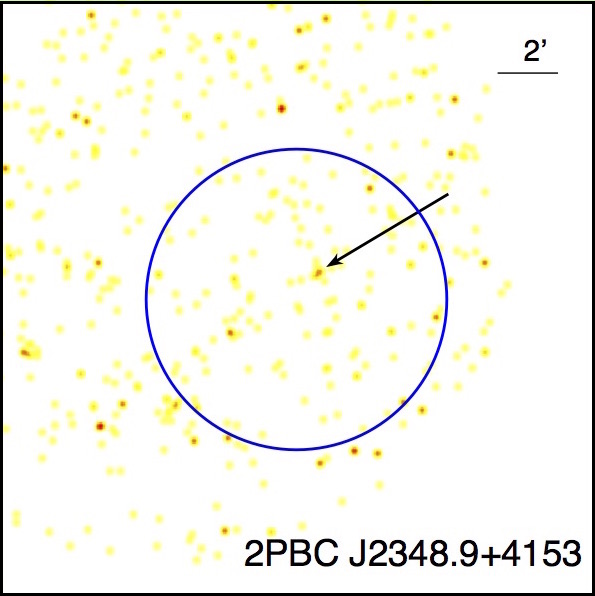}}
\caption{Continued Fig. A.1.} 
\end{figure*}

\newpage

\begin{figure*}[htbp]
\centering
\subfigure{\includegraphics[width=0.25\textwidth]{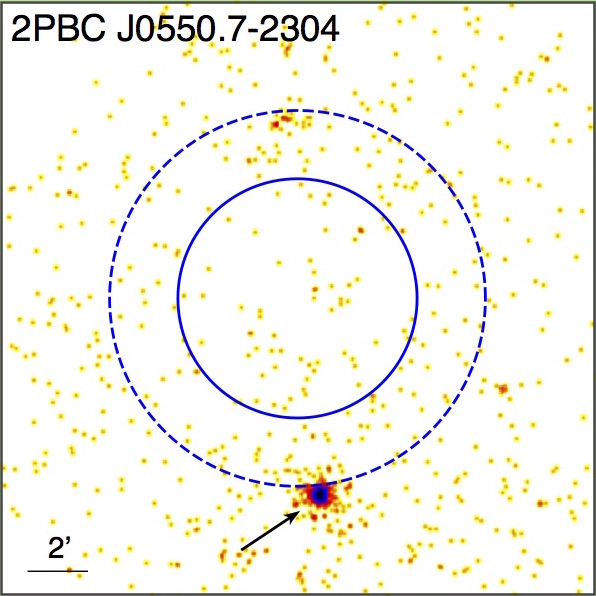}}
\subfigure{\includegraphics[width=0.25\textwidth]{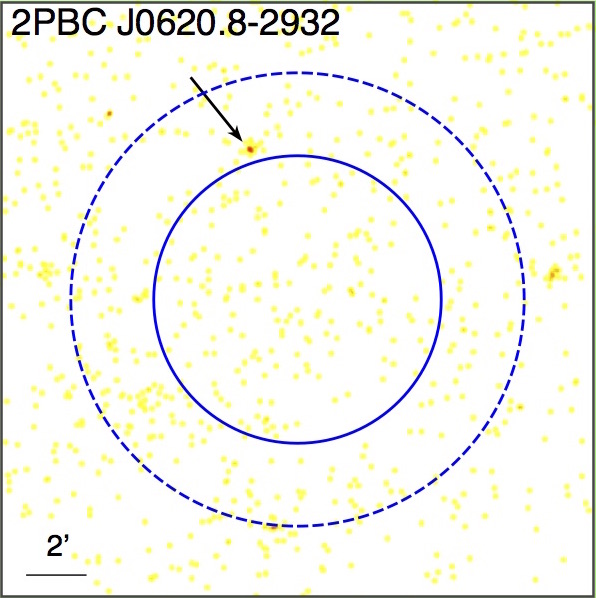}}
\subfigure{\includegraphics[width=0.25\textwidth]{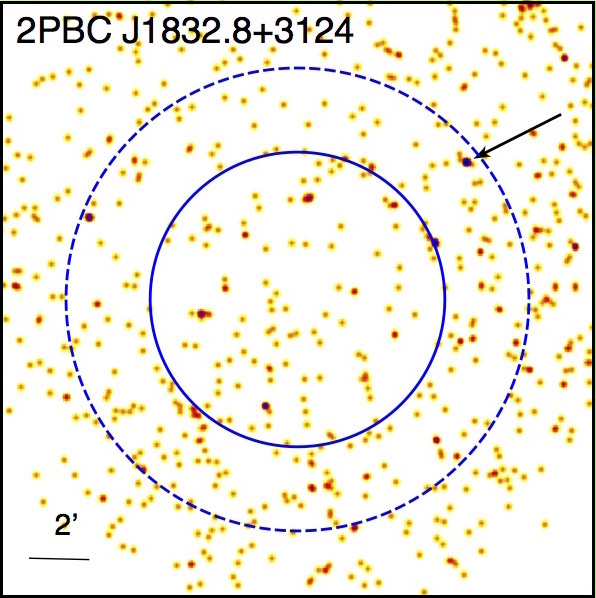}}
\caption{XRT soft X-ray images (0.3 - 10 keV) of the fields of \textit{2PBC} sources for which the corresponding soft X-ray counterpart is outside the 90\% error BAT circle (continour circle, in blue) and inside or at the edge of the 99\% BAT error circle, indicated with dashed blue line. The counterpart identifications for which the optical spectra were acquired are indicated with arrows. Field sizes are 20$'$ x 20$'$. In all cases, north is up and east to the left.}
\end{figure*}

\begin{figure}[htbp]
\begin{center}
\includegraphics[width=0.27\textwidth]{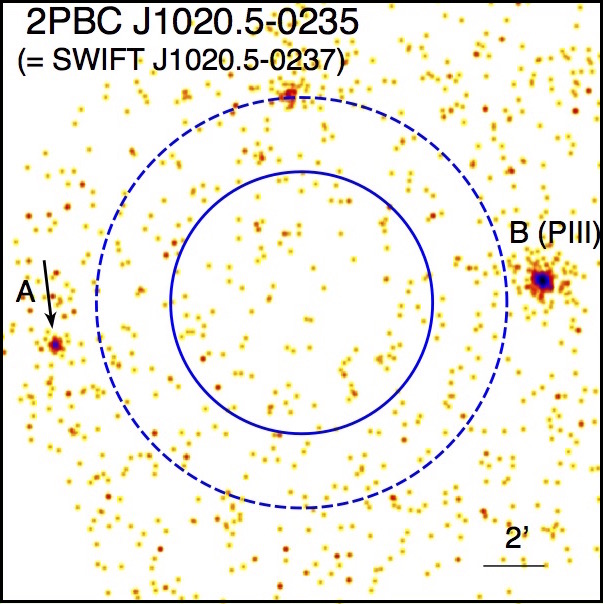}
\begin{spacing}{1}
\caption[0.3\textwidth]{XRT soft X-ray observation of the field of 2PBC J1020.5$-$0235. The 90\% BAT error circle is shown as a continuous blue line while the 99\% error circle is indicated with the dashed blue line. The object for which optical spectroscopy was acquired is indicated with an arrow. Information on object B is presented in PIII. Field size is 20$'$ x 20$'$. North is up and east to the left.}
\end{spacing}
\end{center}
\end{figure}

\begin{figure}[htbp]
\centering
\subfigure{\includegraphics[width=0.27\textwidth]{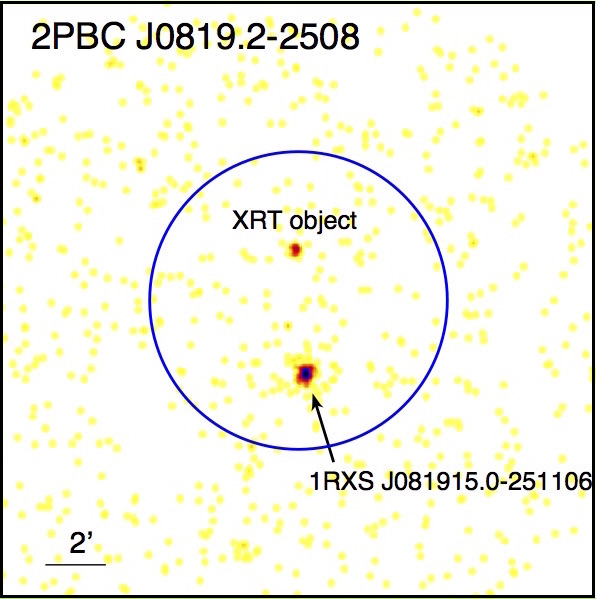}}
\subfigure{\includegraphics[width=0.27\textwidth]{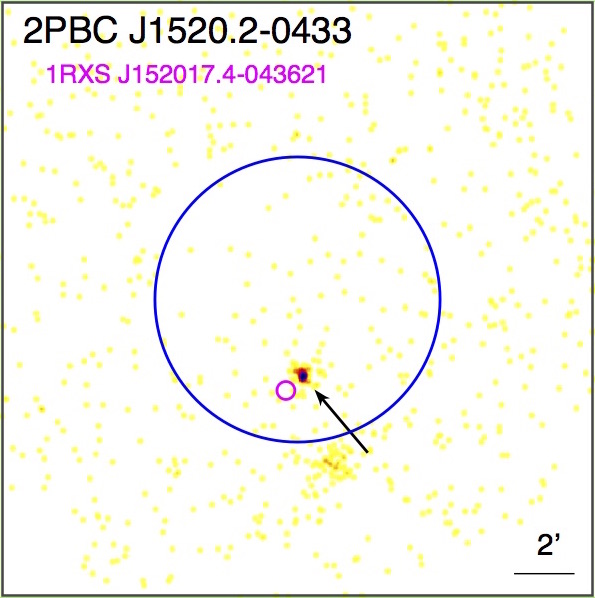}}
\caption{Soft X-ray observations (0.3 - 10 keV) of the field of two \textit{2PBC} sources for which the positions of the soft X-ray counterparts proposed in this work do not coincide with those reported in the Cusumano et al. (2010) and/or Baumgartner et al. (2013) catalogues (see text). The 90\% BAT error circle is shown in blue and our proposed soft X-ray counterpart identifications, for which the optical spectra were acquired, are indicated with arrows. Field sizes are 20$'$ x 20$'$ . In all cases, north is up and east to the left.}
\end{figure}

\newpage
\appendix
\section*{Appendix B: Extra X-ray emitting object in the field of 2PBC J0838.7$+$2612}

\setcounter{table}{0}
\renewcommand{\thetable}{B.\arabic{table}}

\setcounter{figure}{0}
\renewcommand{\thefigure}{B.\arabic{figure}}

As reported in the main text, we note that in the field of 2PBC J0838.7$+$2612 there is another object inside the 99\% BAT error circle (labeled with red tick marks in Fig. A.5 of Appendix A). This object has emission above 3 keV and is actually positionally coincident with two optical sources. \\

The spectra of these two objects were acquired on March 27, 2014 (UT start times at 22:29 and 22:49; exposure time of 2$\times$1200 seconds) with the TNG Telescope in La Palma, Spain and are presented in Fig. B.1; the information about the main optical properties of the LINER is reported in Table B.1.

We found that the pair of objects is composed of two galaxies at the same redshift z = 0.048 $\pm$ 0.001 (D$_{L}$ = 229.7 Mpc). One of them is a LINER and the other one is a normal galaxy; thus, we conclude that the LINER can be the main responsible for the soft X-ray emission. We thus cannot exclude that the hard X-ray emission detected by BAT and labeled as 2PBC J0838.7$+$2612 in Cusumano et al. (2010) can partly be due to the LINER in this galaxy pair, albeit it nominally lies outside the 90\% error circle of the BAT source.

\begin{table}[htbp!]
\caption[]{Main results for the two additional X-ray emission objects observed in the field of 2PBC J0838.7$+$2612.}
\scriptsize
\setlength{\tabcolsep}{6pt} 
\begin{center}
\begin{tabular}{lcccccrccl}
\noalign{\smallskip}
\hline
\hline
\noalign{\smallskip}
\multicolumn{1}{c}{Optical Coordinates} & $F_{\rm H_\alpha}$ & $F_{\rm H_\beta}$ & $F_{\rm [OIII]}$ & Class & $z$ &
\multicolumn{1}{c}{$D_L$} & \multicolumn{2}{c}{$E(B-V)$} &
\multicolumn{1}{c}{$L_{\rm X}$} \\
\cline{8-9}
\noalign{\smallskip}
(J2000) & & & & & & (Mpc) & Gal. & AGN & \\
\noalign{\smallskip}
\hline
\noalign{\smallskip}

Ra = 08:38:59.29 &  20.4$\pm$3 & 1.43$\pm$0.15   &  2.91$\pm$0.12  &  LINER  &  0.048  &  229.7  &  0.047  & 1.68  &  1.74 ( 0.3-10)$^{X}$ \\ 
Dec = +26:08:13.1 & [22.5$\pm$4] & [1.65$\pm$0.12] & [3.52$\pm$0.3] & & & & &  &     \\ 

 \noalign{\smallskip} 
\hline
\noalign{\smallskip} 
\multicolumn{10}{l}{Note: Optical coordinates were extracted from 2MASS catalogue, with an accuracy better than $\sim$ 0".1.} \\
\multicolumn{10}{l}{Emission-line fluxes are reported both as 
observed and (between square brackets) corrected for the intervening Galactic}  \\
\multicolumn{10}{l}{absorption $E(B-V)_{\rm Gal}$ along the object line of sight 
(from Schlegel et al. 1998). Line fluxes are in units of 10$^{-15}$ \textit{erg cm$^{-2}$ s$^{-1}$},} \\
\multicolumn{10}{l}{X--ray luminosities are in units of 10$^{43}$ erg s$^{-1}$,
and the reference band (between round brackets) is expressed in keV.} \\ 
\multicolumn{10}{l}{In the last column, $^{X}$ indicates that the luminosity was calculated using the corresponding X--ray flux measured from $\it{Swift}$/XRT } \\ 
\multicolumn{10}{l}{counts rate assuming a power law model with photon index of $\Gamma$ = 1.8. The typical error of the redshift measurement is $\pm$0.001.} \\
\noalign{\smallskip} 
\hline
\hline
\end{tabular}
\end{center}
\end{table}

\begin{figure}[htbp!]
\centering
\subfigure{\includegraphics[width=0.4\textwidth]{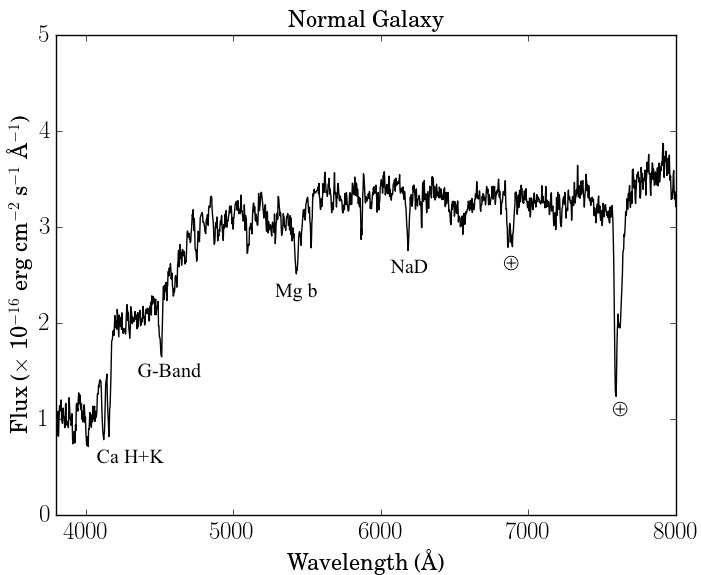}}
\subfigure{\includegraphics[width=0.4\textwidth]{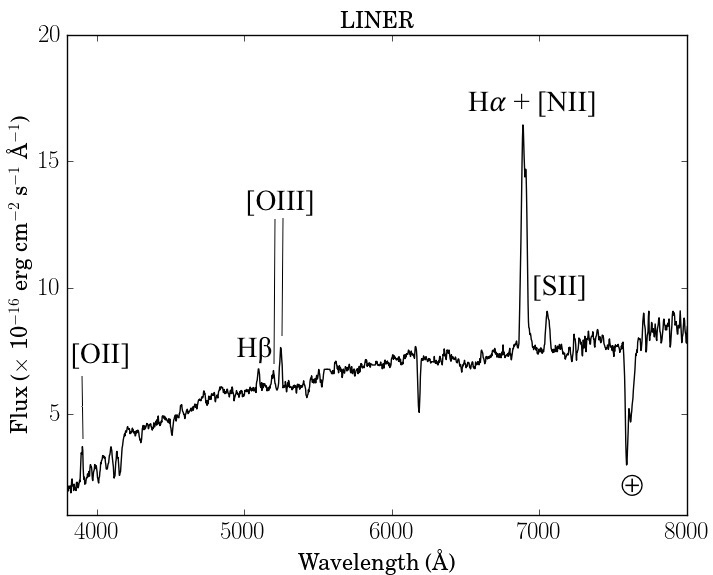}}
\caption{Optical spectra of the putative X-ray emitting objects in the field of 2PBC J0838.7$+$2612, both at the same redshift z = 0.048.}
\end{figure}

\newpage
\appendix
\section*{Appendix C: Information of the 54-month Palermo \textit{Swift}-BAT catalogue from the literature and the Simbad database}
\setcounter{table}{0}
\renewcommand{\thetable}{C.\arabic{table}}

\begin{table}[htbp!]
\caption[]{Additional classification of 38 objects listed in the Cusumano et al. (2010).}
\scriptsize
\setlength{\tabcolsep}{6pt} 
\begin{center}
\begin{tabular}{lcccccrccl}
\noalign{\smallskip}
\hline
\hline
\noalign{\smallskip}
\multicolumn{1}{c}{Object ID} & Class & Information \\
\noalign{\smallskip}
\hline
\noalign{\smallskip}
J0048.7$+$3157   & Sy2    &   	Simbad	 \\	
J0207.0$+$2929   & Sy1      & 		Simbad	 \\	
J0250.3$+$4645   & Sy2  & 	MX	 \\	
J0325.1$+$4042   & Sy2   &		MIX	 \\	
J0328.7$-$2843   & Sy2    	  & 	Buchanan et al. (2006)	 \\	 
J0359.5$+$5058   & QSO     &  	Agudo et al. (2007)		 \\	
J0547.3$+$5040   & Sy2   	&	MX	 \\	
J0600.7$+$0008   & QSO     & 		Simbad	 \\	
J0625.1$+$6450   & Sy1  	  & 		Simbad	 \\	
J0712.1$+$1540   & QSO  	  & 	Simbad	 \\	
J0823.0$-$0454   & Sy2     & 	Landi et al. (2007)	 \\	 
J0902.6$-$4813   & Sy1    &   	Zurita Heras et al. (2009)	 \\	 
J1105.8$+$5854  &  Sy1    &   Simbad	 \\	
J1207.6$+$3353   & Sy2    	  & 	Landt et al. (2010)	 \\	
J1213.0$+$3239   & QSO      & 		Simbad	 \\	
J1346.5$+$1923   & Sy1.2    & 	MIX	 \\	
J1414.2$+$1219   & QSO  	  & 	Simbad	 \\		
J1745.7$+$2907   & Sy1      & 		Burenin et al. (2008)	 \\	
J1826.0$-$0708   & Sy1    &   	Burenin et al. (2009)	 \\	
J2021.7$+$4359   & Sy2    & 	Bikmaev et al. (2008)	 \\	
\hline 
J0207.9$-$7425   & HXB      & 		Kahabka \& Hilker  (2005)	 \\	
J0451.1$-$6948   & HXB   	  &   Klus et al. (2013)	 \\	
J0537.7$+$2106   & Star    & 	Simbad	 \\	
J0845.3$-$5227   & Active star   & 	Simbad  	 \\	
J1320.3$-$7013   & Star    & 	 Riaz et al. (2006)	 \\	
J1400.7$-$6323   & SNR  	  & 		Renaud et al. (2010)	 \\	
J1632.7$-$4727  &  HXB 	  & 	 Coleiro et al. (2013)	 \\	
J1711.8$-$3933  &  SNR 	  & 		Horan \& Weekes et al. (2004)	 \\	
J1735.4$-$3256   & HXB 	  & 	 Coleiro et al. (2013)	 \\	
J1747.2$-$2721   & LXB  	  & 	Altamirano et al. (2010)		 \\	
J1758.5$-$2122   & HXB  	  & 	Coleiro et al. (2013)	 \\	 
J1810.3$-$1906   & LXB    &   Markwardt et al. (2008)	 \\	
J1821.4$-$1318   & HXB  	  & 	Butler et al. (2009)	 \\	
J1848.8$-$0002  &  SNR  	  & 	Gotthelf et al. (2011)	 \\	
J1929.9$+$1818  &  HXB   	  & 	Rodriguez et al. (2009)	 \\	
J2218.4$+$1925   & CV      & 	Thorstensen et al. (2013)	 \\	 
J2337.7$+$4309   & CV   	  & 	Simbad	 \\	
J2341.0$+$7642   & CV  	  & 	Lutovinov et al. (2012)	 \\	 
 \noalign{\smallskip} 
\hline
\noalign{\smallskip} 
\multicolumn{3}{l}{Classification of sources listed in the 54-month Palermo \textit{Swift}-BAT catalogue performed by } \\
\multicolumn{3}{l}{other authors and reported in the literature and/or Simbad database in Figure 11 (grey bar).} \\
\multicolumn{3}{l}{The extragalactic objects are shown in the upper part of the table, whereas the 18 galactic objects in the lower part.} \\
\noalign{\smallskip} 
\hline
\hline
\end{tabular}
\end{center}
\end{table}

\end{document}